\documentclass[aps,prb,twocolumn,superscriptaddress,citeautoscript,amsmath,amssymb]{revtex4-2}
\usepackage{CJK}
\usepackage{setspace}
\usepackage{multibib}
\usepackage{color}
\usepackage{hyperref}
\usepackage{booktabs}
\usepackage{blindtext}
\usepackage{xr}
\usepackage{indentfirst}
\usepackage{multirow}
\usepackage{chemformula} 
\usepackage{orcidlink}
\newcommand\ce[1]{\ensuremath{\mathrm{#1}}}
\newcommand*{\nolink}[1]{%
  {#1}%
}

\begin{document}

\setcitestyle{super} 
\begin{CJK*}{UTF8}{gbsn}

\title{Effects of Grain Boundaries and Surfaces \\ on Electronic and Mechanical Properties of Solid Electrolytes}

\author{Weihang Xie (谢维航)~\orcidlink{0000-0002-6498-2328}}
\affiliation{Department of Materials Science and Engineering, National University of Singapore, 9 Engineering Drive 1, 117575, Singapore}
\author{Zeyu Deng (邓泽宇)~\orcidlink{0000-0003-0109-9367}}
\affiliation{Department of Materials Science and Engineering, National University of Singapore, 9 Engineering Drive 1, 117575, Singapore}
\author{Zhengyu Liu (刘政宇)}
\affiliation{Department of Materials Science and Engineering, National University of Singapore, 9 Engineering Drive 1, 117575, Singapore}
\author{Theodosios Famprikis~\orcidlink{0000-0002-7946-1445}}
\affiliation{Faculty of Applied Sciences, Delft University of Technology, 2628 Delft, The Netherlands}

\author{Keith T. Butler~\orcidlink{0000-0001-5432-5597}} 
\affiliation{Department of Chemistry, Kathleen Lonsdale Building, London WC1E 6BT, United Kingdom}

\author{Pieremanuele Canepa~\orcidlink{0000-0002-5168-9253}}
\email{pcanepa@uh.edu}
\affiliation{Department of Electrical and Computer Engineering, University of Houston, Houston, TX 77204, USA}
\affiliation{Texas Center for Superconductivity, University of Houston, Houston, TX, 77204, USA}
\affiliation{Department of Materials Science and Engineering, National University of Singapore, 9 Engineering Drive 1, 117575, Singapore}

\begin{abstract}
Extended defects, including exposed surfaces and grain boundaries, are critical to the properties of polycrystalline solid electrolytes in all-solid-state batteries (ASSBs). 
These defects can significantly alter the mechanical and electronic properties of solid electrolytes, with direct manifestations on the performance of ASSBs. 
Here, by building a library of 590 surfaces and grain boundaries of 11 relevant solid electrolytes ---including halides, oxides, and sulfides--- their electronic, mechanical, and thermodynamic characteristics are linked to the functional properties of polycrystalline solid electrolytes.  
It is found that the energy required to mechanically ``separate'' grain boundaries can be significantly lower than in the bulk region of materials, which can trigger preferential cracking of solid electrolyte particles in the grain boundary regions. 
The brittleness of ceramic solid electrolytes, inferred from the predicted low fracture toughnesses at the grain boundaries, contributes to their cracking under local pressure imparted by Lithium or Sodium penetration in the grain boundaries. 
Extended defects of solid electrolytes introduce new electronic ``interfacial'' states within bandgaps of solid electrolytes. 
These interfacial states alter and possibly increase locally the availability of free electrons and holes in solid electrolytes. Factoring effects arising from extended defects appear crucial to explain electrochemical and --mechanical observations in ASSBs. 

\end{abstract}


\maketitle

\end{CJK*}
\section{Introduction}
\label{sec:intro}
\noindent With the widespread adoption of rechargeable batteries in electric vehicles, laptops, and mobile phones,  energy storage devices with high energy and power densities are currently needed.\cite{vannoordenRechargeableRevolutionBetter2014,liuPathwaysPracticalHighenergy2019,deysherTransportMechanicalAspects2022}  
For example, in ASSBs, the energy density can be potentially increased by replacing graphite with a metallic anode, such as lithium (Li) or sodium (Na).\cite{albertusStatusChallengesEnabling2017}  
Moreover, in ASSBs, the safety of the electrochemical cell is increased by replacing the flammable liquid electrolyte with non-flammable solid electrolytes (SEs).\cite{goodenoughChallengesRechargeableLi2010,famprikisFundamentalsInorganicSolidstate2019,bachmanInorganicSolidStateElectrolytes2016}

In the context of ASSBs, considering that the bulk modulus of ceramic SEs is typically at least twice that of metals (Li or Na), it was postulated that Li or Na metal penetration can be suppressed by sufficiently stiff ceramic SEs.\cite{wolfenstineMechanicalBehaviorLiionconducting2018,famprikisFundamentalsInorganicSolidstate2019,monroeImpactElasticDeformation2005} 
However, this criterion is not strictly applicable to polycrystalline SEs where Li filament propagation is still observed.\cite{kasemchainanCriticalStrippingCurrent2019,hanHighElectronicConductivity2019,kazyakLiPenetrationCeramic2020} In this work, the metal deposition inside SEs will be referred to as Li ingress/penetration or Li filament.
Furthermore, the critical current density ---the largest  current density before Li ingress is triggered--- of SEs are still not satisfactory compared with commercial liquid electrolytes.\cite{qianHighRateStable2015,garcia-mendezEffectProcessingConditions2017,hanSuppressingLiDendrite2018,porzMechanismLithiumMetal2017,hanHighElectronicConductivity2019,takadaRecentProgressInterfacial2015}

Extended defects, such as exposed surfaces created by microcracks or pores, as well as grain boundaries (GBs), are ubiquitous in ceramic polycrystalline SEs.\cite{ricePoresFractureOrigins1984,mecholskyFractureSurfaceAnalysis1976,famprikisFundamentalsInorganicSolidstate2019,kimSolidStateLi2021,milanRoleGrainBoundaries2023}  
Therefore, understanding complex phenomena involving surfaces and interfaces in ASSBs appears crucial. 
Nevertheless, the buried nature of GBs and surfaces in fully assembled ASSBs complicate enormously their investigations, requiring the implementation of specialized tools that can provide spatially resolved information on interfacial phenomena, which are often masked by overwhelmingly strong signals from the bulk and the other battery components.\cite{tanTailoringUniformOrdered2021,vahidiReviewGrainBoundary2021,huCarrierGrainBoundary2022,jalemLithiumDynamicsGrain2023}

First-principles models as a base of multiscale models can provide an informed assessment of extended defects in SEs mimicking the conditions of ASSBs. For example, modeling elucidates the causal link between the observed electronic conductivity and the availability of GBs, interfaces, and exposed surfaces in SEs.\cite{dengAutonomousHighthroughputMultiscale2022,butlerDesigningInterfacesEnergy2019,dawsonGoingGrainAtomistic2023}

Microcracks in SEs can lead to a concentration of stresses, while GBs are believed to be the source of initial cracks due to their low fracture toughness compared to their respective bulks.\cite{monismithGrainboundaryFractureMechanisms2022,ningDendriteInitiationPropagation2023} 
For example, Vishnugopi \emph{et al.} \cite{vishnugopiMesoscaleInterrogationReveals2022} revealed that a steep increase of local strain near GBs can cause mechanical failure of SEs. 
Using a phase-field model, Yuan \emph{et al.}\cite{yuanCoupledCrackPropagation2021} investigated Li penetration in the SE \ce{Li_7La_3Zr_2O_{12}} (LLZO) and claimed that the mechanical forces (from stacking pressure) and the driving forces of mechanical and electrochemical origin may promote Li penetration into GB  or pre-existing cracks in the SE. 
Ning \emph{et al.}\cite{ningVisualizingPlatinginducedCracking2021} claimed to observe the propagation of Li filament in argyrodite, suggesting that the ingress of \ce{Li} into the available pores may widen and propagate cracks in the SE.

The unwanted problematic nucleation and growth of Li(Na) filament have been tentatively linked to changes in electronic leakage in SEs, which could be caused by inhomogeneous electronic conductivity in SEs' microstructure, \emph{e.g.} changes in electronic conductivity from bulk to grain boundaries and exposed surfaces.\cite{tianInterfacialElectronicProperties2019,zhuUnderstandingEvolutionLithium2023,porzMechanismLithiumMetal2017,kazyakLiPenetrationCeramic2020,dawsonGoingGrainAtomistic2023}
At the electronic structure level of SE materials, the availability of GBs or exposed surfaces may introduce interfacial states contributing to local changes of electronic conductivity.\cite{fuchsCurrentDependentLithium2023,tianInterfacialElectronicProperties2019,gaoSurfaceDependentStabilityInterface2020,quirkDesignPrinciplesGrain2023,dawsonGoingGrainAtomistic2023} 
The undesired electronic conductivity may cause Li/Na ion in SE to be reduced to metal Li/Na, leading to local metal nucleation, and subsequent growth and propagation.\cite{kasemchainanCriticalStrippingCurrent2019,songProbingOriginElectronic2019,hanHighElectronicConductivity2019,porzMechanismLithiumMetal2017} 
For example, Zhu \emph{et al.} \cite{zhuUnderstandingEvolutionLithium2023} claimed to have observed the growth of Li filament in the GB regions of Aluminum (Al)--stabilized LLZO, which they explained by an increase of the electron concentration gradient in the GB regions which, in turn, could ``attract'' Li-ions nucleating Li-metal filaments in these regions. 
Through scanning electron microscopy, energy-dispersive X-ray spectroscopy, and Auger spectroscopy, Cheng \emph{et al.}\cite{chengIntergranularLiMetal2017} claimed to directly observe preferential deposition of Li metal of GBs in Al-doped LLZO.
Using \emph{in situ} electron energy loss spectroscopy  Liu \emph{et al.} \cite{liuLocalElectronicStructure2021} studied the GB region of LLZO, and observed a reduction of the electronic bandgap around the GB region to $\sim$1~eV. 
The measured bandgap appears as low as  16\% of that of LLZO bulk,\cite{liuLocalElectronicStructure2021} suggesting that Li may reduce around the more electronically conductive GBs and even initiate nucleation of Li filament.\cite{porzMechanismLithiumMetal2017} 
Han \emph{et al.}\cite{hanHighElectronicConductivity2019} suggested that changes in the electronic conductivity of LLZO are the reason leading to metal filament formation, and pointed out the importance of investigating extended defects as possible origins for the high electronic conductivity.

Therefore, understanding the effects of concentrations of extended defects,  such as GBs and surfaces in SEs,  on their macroscopic properties appears crucial to ({\emph{i}}) address the mechanical stability of polycrystalline SEs, ({\emph{ii}}) understand the root-cause of electronic conductivity in SE materials,\cite{kalnausSolidstateBatteriesCritical2023} and (\emph{iii}) explain how these extended defects impact the nucleation and growth of Li(Na) filaments. 
To address the effects of extended defects on the mechanical and electronic properties of polycrystalline SEs, we build accurate surface and GB models of topical SEs, including oxides, sulfides, phosphates, and halides. 
These models provide insights into the variation of mechanical and electronic properties introduced by the interfacial regions of SEs.
    
From our comprehensive library of 590 surface and GB models of SEs, we conclude that the energy to mechanically ``separate'' GBs can be significantly lower than their bulk analogs, which can trigger preferential cracking of SE particles at their GB regions.
Compared to the respective bulks, the excess volumes of GBs arise from the different coordination environments experienced by the chemical species expressed in these defects.  
From our models, we observed that GBs and surfaces of SEs will introduce new electronic ``interfacial'' states within the bulk SE gap. 
These interfacial states alter and possibly increase the availability of free electrons and holes in SEs. 
An imbalance of holes and electrons near these extended defects can potentially increase the local electronic conductivity of SEs, and thus facilitate failure-inducing metal plating.

\section{Results}\label{sec:results}
\noindent To establish a comprehensive overview of the effect of extended defects in SEs, we investigate representative electrolyte chemistries.  
We investigate the orthorhombic $\gamma$-\ce{Li_3PO_4},\cite{zhaoHighLithiumIonic2020} and its thiophosphate analog $\beta$-{\ce{Li_3PS_4}}.\cite{hommaCrystalStructurePhase2011}  
For a comparison between Li and Na thiophosphates, we considered \ce{Na_3PS_4} (both the cubic and the tetragonal phases).\cite{berngesScalingRelationsIonic2022,seidelPolymorphsNaIon2020,krauskopfLocalTetragonalStructure2018,famprikisPressureMechanochemicalEffects2020,krauskopfComparingDescriptorsInvestigating2018} We included the argyrodite \ce{Li_6PS_5Cl} as an example of halogenated thiophosphate SEs.\cite{goraiDevilDefectsElectronic2021,deiserothLi7PS6Li6PS5XCl2011,rayavarapuVariationStructureLi2012,kraftInfluenceLatticePolarizability2017} 
Among the oxide ceramics, we included the garnet \ce{Li_7La_3Zr_2O_{12}} (LLZO).\cite{muruganFastLithiumIon2007,gaoSurfaceDependentStabilityInterface2020}
We investigate the Natrium SuperIonic CONductor (NaSICON)  with composition \ce{Na_3Zr_2Si_2PO_{12}} as a representative mixed phosphate/silicate.\cite{dengPhaseBehaviorRhombohedral2020,hongCrystalStructuresCrystal1976,goodenoughFastNaIon1976} 
Prototypical halide SEs were described by  \ce{Li_3YCl_6},  which is claimed to be moisture insensitive and stable against high voltage positive electrode materials.\cite{asanoSolidHalideElectrolytes2018,wangLithiumChloridesBromides2019,liProgressPerspectivesHalide2020,sebtiStackingFaultsAssist2022} 
Binary compounds including \ce{Li_2S}, \ce{Li_2O}, \ce{Na_2S}, and \ce{LiCl}, whose reduced structural complexities enable a systematic exploration of their surfaces and GB models were also included in this analysis. 
Moreover, these compounds are also commonly found as primary decomposition products of a broad range of oxide and sulfide SEs at specific potentials vs.\ Li/Li$^{+}$  (or Na/Na$^{+}$).\cite{zhuOriginOutstandingStability2015,richardsInterfaceStabilitySolidState2016,famprikisFundamentalsInorganicSolidstate2019,goraiDevilDefectsElectronic2021} 
Characteristics of SEs investigated in this work are in Table~\ref{tab:materials_involved} of the supporting information (SI). 

\subsection{Construction of Surfaces and Grain Boundaries of Solid Electrolytes} 
Given the complexities associated with the study of extended defects,\cite{dengAutonomousHighthroughputMultiscale2022} such as GBs and surfaces of multi-component polycrystalline ceramic materials, the existing literature appears scarce, with this investigation aiming to close this knowledge gap.\cite{quirkDesignPrinciplesGrain2023,yuGrainBoundaryContributions2017,baraiMechanicalStressInduced2019,ningDendriteInitiationPropagation2023,kalnausSolidstateBatteriesCritical2023}

Starting from the experimentally reported bulk structures (organized in Table~\ref{tab:materials_involved} of the SI), electrostatically sound slab models of distinct Miller indices surfaces were built.\cite{butlerDesigningInterfacesEnergy2019}
All slab models are charge-neutral and symmetric and do not present intrinsic electrical dipole moments.\cite{taskerStabilityIonicCrystal1979,butlerDesigningInterfacesEnergy2019}  
To downsize the vast set of possible SE surfaces, only a subset of surfaces up to Miller index~$=$~3, or 2 for LLZO, NaSICON, and \ce{Li_3YCl_6} was explored. 
In polyanion systems, such as \ce{Li_3PS_4} and NaSICON, all the \ce{PS_4^{3-}}, and  \ce{PO_4^{3-}} and \ce{SiO_4^{4-}}  moieties must be strictly preserved, which avoids alteration of the underlying chemistry of these SEs.\cite{maranaComputationalCharacterizationVLi3PS42022} To balance the charge after preserving the integrity of polyanion moieties, such as \ce{PO_4^{3-}} or \ce{SiO_4^{4-}},  some surface atoms (Li, Na, O, or S) have to be removed, thus introducing off-stoichiometry. This criterion also ensures that only weak Li(Na)--S or --O bonds are broken, leading typically to low surface energy models.\cite{camacho-foreroElucidatingInterfacialPhenomena2020,tianInterfacialElectronicProperties2019} For each Miller index, several distinct surfaces, \emph{i.e.}{\/} chemically different terminations can exist. 
For example, in LLZO, surface cuts along specific Miller planes may differ the exposed species, \emph{e.g.}~O, La, Li, and Zr, with different stabilities (reactivities).\cite{canepaParticleMorphologyLithium2018}
  
While implementing these strategies downsizes the total number of surface cuts, still $~10^5$ distinct surface models may be found. To circumvent this limitation, only representative surface models carrying the lowest classical Ewald energy were selected for subsequent GB construction.\cite{ewaldBerechnungOptischerUnd1921,ongPythonMaterialsGenomics2013a}  
As surfaces of SEs are the starting block to build GB models,  Figure~\ref{fig:wulff1},  Figure~\ref{fig:wulff2}, and Figure~\ref{fig:wulff3} of the SI show the predicted crystal morphologies of each SEs, in terms of their Wulff shapes.

In polycrystalline SEs, several types of GBs can be envisioned, tilting, twinning, \emph{etc.}, which are formed from the interface of multiple chemically terminated grains of SEs. Hence, to reduce the types of GBs, we have considered twinning types of a selection of chemically meaningful terminations of SEs,\cite{bozzoloViewpointFormationEvolution2020} including non-stoichiometric cases. 
It has been shown that low-index twinning GBs tend to possess low energy,\cite{zhengMobilePinnedGrain2020} for example, $\Sigma$3(111) in \ce{Li_3ClO}\cite{dawsonAtomicScaleInfluenceGrain2018} (0.34~$\mathrm{J\; m^{-2}}$) or  $\Sigma$2(110) in \ce{Li_{0.16}La_{0.62}TiO_3}\cite{symingtonElucidatingNatureGrain2021} (0.30~$\mathrm{J\; m^{-2}}$).
Experimentally, twin boundaries have been observed in various perovskite samples including  \ce{BaTiO_3}, \ce{SrTiO_3}, \ce{BaZrO_3} and \ce{CaMnO_3}.\cite{iguchiInfluenceGrainStructures2006,vonalfthanStructureGrainBoundaries2010,ernstPreferredGrainOrientation2004} 
 
Twinning-type GBs are formed by placing together one slab model of a selected Miller index (representing the bottom grain) and its mirror image along the cutting plane (as the top grain), with the exposed surfaces of the two grains in parallel.
The degrees of freedom explored in these GBs are discussed in Section~\ref{sec:wulff} of the SI. 
Therefore, the top grain is rotated by an angle of 180~$^{\circ}$ to the bottom grain. 
This rotation introduces a misorientation between grains. In this paper, Miller indices of surfaces and the twin GBs are indicated as  $\left(h, k, l\right)$ and  $\left\{h, k, l\right\}$, respectively.
To reduce the number of degrees of freedom arising from the rigid translations between two grains (slabs), we searched the ``global'' minimum of the potential energy surface ---the $\Sigma$-surface,\cite{kurtzEffectsGrainBoundary2004} which is the energy hyperplane formed to in-plane rigid translations of each grain (slab) of the GB model. 
Details are discussed in Section~\ref{sec:dof_gb} of the SI. 

\subsection{Mechanical Properties of Grain Boundaries} 

\begin{figure*}[!ht]
    \centering
    \includegraphics[width=1\textwidth]{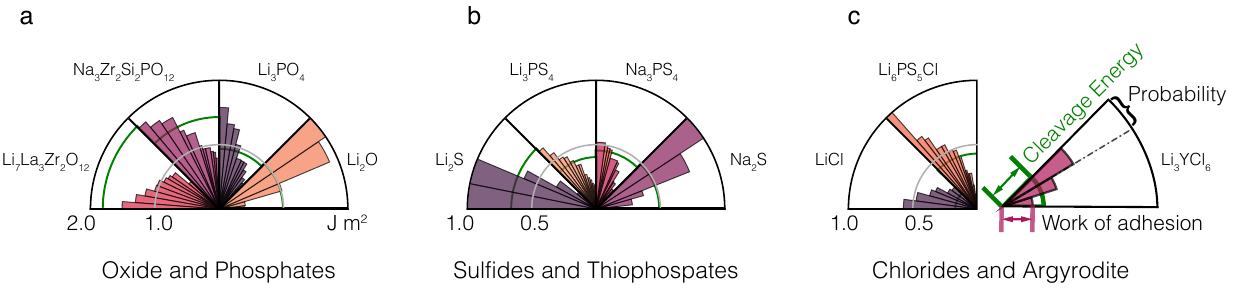}
    \caption{Predicted work of adhesion, $W_{ad}$ of various SEs{\textquotesingle} grain boundaries. \ce{Li_7La_3Zr_2O_{12}} (LLZO), and \ce{Na_3Zr_2Si_2PO_{12}} (NaSICON). The length of each sector is the $W_{ad}$ (Eq.~\ref{eq:woh})  of each GB. The width of each sector is proportional to the relative thermodynamic abundance of each GB evaluated from a Boltzmann distribution of the formation energy  (Eq.~\ref{eq:grainboundaryformation}) of all GBs of each SE. Each GB model may show different chemical terminations and different $\sigma$; here, we only report models with the lowest value of $\sigma$ (see Eq.~{\ref{eq:sigma}}) are reported. Green lines are the fracture (cleavage) energies of SEs in  J\;m$^{-2}$. The maximum scale of $W_{ad}$ and {$W_f$} are: $\mathrm{2\ J\; m^{-2}}$ for oxides/phosphates, $\mathrm{1\ J\;
    m^{-2}}$ for sulfides/thiophosphates, \ce{LiCl} and the argyrodite, and  $\mathrm{0.5\ J\;m^{-2}}$ for \ce{Li_3YCl_6}. The data graphed in this figure is tabulated in Section~\ref{sec:data_each_grain_surface} of Supporting Information.
    }
    \label{fig:pieviolin}
\end{figure*}

Quantifying the mechanical properties of SEs, in particular, those of air-sensitive materials represents a major experimental endeavor.\cite{kalnausSolidstateBatteriesCritical2023} 
Important thermodynamic quantities qualifying the surfaces and GBs include: (\emph{i}) The surface energy, $\gamma$, of Eq.~{\ref{eq:gamma}} is the energy to cut out a surface from the bulk SE. (\emph{ii}) The GB excess energy, $\sigma$, of Eq.~{\ref{eq:sigma}} is the excess energy of GB per unit area compared with the bulk. (\emph{iii}) The work of adhesion, $W_{ad}$, of Eq.~{\ref{eq:woh}} that is the energy absorbed per unit area when separating a GB into two surfaces. 
(\emph{iv}) The cleavage energy, $W_{f}$, of Eq.~{\ref{eq:cleavageenergy}} is the energy absorbed per unit area to separate the bulk into two surfaces. $W_{f}$ is twice the surface energy, $\gamma$ of the stochiometric surface with the lowest $\gamma$ (\emph{i.e.}{\/} the most stable surface cut).
(\emph{v}) The GB formation energy, $E_f$, of Eq.~{\ref{eq:grainboundaryformation}} is the excess energy of GB per atom compared with the bulk.  Thus, $E_f$ quantifies the thermodynamic stability of a GB model. 

 Figure~\ref{fig:pieviolin} shows the general trend of computed work of adhesion ($W_{ad}$,  in Eq.~{\ref{eq:woh}}) of several GBs in \ce{Li_2O}, \ce{LiCl}, \ce{Li_2S}, \ce{Na_2S}, \ce{Li_3PO_4}, \ce{Li_3PS_4}, \ce{Na_3PS_4},  \ce{Li_3YCl_6},  \ce{Li_6PS_5Cl}, \ce{Li_7La_3Zr_2O_{12}}, and \ce{Na_3Zr_2Si_2PO_{12}}.
The width of each sector is proportional to the Boltzmann distribution of the formation energy, which provides a visual measure of the thermodynamic abundance of specific GBs.

The radius of each sector is the work of adhesion, $W_{ad}$. 
Figure~\ref{fig:pieviolin}  reports only the GBs with the lowest values of $\sigma$.

Evaluating the distribution of work of adhesions,  $W_{ad}$s for several SEs and comparing these values with their $W_f$s provides insights into the mechanical stability of the SEs' GBs. 
It was proposed that the population of GB is inversely proportional to values of excess energy, $\sigma$.\cite{rohrerGrainBoundaryEnergy2011, ratanaphanGrainBoundaryCharacter2017,ernstPreferredGrainOrientation2004} 
GB models of each SE chemistry are sorted by a Boltzmann probability distribution $p(N,\,V,\,T)\propto e^{\frac{-E_f}{kT}}$ of the GBs{\textquotesingle} formation energies $E_f$, with {\emph{k}} the Boltzmann constant, $E_f$ the formation energy defined in Eq.~{\ref{eq:grainboundaryformation}}, and {\emph{T}} the reported synthesis temperature of each SEs (see Section~\nolink{\ref{sec:synthesiscondition}} in Supporting Information). 
For  \ce{Li_2S}, \ce{LiCl}, \ce{Li_2O} and \ce{Na_2S}, we set $T = 273$~K.

In general, the range of computed work of adhesion, $W_{ad}$ of Figure~\ref{fig:pieviolin} appears very broad. 
For example, in \ce{Li_2S}, the work of work of adhesion,  $W_{ad}$  can be as low as $28\%$ ($\sim$0.2~$\mathrm{J\; m^{-2}}$) of its $W_f$ ($\sim$0.7~$\mathrm{J\; m^{-2}}$). 
The surface energies $\gamma$ of oxide SEs are generally larger than $\mathrm{0.5~J~m^{-2}}$, whereas for sulfides and chlorides, both the GB excess energy, $\sigma$, and the surface energy, $\gamma$ are generally smaller than $\mathrm{0.5~J~m^{-2}}$.  $\gamma$ and $\sigma$ are related in Figure~\ref{fig:gamma_sigma} of the Supporting Information.

It is important to analyze trends of  GB work of adhesion, $W_{ad}$ for simple Li binary systems, \emph{i.e.} {\ce{Li_2O}}, {\ce{Li_2S}} and {\ce{LiCl}}. By comparing the cleavage energy, $W_{f}$ and the work of adhesion, $W_{ad}$ of each Miller index of the  {\ce{Li_2O}}, {\ce{Li_2S}} and {\ce{LiCl}},  their mechanical stabilities  follows the order  \ce{Li_2O}~$>$~\ce{Li_2S}~$>$~\ce{LiCl}.
Values of work of adhesion, $W_{ad}$ in Figure~{\ref{fig:pieviolin}} suggest that ternary compounds, such as {\ce{Li_3PO_4}}, {\ce{Li_3PS_4}} and {\ce{Na_3PS_4}}  are weaker than binary compounds ({\ce{Li_2O}}, {\ce{Li_2S}}, and {\ce{Na_2S}}), which are  explained by the inductive effects of the polyanionic moieties that  weaken the Li(Na)--O(S) bonds.\cite{liuUnderstandingElectrochemicalPotentials2016,famprikisInsightsRichPolymorphism2021,krauskopfBottleneckDiffusionInductive2018,culverEvidenceSolidElectrolyteInductive2020}
In Figure~{\ref{fig:pieviolin}}, \ce{Li_3PO_4} (0.9~{$\mathrm{J\; m^{-2}}$}) has about twice the value of cleavage energy, $W_f$ of {\ce{Li_3PS_4}} and {\ce{Na_3PS_4}} (both $\sim$0.4 {$\mathrm{J\; m^{-2}}$}). 
The comparison of \ce{Na_3PS_4} with \ce{Li_3PS_4} is less immediate, as these two SEs have different crystalline structures.\cite{hommaCrystalStructurePhase2011,zhaoHighLithiumIonic2020}
Previous work suggested a high similarity between the cubic and tetragonal phases of \ce{Na_3PS_4},\cite{famprikisInsightsRichPolymorphism2021} which explains the similarities in the work of adhesion, $W_{ad}$ of these polymorphs. 
   
The construction of GB models of  \ce{Li_7La_3Zr_2O_{12}} (LLZO) requires breaking the interconnected oxide frameworks which may lead to relatively high surface energies, $\gamma$s (at least $\sim$0.9~$\mathrm{J\; m^{-2}}$ for stochiometric cases), and therefore, LLZO shows the largest value of cleavage energy, $W_f$ (1.8 $\mathrm{J\; m^{-2}}$) among all SEs investigated here. 
As a comparison, in \ce{Li_3PO_4} the {\ce{PO_4^{3-}}} tetrahedra do not share corners/edges, therefore the energy for breaking the Li--{\ce{PO_4^{3-}}} bonds appears low (the lowest $\gamma \approx$~0.5~$\mathrm{J\; m^{-2}}$). Looking more closely at LLZO, the estimated excess energies $\sigma$ range between 0.6 to 1.4~$\mathrm{J\; m^{-2}}$.\cite{yuGrainBoundaryContributions2017}  
Off-stoichiometric surfaces and GBs are expected to dominate the LLZO grains.\cite{canepaParticleMorphologyLithium2018,thompsonElectrochemicalWindowLiIon2017} 
A nonstoichiometric GB of LLZO $\mathrm{\{11\bar{1}\}}$ (\emph{i.e.}{\/} \ce{28Li_7La_3Zr_2O_{12}} + \ce{2La_2O_3} + \ce{14Li_2O}) with excess O, La, and Li displays the lowest excess energy, $\sigma$  (0.6~$\mathrm{J\; m^{-2}}$). Unsurprisingly, a stochiometric Zr-terminated $\mathrm{\{001\}}$ GB of LLZO has the largest $\sigma$ (1.4~$\mathrm{J\; m^{-2}}$). This implies that Li-rich GBs and surfaces,\cite{canepaParticleMorphologyLithium2018} may be preferred and act as a natural Li-ion reservoir. 
Values of work of adhesion, $W_{ad}$s  are similar for the Li or La terminated surfaces (0.8--1.2 $\mathrm{J\; m^{-2}}$), but remain significantly lower than for the Zr terminated surfaces (1.5--1.7 $\mathrm{J\; m^{-2}}$) in LLZO.

Similar to LLZO, {\ce{Na_3Zr_2Si_2PO_{12}}} (NaSICON) displays a 3D rigid framework. 
The construction of  NaSICON surfaces also involves cutting Zr-O bonds, which leads to high surface energies, $\gamma$s ranging from 0.9 to 1.5~{$\mathrm{J\; m^{-2}}$}. 
These surfaces are generally terminated by sodium, and oxygen atoms, the latter taking part into {\ce{PO_4^{3-}}} and {\ce{SiO_4^{4-}}} moieties. 
There is no significant difference in work of adhesion, $W_{ad}$ (ranging between 0.9 to 1.8~{$\mathrm{J\; m^{-2}}$}) between stoichiometric and nonstoichiometric GB models in NaSICON, while  $\sigma$ of nonstoichiometric grain boundary (0.9--1.5~$\mathrm{J\; m^{-2}}$) are generally larger than the stoichiometric GBs (0.1--0.5 $\mathrm{J\; m^{-2}}$).

Compared with other SEs of Figure~\ref{fig:pieviolin}, \ce{Li_3YCl_6} shows low values of work of adhesion, $W_{ad}$ and cleavage energy, $W_f$ ($\sim$0.13~$\mathrm{J\; m^{-2}}$). This indicates the softer nature of  \ce{Li_3YCl_6} and chloride materials, as well as their low mechanical stability.
This finding is further confirmed by the ease of forming stacking faults in \ce{Li_3YCl_6} and similar halide SEs.\cite{sebtiStackingFaultsAssist2022} By inspecting the most stable surface cuts of \ce{Li_3YCl_6}, we observed that the \ce{YCl_6^{3-}} moieties are preserved at the expensed of easier-to-cleave Li--Cl bonds.

The argyrodite \ce{Li_6PS_5Cl} can be expressed as a 1:1:1 mixture of \ce{Li_2S}, \ce{LiCl} and \ce{Li_3PS_4}. 
In \ce{Li_6PS_5Cl}  GBs and surfaces show either \ce{PS_4^{3-}} moieties or dangling Li--Cl and Li--S bonds.
Remarkably, the surface energies of Argyrodite SEs sit in between that of \ce{Li_2S}, \ce{LiCl} and \ce{Li_3PS_4} in agreement with Ref.~\citenum{damoreSymmetryBreakingBulk2022}. 
Synthesis conditions and working conditions of \ce{Li_6PS_5Cl} will set  S--rich or S--poor environments, which are tuned by the sulfur chemical potential S (derived from our phase diagrams of the Supporting Information).\cite{kraftInfluenceLatticePolarizability2017}  
Unsurprisingly, in an S-rich environment, the excessive energy, $\sigma$ of nonstoichiometric S-rich GBs is significantly lower compared with S-poor environments. 
For example, $\sigma$ of a nonstochiometric \{\ce{11\bar{1}}\} GB in a  S-rich condition is $\sim$0.23~{$\mathrm{J\; m^{-2}}$} in S-rich environment, and $\sim$0.57~{$\mathrm{J\; m^{-2}}$} in a S-poor environment. 
From Eq.~\ref{eq:sigma}, the synthesis conditions of these SEs control the excess energy of GBs, and therefore also the population of GBs, as the excess energy, $\sigma$ appears linearly proportional to $E_f$, whereas $E_f$ is inversely related to the GBs' Boltzmann population. 
This implies that in S--rich synthesis conditions, S-rich grain boundaries are formed. 
We also found that the S-rich GBs have higher work of adhesion, $W_{ad}$ compared with S--poor grain boundaries, possibly due to segregated \ce{Li_2S} being ``stronger'' than \ce{LiCl} in S--poor grain boundaries. 
This observation suggests that the mechanical stability of argyrodite particles can be controlled by appropriate synthesis conditions, and the mechanical stability of GB  can be achieved with excess sulfur in the synthesis environment. \cite{kraftInfluenceLatticePolarizability2017}


\subsection{Volume Changes in Grain Boundaries of Solid Electrolytes}

\begin{figure}[!ht]
    \centering
    \includegraphics[width=1.\columnwidth]{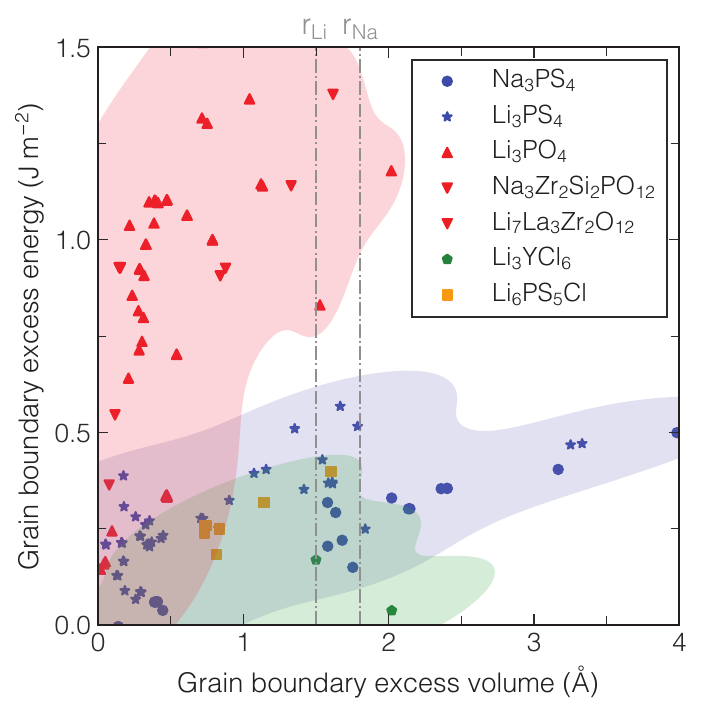}
    \caption{Relationship between the excess volume $\Omega$ (in \AA) and excess energy $\sigma$ (in J~m$^{-2}$) of GB models of different SEs.  Each data point corresponds to a GB model. Grey lines mark the atomic radii of Li and Na.  Colors categorize GBs that share ``similar'' chemistries.
}
    \label{fig:volume}
\end{figure}

During the formation of  GBs from the bulk, volume changes are introduced in the GB regions to accommodate the different coordination environments of atoms. These volume changes are referred to as GB excess free volumes (BFVs) per area,
\begin{equation}
    \label{eq:bfv}
    \Omega=\frac{1}{2S}\cdot \left[V_\mathrm{GB}-2N_\mathrm{slab}V_\mathrm{bulk}\right]\, ,
\end{equation}
where $V_\mathrm{GB}$ is the volume of the GB model of SE, and $V_\mathrm{bulk}$ is the volume per formula unit of the bulk structure of SE.  
$\Omega$  can be understood as the absolute values of how much the normal axis to the GB plane varies in the GB region compared to the reference bulk lattice.\cite{shvindlermanUnexploredTopicsPotentials2006} 
Figure~\ref{fig:volume} shows the relationship between the GB excess energy, $\sigma$, and the excess volume, $\Omega$ of GB models. 
Overall, it appears that the GBs with large excess volume tend to have high excess energy. 
It is important to understand the connection between the excess volume, $\Omega$, and the excess energy, $\sigma$ of GBs.
Densifications through spark-plasma sintering imply pressure in the order of GPa's. Assuming 1~Gpa of applied pressure to a polycrystalline SE carrying an excess volume of $\sim$1.0~\AA, an additional excess energy on the GB compared with the bulk can be estimated as 1.0~GPa~$\times$~1.0~\AA~=~0.1~J~m$^{-2}$. This implies that the GBs with high excess volume can be suppressed due to additional excess energy induced by pressure, which is consistent with the effect of densification, \emph{i.e.}\ sintering. In Figure~{\ref{fig:volume}}, {\ce{Li_3PS_4}} and {\ce{Na_3PS_4}} feature GBs with  high-excess volumes. Under  1.0~GPa of applied pressure, their excess energies would be nearly doubled, hence the substantial effects that hot-pressing can impart on these thiophosphates.

\subsection{Electronic Structure of Extended Defects}

\begin{figure*}[!ht]
    \centering
    \includegraphics[width=1\textwidth]{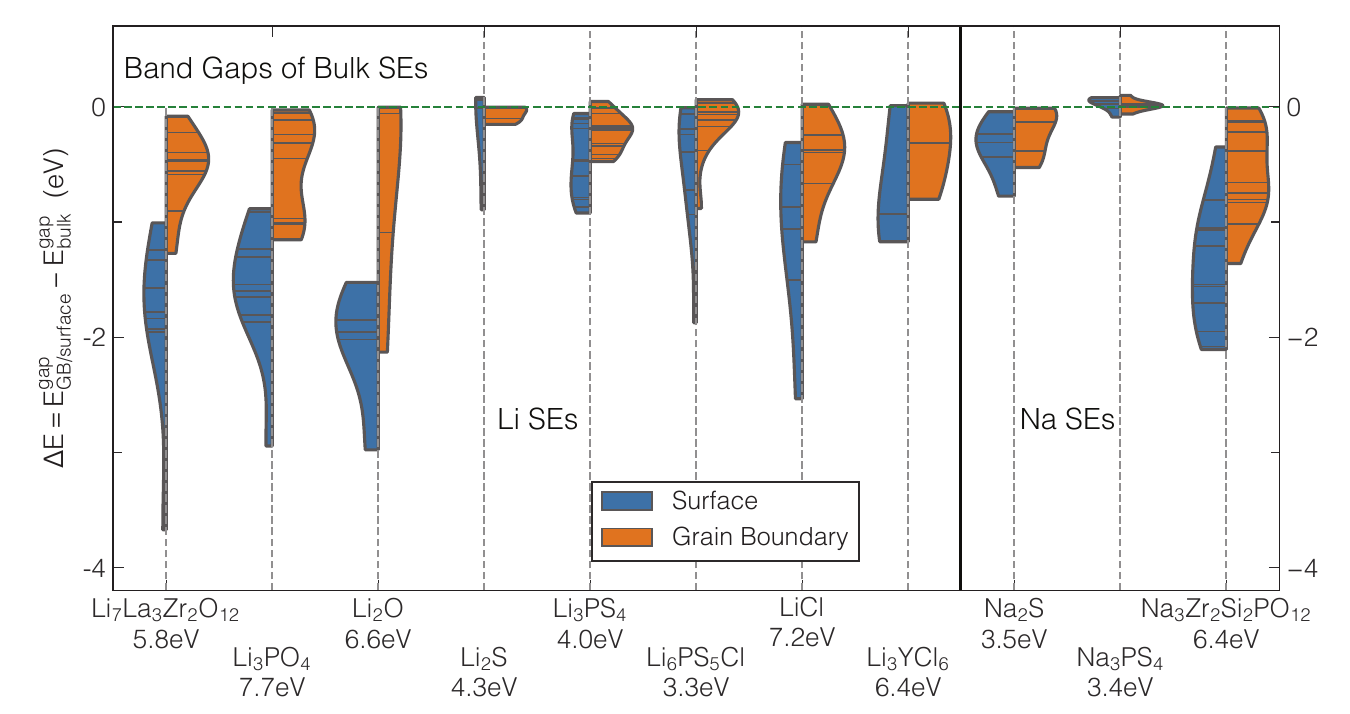}
    \caption{Variation in simulated bandgaps (in eV) in GB and surface models of SEs compared to their bulk analogs. Orange shapes show the distribution in bandgap change of grain GBs to their bulks, while the blue shapes are for surface (slab) models.  Each horizontal line of violin shapes represents the most stable GBs (or surfaces) models of various crystallographic directions, and thus violine shapes with large widths indicate stable GBs or surfaces with high probability. The horizontal dashed line sets the bulk reference. The bandgap change is calculated based on the semi-local GGA (PBE) bandgap of each SE. Accurate hybrid functional bandgaps (HSE06) of bulk SEs are reported as a reference under the respective chemical formula of SEs.
    }
    \label{fig:bandgap}
\end{figure*}

Structural changes introduced by the presence of extended defects are the results of the combination of chemical processes where existing bonds are broken and new bonds are formed. 
These chemical processes will unavoidably alter the underlying electronic structure (the band structure) of GBs and the surfaces of SEs compared to their bulks.

Figure~\ref{fig:bandgap} emphasizes the change in the ``local'' bandgap of several SEs caused by extended defects, such as surfaces and GBs. 
In general, compared to GB models, surfaces of SEs show a greater decrease of bandgaps from the respective bulk structures.  
Cleavage of surfaces of SEs implies breaking selected bonds and a subsequent reduction of species' coordination number, which explains the large extent of reduction of bandgap compared to SEs due to surface states.

Variation of electronic structure properties is particularly accentuated in oxide-based SEs. 
A bandgap decrease can be as high as $\sim$3.0 eV in the case of \ce{Li_3PO_4} and \ce{Li_2O} surfaces, which implies that breakage of Li--O bonds is similar to breaking the Li--\ce{PO_4} bonds. 
For LLZO, the formation of extended defects involves breaking  Li--O, La--O, and Zr--O bonds, with bandgap changes in GB models as high as 1.3 eV, whereas the bandgap change can go up to $\sim$3.7 eV in the $(001)$ Zr-terminated surface.
We recalculated the band structure of this surface model by HSE06 hybrid functional and the bandgap was only 1.8~eV.
The bandgap changes appear somewhat smaller in  NaSICON GBs and surfaces  (up to 2.2 eV) compared to a pure complex oxide, such as LLZO.

 The electronic structures of sulfides and thiophosphate materials appear less sensitive to the introduction of GB and surface defects compared to oxides and phosphates. 
 The largest bandgap decrease of $\sim$1.0~eV about their bulks is reported for GBs and surfaces of \ce{Li_2S}, \ce{Na_2S} and \ce{Li_3PS_4}.  
 Many surface models display bandgap changes from their bulk references that are close to 0~eV. This is the case in the \ce{Li_2S} $(210)$, the \ce{Na_2S} $(111)$, and the \ce{Li_3PS_4} $(001)$ surfaces. 
 The same observation can be extended to \ce{Na_3PS_4}, where there is almost no bandgap change upon the formation of surface and GBs. 
 Chloride SEs show a moderate change (up to 2.5 eV from the bulk reference) in the bandgap. 
 Unsurprisingly, the bandgap drop of surfaces and GBs of argyrodites lie between \ce{LiCl} and \ce{Li_3PS_4}.

A closer inspection through a comparison of the electronic structures of bulk and GBs or surfaces of SEs reveals that extended defects introduce additional states in the bulk bandgap region rather than merely moving closer valence and conduction bands. 
The position of new energy levels ---the interfacial states--- introduced by extended defects in the bandgap of bulk SEs does affect the electronic conductivity of the material if these levels are sufficiently close to the band edges, that is, the valence band maximum (VBM) and conduction band minimum (CBM). 
In contrast, deep defect levels that appear farther away from the band edges will form localized states that will negligibly affect the electronic conductivity of SEs.    

To quantify the extent of localization of defects introduced by GBs and surfaces, the inversed participation ratio (IPR) is used.\cite{bakirkandemirModelingAtomicStructure2014} The IPR is defined as $\frac{N\sum^N_{i=1}[p_i]^2}{(\sum^N_{i=1}[p_i])^2}$, where  $p_i$ is the electron density at site $i$, and $N$ the total number of sites (atoms). For completely localized states IPR~=~1.0, whereas for evenly unlocalized states $\mathrm{IPR}=\frac{1}{N}$. For example, Section~\ref{sec:llzodos} of the Supporting Information compares the density of states and IPR plots of bulk LLZO with those of the Zr-terminated $(001)$ surface, and the corresponding GB. In contrast, the conduction band and valence band of the bulk structure are dense and continuous  (Figure~\ref{fig:iprbulk}),  interfacial states introduced by surface (Figure~\ref{fig:ipr}) and GB (Figure~\ref{fig:iprgb}) are sharp and isolated in the bandgaps.  Such states are generally localized as inferred by the high values of IPR.  Due to a larger density of broken bonds in surface models, more interfacial states are introduced in the gap compared to GBs. However, localized states may not contribute to the electronic conductivity, but may promote local polarizations if localized states are near the surfaces of SEs. 

As shown in Section~\nolink{\ref{sec:excess_energy_bandgap}} in Supporting Information, GBs with high values of $\sigma$ which are less thermodynamically stable, tend to show larger variations in their bandgaps compared to their reference bulks.  Unstable GBs exhibit extended structural reconstruction due to the higher densities of broken bonds and significantly reduced coordination environments (of specific atoms) compared to their bulk counterparts. This leads to a notable alteration in the electronic structures of these GBs.

\subsection{Stability of Solid Electrolytes Surfaces to  Metal Anodes}

\begin{figure*}
    \centering
    \includegraphics[width=1\textwidth]{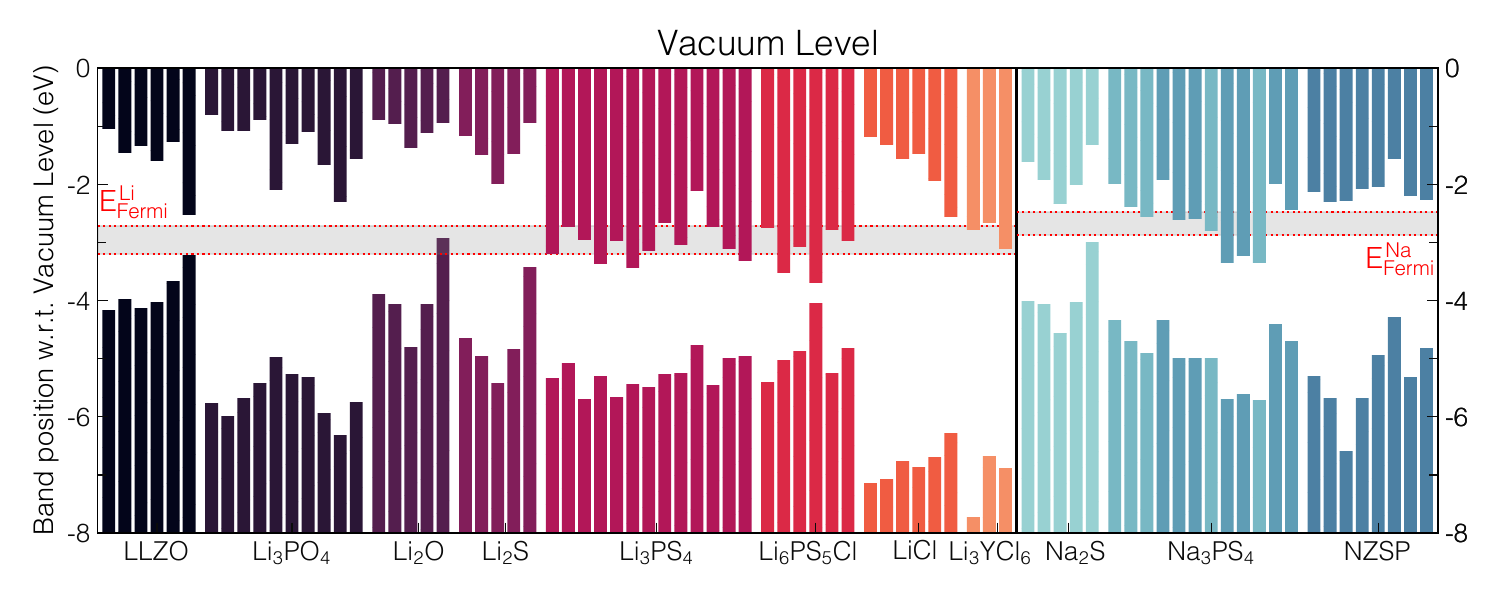}
    \caption{Alignment of SE surfaces vs.\ Li and Na metal references indicated by the gray areas enclosed in red dotted borders. The $y$-axis is the energy level relative to the vacuum level of each SE, with $y$~=~0 indicating the vacuum level. The upper bars are the conduction band minima, and the bottom bars are the valence band maxima of each SE.  NZSP refers to NaSICON \ce{Na_3Zr_2Si_2PO_{12}}.  }
    \label{fig:aligned_bandgap}
\end{figure*}

Through a procedure of band alignment, we explored the electronic collocation of different SE surfaces to those of Li and Na metals. The work function of Li  ($-E_{fermi}$ in Figure~\ref{fig:aligned_bandgap})   lies in the range 2.7 to 3.2~eV, which is in good agreement with previous experimental work.\cite{andersonWorkFunctionLithium1949,CRCHandbookChemistry2008} The work function for Na ranges from 2.5 to 2.9~eV.\cite{CRCHandbookChemistry2008}

In Figure~\ref{fig:aligned_bandgap} tops of the bottom bars and bottoms of the top bars are the band edges, that is the VBM and the CBM of each surface model of the SE. 
In practice, the CBMs represent the electron affinities of surface models of each SE, whereas the VBMs represent the ionization energies.  
An overlap of the conduction band in SEs with the Fermi energies of metal electrodes is a signature of the chemical reactivity of these materials. 

Unsurprisingly, binary compounds, such as \ce{Li_2O}, \ce{Li_2S}, \ce{Na_2S}, and \ce{LiCl} show electron affinities values far apart from the potentials (Fermi energies) of Li (or Na), which explains the intrinsic stability of these binary compounds when in contact with metal electrodes. Thiophosphates,\cite{katoXPSSEMAnalysis2018,zhuOriginOutstandingStability2015,lepleyStructuresLi2013} chlorides,\cite{nikodimosHalideSolidStateElectrolytes2023,nikodimosHalideSolidStateElectrolytes2023,jiKineticallyStableAnode2021} and as a mixture of chloride, thiophosphate, and sulfide, argyrodites are likely to be reduced by metallic Li.\cite{zhuOriginOutstandingStability2015,rieggerLithiumMetalAnodeInstability2021,nikodimosHalideSolidStateElectrolytes2023,wenzelInterfacialReactivityInterphase2018,kasemchainanCriticalStrippingCurrent2019}
Furthermore, the electron affinity of different SEs varies as a function of the Miller index. 
In \ce{Li_3PS_4} the  $(112)$ surface has the largest large electron affinity ($\sim$3.4~eV), which makes it prone to react with Li metal, whereas the  $(011)$ surface with an electron affinity $\sim$2.1~eV seems to be stable against Li metal.

Despite the large bandgap drop of oxide-based SEs, the electron affinities of oxides are smaller than the work function of the corresponding metal (Li or Na).\cite{lepleyStructuresLi2013}
Although \ce{Li_3PO_4} appear stable against the Li-metal anode,\cite{chengUnveilingStableNature2020} NaSICON is thermodynamically unstable against Na metal,\cite{goodwinProtectiveNaSICONInterlayer2023,ortmannKineticsPoreFormation2023}  which Figure~\ref{fig:aligned_bandgap} does not correctly capture.
Indeed, the surface terminations appear to play an important role in these systems.  
Although we observed a large bandgap drop in a Zr-terminated $(001)$ surface, the conduction band of LLZO still lies slightly above the Li metal Fermi energy, which agrees with previous results.\cite{thompsonElectrochemicalWindowLiIon2017}

\section{Discussion}\label{sec:discussion}

Using a thermodynamic model established through accurate first-principles calculations of several topical oxide, sulfide, and halide SEs, we reveal the effects introduced by extended defects, such as grain boundaries (GBs) and surfaces on the thermodynamic, mechanical, and electronic properties of polycrystalline SEs.

The work of adhesion, $W_{ad}$ was identified as a major descriptor for the mechanical strength of polycrystalline SEs, and thus the characterization of their GBs.  The work of adhesion, $W_{ad}$, and other mechanical properties consolidated here provide an accurate set of parameters for the parametrization of continuum models.
Irrespective of the type of chemistry, ternary lithium (and sodium) phosphates, thiophosphates, and chlorides show inferior mechanical properties than their respective binary compounds (\ce{Li_2O}, \ce{Li_2S}, and \ce{LiCl}).

SEs displaying rigid frameworks, such as LLZO and NaSICONs provide larger values of work of adhesion, $W_{ad}$, and cleavage energy, $W_f$ than phosphates, thiophosphates, or chlorides. 
In LLZO GBs, values of work of adhesion, $W_{ad}$ are systematically larger than sulfides and chloride SEs of Figure~{\ref{fig:pieviolin}}, but smaller than {\ce{Li_2O}}, {\ce{Li_3PO_4}} and NaSICON. 
This is due to the high-grain boundary excess energy of LLZO, considering its negative contribution to the work of adhesion,$W_{ad}$ (as defined in Eq.~\ref{eq:woh}).

At the low potential set by Li-metal, \ce{Li_2S} and \ce{Li_3P} are common decomposition products of sulfide SEs (\emph{e.g.}\/{} \ce{Li_3PS_4}).\cite{zhuOriginOutstandingStability2015,richardsInterfaceStabilitySolidState2016,wangResistiveNatureDecomposing2022} These binaries compounds display values of  cleavage energies, $W_f$ ($\sim$0.7~{$\mathrm{J\; m^{-2}}$} for \ce{Li_2S}, and 1.0~{$\mathrm{J\; m^{-2}}$} for \ce{Li_3P}) higher than \ce{Li_3PS_4} ($\sim$0.4~{$\mathrm{J\; m^{-2}}$}).\cite{wangResistiveNatureDecomposing2022}
Although it is tempting to claim an increase in fracture toughness of the ``composites'' formed by a SE and its decomposition products, due to the higher toughness of the latter compared to the SE alone, this is probably incorrect as ``the fracture toughness of the composite'' will depend on the weakest link, among the fracture toughness of the SEs (cleavage energy, and GB work of adhesion) and the delamination of binary interfaces (works of adhesion of the composite-interface). Understanding the interplay of these forces remains an extremely complicated task, which requires a complete understanding of the microstructures, particle size distributions, morphologies, and porosities of such composites.

The relative strength of GBs vs.\ bulk of a SE may drive the type of fracture pattern as either intergranular fracture ---it occurs in the grain boundary--- or as transgranular  ---it occurs in the grain. 
The fracture energy ratio $R_{\rho p}$ between the work of adhesion $W_{ad}$ and the cleavage energy $W_f$ is often used as a descriptor to discriminate the type of fracture pattern in ceramics.\cite{linRoleCohesiveZone2017,kraftStatisticalInvestigationEffects2008,evansPerspectiveDevelopmentHighToughness1990,becherMicrostructuralDesignToughened1991,holmSurfaceFormationEnergy1998,shenderovaAtomisticModelingFracture2000}
A simulation study on \ce{Al_2O_3} suggested that intergranular fracture is preferred when $R_{\rho p} < 0.5$.\cite{linRoleCohesiveZone2017}
In addition, the fracture mechanism also depends on the portion and distribution of GBs in the materials,\cite{kraftStatisticalInvestigationEffects2008} the temperature, the strain rate,\cite{kobayashiDynamicFractureCeramics1991} and the grain sizes.{\cite{riceCeramicFractureModeintergranular1996,holmCriticalManifoldsPolycrystalline2004}} 
Furthermore, the ``wettability'' of the SE by the metal anode,  namely, the interfacial energy between the anode metal and the SE may affect the behavior of the fracture. It is also possible that the active metal (Li or Na) penetrates the SEs simultaneously as the crack propagates, which is different from the current assumption of this work envisaging the propagation of the crack front before the metal ingress.\cite{ningVisualizingPlatinginducedCracking2021} In this mechanism, instead of introducing voids between cracked grain, new interfaces between metal and SE are formed. The energy change of this process can be measured as the interfacial formation energy between the SE and the metal,\cite{wangResistiveNatureDecomposing2022} instead of $W_{ad}$ and  $W_f$ here. However, strained epitaxial interfaces remain ambiguously defined and probably very unlikely in a polycrystalline material.

Here, the simulated thermodynamic properties of these SEs are integrated with elements of fracture mechanics. 
The fracture toughness, $K_{Ic}$ under the crack opening, illustrative of brittle ceramics is defined in Eq.~\ref{eq:critical}.\cite{tromansFractureToughnessSurface2004,broekElementaryEngineeringFracture1982,lawnContinuumAspectsCrack1993}
\begin{equation}
\label{eq:critical}
K_{Ic}=\frac{ \sqrt{{E \, G_c}}}{\sqrt{(1-\nu^2)}}\, .
\end{equation}
where for a SE, $E$ is its Young modulus, $\nu$ is its Poisson ratio, and $G_c$ is the critical energy release rate required for crack propagation.
In most derivations of Eq.~{\ref{eq:critical}}, $\sqrt{(1-\nu^2)}$ is omitted, as this value typically approaches unity.\cite{tromansFractureToughnessSurface2004} 
Considering LLZO a  hard brittle ceramic,\cite{zhangDurableSafeSolidstate2018} we approximate {$G_c$} by the computed cleavage energy, $W_f$ of $\sim$1.8~{$\mathrm{J\; m^{-2}}$},{\cite{tromansFractureToughnessSurface2004}} an approach that appears accurate to investigate materials in their linear-elastic regime. 
Here, using Young's modulus of 163 GPa and a Poisson ratio of 0.26,{\cite{yuGrainBoundarySoftening2018,wolfenstineMechanicalBehaviorLiionconducting2018,niRoomTemperatureElastic2012}} a fracture toughness of $\sim$0.56$~\mathrm{MPa \cdot \sqrt{m}}$ is derived.
Our computed value ($\sim$0.56$~\mathrm{MPa \cdot \sqrt{m}}$) appears in excellent agreement with previous experimentally results of 0.60$~\mathrm{MPa \cdot \sqrt{m}}$.\cite{sharafiControllingCorrelatingEffect2017} 
Notably, the experimental values of $K_{Ic}$ for LLZO demonstrate a wide distribution, ranging from 0.44 to 1.63~$\mathrm{MPa \cdot \sqrt{m}}$.\cite{wolfenstineMechanicalBehaviorLiionconducting2018,wolfenstinePreliminaryInvestigationFracture2013,niRoomTemperatureElastic2012,guoAchievingHighCritical2021,huElasticModulusHardness2021,sharafiControllingCorrelatingEffect2017} Therefore, thermodynamic stable GB (twin-type) set a lower bound in the spread of expected fracture toughness, $K_{Ic}$.

Several factors may contribute to the spread of values of $K_{Ic}$, for example: 
the occurrence of dislocations in regimes of plastic deformation;{\cite{suttonAnalyticModelGrainboundary1991}} 
the anisotropic behavior of polycrystalline specimens;{\cite{lawnContinuumAspectsCrack1993}} different experimental measurements (\emph{e.g.}~nanoindentation \emph{vs.}~microindentation); sample conditions, including grain size, porosity, and the availability of impurities; and the singularity at the crack field.\cite{lawnContinuumAspectsCrack1993,suttonAnalyticModelGrainboundary1991}  For the reasons discussed above,  it remains difficult to reproduce experimentally measured values of some SEs.

From this analysis, it emerges that the mechanical strength of SEs can be improved through their amorphization or by doping.
Amorphization of materials reduces the density of GBs, and may be beneficial to improve the overall mechanical strength of SEs. 
It has been discussed that the amorphous sulfide SEs possess higher formability than their crystalline counterparts due to the isotropic nature of amorphous solids and larger molar volume per atom.\cite{kimSolidStateLiMetal2021,hayashiEffectHydrothermalTemperature2018} Recently, the group of Rupp has extended amorphization procedures to hard oxides, in particular, LLZO.\cite{pfenningerLowRideProcessing2019}

From our simulations and previous experiments, it is observed that Li-ions (and Na-ions) tend to aggregate near the facets (surfaces) of SE particles. The bond strength imparted by Li/Na-ions in between two grains is minimal because of the high mobility (low bonding strength) of these ions. In addition, the most stable coordination environment of Li-ion is 4, hence minimizing the number of bonds with surrounding anion species. Therefore, adding non-redox active multivalent dopants with a preferred coordination number larger than 4 and/or a higher bond strength with cation can strengthen the texture of the grain boundary by increasing the bonding between grains. 

However, as cycling stresses are applied to polycrystalline SEs during repeated charging/discharging,\cite{koerverCapacityFadeSolidState2017,kalnausSolidstateBatteriesCritical2023} the density of accumulated dislocations arising from plastic deformation of SEs (and electrodes) may contribute to the degradation of SEs, and even induce fractures.{\cite{marinescuDeformationFractureCeramic1999,kalnausSolidstateBatteriesCritical2023}} 
From this perspective, linear-elastic measures of cleave energy, {$W_f$} and work of adhesion, $W_{ad}$ can be a conservative, but safe approximation to evaluate the strength of solid-state electrolytes instead of $G_c$. 
In particular, a recent experimental investigation of Li penetration in LLZO demonstrated that the observed stress field of the SE shows an elastic (linear) behavior, and thus in agreement with our approximation of cleavage energy, $W_f$ of  Eq.~\ref{eq:critical}.{\cite{athanasiouOperandoMeasurementsDendriteinduced2023d}}  
If {$G_c$} is taken as a measure of mechanical stability, Jokl et al.\cite{joklMicroscopicTheoryBrittle1980} proposed that $G_c$ is a monotonic function of the cleavage energy, $W_f$. This suggests that GBs with work of adhesion, $W_{ad}$ smaller than their bulk cleavage energies should show smaller values of critical energy release rate. 
Meanwhile, Young's moduli of SEs are expected to decrease near GBs' regions.\cite{zhangElasticStiffnessGrain1992,baraiMechanicalStressInduced2019,tingdongElasticModulusGrainboundary2004,yeheskelElasticModuliGrain2005} 
For example, Yu et al.\cite{yuGrainBoundarySoftening2018} demonstrated that in the $\Sigma5$ GB of LLZO, the moduli (including the Young modulus) can decrease as much as 50\% (from that of bulk), which leads to smaller values of fracture toughness of this SE. 
This implies that GBs in SEs are an obvious source of crack initiation as GBs are prone to fracture from the induced stresses of penetrating Li, perhaps as Li filaments during specific electrochemical conditions.

\begin{figure}[!ht]
    \centering
    \includegraphics[width=1.\columnwidth]{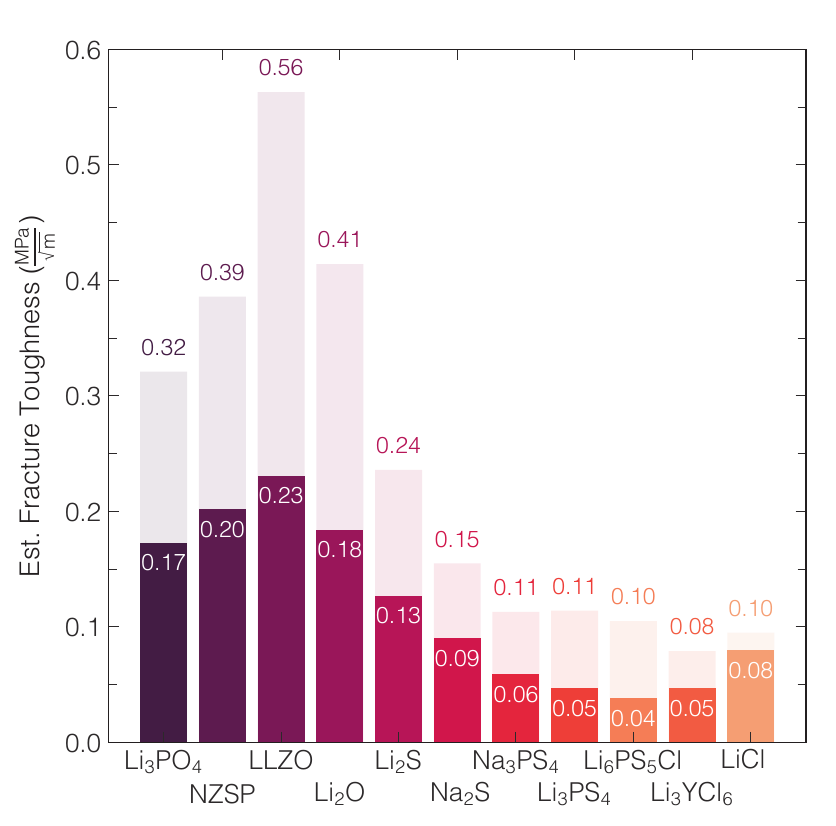}
    \caption{Estimated fracture toughness $K_{Ic}$s from Eq.~\ref{eq:critical} of SEs derived from computed cleavage energy, $W_f$ and work of adhesion, $W_{ad}$, respectively. LLZO is for \ce{Li_7La_3Zr_2O_{12}}, and NZSP is for \ce{Na_3Zr_2Si_2PO_{12}}. Details used in calculations of Figure~\ref{fig:fracturetoughness} are in Section~\nolink{\ref{sec:estimated_fracture}} of the SI.}
    \label{fig:fracturetoughness}
\end{figure}

Although using the computed cleavage energy, $W_f$ to approximate the $G_c$ tends to underestimate measured values of $K_{Ic}$  fracture toughness of SEs, it is still informative to extend the estimation of $K_{Ic}$ to all the SE materials investigated in this work. Figure~\ref{fig:fracturetoughness} summarizes the estimated fracture toughness of the bulk region and GB region.
Solid bars of Figure~\ref{fig:fracturetoughness} show an estimated lower limit of fracture toughness using the weakest GB model (see Figure~\ref{fig:pieviolin}) of each SE,  where the critical energy release rates, $G_c$s of Eq.~\ref{eq:critical} are estimated by the minimum value of work of adhesion, $W_{ad}$ while assuming a 50\% reduction of the SE Young's modulus.\cite{yuGrainBoundarySoftening2018} 
Translucent bars of Figure~\ref{fig:fracturetoughness} are estimated upper limits of fracture toughness of the bulk SEs, where $G_c$ is estimated by the cleavage energy, $W_f$ of bulk SE. 

In Figure~{\ref{fig:fracturetoughness}} computed values of fracture toughnesses $K_{Ic}$s  of oxides-like SEs are systematically larger than sulfides, and chlorides. Therefore, our model correctly captured the general trend of fracture toughness of SEs, with oxides~$\gg$~sulfides~$>$~chlorides.
Unsurprisingly, LLZO displays the highest fracture toughness for both GBs (solid bar) and bulk (translucent bar), due to the high cleavage energy $W_f$, work of adhesion $W_{ad}$, and Young's modulus. Argyrodite {\ce{Li_6PS_5Cl}} possesses the smallest fracture toughness of GBs, while {\ce{Li_3YCl_6}} shows the smallest bulk fracture toughness among all SEs considered.
In general, the fracture toughness of bulks is approximately twice that of GBs.{\cite{kalnausSolidstateBatteriesCritical2023}} 
An anomaly to this trend is  {\ce{LiCl}} as it is difficult to identify GB models with work of adhesion, $W_{ad}$ smaller than the cleavage energy, $W_f$. 

Recently, Ning \emph{et al.}\cite{ningDendriteInitiationPropagation2023} claimed that the initiation of a Li(Na) deposition can be linked to the fracture toughness of GBs in SEs, and proposed a numerical model to evaluate directly the critical current density before Li can start depositing. 
Upon evaluation of the $J$-integral associated with SEs' fracture under external stress, for example, that of inserting Li, it was hypothesized that larger values of fracture toughness lead to higher critical Li current densities. 
This highlights the importance of GB's fracture strength instead of macroscopic fracture toughness and emphasizes the importance of GB engineering for SE synthesis.  

The fracture stress, that is the critical tensile stress for crack propagation of a material can be calculated by $\zeta_c=\frac{K_{Ic}}{\sqrt{Ya}}$,\cite{barsoumFundamentalsCeramics2019}  where $K_{Ic}$ is the fracture toughness, $Y$ is the shape factor and $a$ is the flaw size. 
The stress required for the crack propagation will be small at the GB region compared to the bulk region. 
This investigation provides the simulation data for the fracture toughness, which can be directly linked to the fracture when a flaw size is hypothesized. 
Several studies suggest that pure Li metal with a diameter of 76~nm can support up to 244~MPa under uniaxial compression. 
This raises new challenges for designing more ductile, fracture-resistant SEs.{\cite{zhangLithiumWhiskerGrowth2020,kalnausSolidstateBatteriesCritical2023}}   
Ning \emph{et al.}\cite{ningVisualizingPlatinginducedCracking2021} claimed that metal filaments can open the crack and penetrate through \ce{Li_6PS_5Cl} SE. 
In light of this experimental observation, it can be expected that SEs whose GBs are weaker than that of \ce{Li_6PS_5Cl} will be less resistant to the stress imparted by Li-metal filaments. 
From our analysis chloride SEs, such as \ce{Li_3YCl_6} are expected to show poor electromechanical resistance towards Li-metal ingress.

In contrast to oxide SEs (LLZO, NaSICON and \ce{Li_3PO_4}), the large excess volume estimated (see Figure~\ref{fig:volume}) in sulfide and chloride SEs implies abundance of free space for interstitial atoms, which result in low segregation energy,\cite{leungSpatialHeterogeneitiesOnset2017} especially of small atoms, such as Li ions. \cite{dongFastHydrogenDiffusion2017}
Speculatively, the large excess volumes observed for halide and thiophosphate SEs might be the reason that their GBs do not largely affect the ionic conductivity of these SEs. However, the large excess volume may also facilitate metal deposition. 

It was proposed that the GB excess energy is controlled by their elastic energy, which can be estimated as the product of the SE shear modules, the excess volume, and the energy contribution arising from atomic bonding.\cite{uesugiFirstprinciplesCalculationGrain2011} 
The GB elastic energy arising from excess volume can be estimated via $\gamma _s = \frac{2}{3} \Omega G$, where the $\gamma _s$ is the GB elastic energy and $G$ is the shear modulus. 
Except for the linear behavior of \ce{LiCl} and \ce{Na_3PS_4}, most SEs do not follow a linear relationship between the excess energy $\sigma$ and $\Omega$ (Figure~\ref{fig:volume}). 
However, even in \ce{LiCl} and \ce{Na_3PS_4}, the elastic energy appears twice (and even 3 times) larger than the excess energy $\sigma$. 
The lack of linear relationship between $\gamma _s$ and  $G$ has been previously discussed in Ref~\citenum{palSpectrumAtomicExcess2021}; the authors suggested that the excess volume is not a reliable indicator for the excess energy, especially in hard-ceramic systems as many SEs (Figure~\ref{fig:fracturetoughness}).

Knowledge of relevant thermodynamically sound GB models of SEs and their surfaces is extremely useful for investigating the variation of electronic properties introduced by these types of extended defects. 
Figure~\ref{fig:bandgap} summarizes the bandgap change of surfaces and GBs of SE we studied, as an indicator of change in electronic structure.  
The formation of surfaces and GBs of SEs causes the removal and formation of new bonds, resulting in the introduction of ``interfacial states'' in these materials.\cite{goraiDevilDefectsElectronic2021}  
In practice, these interfacial states may trap excess electrons,\cite{zhuUnderstandingEvolutionLithium2023} and can cause a sudden reduction of SE's bandgaps.
Changes in bandgaps can have an immediate impact on the electronic conductivity of SEs.\cite{goraiDevilDefectsElectronic2021} 
For example, a recent investigation via advanced transmission electron microscopy detected a sudden decrease of the bandgap in the GB region of LLZO, which was linked to a potential cause for Li filament growth.\cite{liuLocalElectronicStructure2021}

\section{Conclusion}\label{sec:conclusion}
The functional properties of polycrystalline materials, including solid electrolytes are modulated by surfaces and grain boundaries occurring between grains. Through a unified methodology, relying on first-principles calculations, an extensive library of thermodynamic and electronic properties of surfaces and grain boundaries of solid electrolytes for a total of 590 models. Structure-property relationships between extended defects, such as surfaces and grain boundaries of solid electrolytes, and their mechanical and electronic properties are established. 

Twinning-type grain boundaries show low excess energy indicating that they are representative of grain boundaries in polycrystalline SE materials. The low work of adhesion of grain boundaries compared to the cleavage energy of bulk indicates that grain boundaries can be the source of crack initiation in polycrystalline solid electrolytes. The brittleness of many oxides solid electrolytes, such as \ce{Li_7La_3Zr_2O_{12}} and \ce{Na_3Zr_2Si_2PO_{12}} contribute to their cracking under local pressure exerted by Li/Na ingress in their grain boundaries. 
Therefore strategies of grain boundary engineering, doping or amorphization of these solid electrolytes can help decrease the brittleness and increase the plasticity of these materials, and hence increase solid electrolytes' compliance to exerted stresses caused by Li(Na)--metal penetration.
 
There is not a simple relationship between grain boundary energy and excess volume. Pressure optimization may be used to control the distribution of grain boundaries and their properties.

It is demonstrated that the occurrence of grain boundaries and surfaces alter significantly the electronic structure of solid electrolytes. 
The electronic properties of sulfur-containing solid electrolytes appear less sensitive than oxygen-based materials, which are much more prone to bandgap closure at grain boundaries. 
Both grain boundaries and surfaces introduce localized donor or acceptor states in the bandgap, which may contribute to local variations of electronic conductivities in solid electrolytes. 

This analysis suggests that bulk properties alone are not sufficient for a complete assessment and engineering of solid electrolyte materials. Variations of mechanical and electronic properties in grain boundaries and exposed surfaces are of primary importance and should be addressed when evaluating new solid electrolytes and their devices.

\section{Methodology}
\label{sec:method}


\subsection{Energetics of Grain Boundaries and Surfaces} 

The excess energy of the surface (which we model as a slab) compared to the bulk is measured by the surface energy, $\gamma$ of  Eq.~\ref{eq:gamma}.\cite{canepaParticleMorphologyLithium2018,butlerDesigningInterfacesEnergy2019} 
    \begin{equation}
        \label{eq:gamma}
        \gamma = \frac{1}{2S}\cdot \left[E_\mathrm{slab}-N_\mathrm{slab}E_\mathrm{bulk}-\displaystyle\sum_i ^\mathrm{species}\Delta n_i\mu_i \right]\, ,
    \end{equation}
    where $N_\mathrm{slab}$ is the number of formula units in the slab model, $E_\mathrm{slab}$ and $E_\mathrm{bulk}$ are approximated by the DFT total energy of the slab model and the bulk structure. $S$ is the surface area of the model slab. For off-stochiometric slabs, such as argyrodite, LLZO, \ce{Li_3YCl_6} and NaSiCON, $\mu_i$ is the chemical potential of species \emph{i} that is added (or removed) in quantity $n_i$ in (from) the slab model. Values of $\mu_i$ are derived from the computed multidimensional phase diagrams of each SE. Details are discussed in Section~\nolink{\ref{sec:phase_diagram}} in Supporting Information.     
    Eq.~\ref{eq:cleavageenergy} defines the cleavage energy $W_f$, which is the energy required to crack the bulk SE and form two identical surfaces.    In Eq.~\ref{eq:cleavageenergy}, $W_f$ amounts to twice the lowest surface energy $\gamma_{min}$  among all the stoichiometric surface cuts of a SE.\cite{tianInterfacialElectronicProperties2019}
       \begin{equation}
        \label{eq:cleavageenergy}
        W_f = 2\, \cdot \, \gamma_{min} \, , 
    \end{equation}
$\sigma$ in  Eq.~\ref{eq:sigma} defines the  excess energy per unit area  of a GB  compared to its bulk,
    \begin{equation}
        \label{eq:sigma}
        \sigma=\frac{1}{2S} \, \cdot \,  \left[E_\mathrm{GB}-N_\mathrm{GB}E_\mathrm{bulk}-\displaystyle\sum_i^\mathrm{species}\Delta n_i\mu_i\right] \, ,
    \end{equation}
    where $E_\mathrm{GB}$ is the DFT energy of the grain boundary model which contains twice as many formula units (f.u.) as the corresponding slab model. $N_\mathrm{GB}$ is the number of f.u.\ in the GB model, and $S$ is the grain boundary area. For some off-stochiometric cases in argyrodite, LLZO, \ce{Li_3YCl_6} and NaSiCON materials, $\mu_i$ is the chemical potential of the specie \emph{i}, and as defined in Eq.~\ref{eq:gamma}. 
    The formation energy of a GB, $E_f$ is the GB excess energy per atom as defined in Eq.~\ref{eq:grainboundaryformation}. 
    \begin{equation}
        \label{eq:grainboundaryformation}
        \begin{aligned}
        E_f & =\frac{1}{N_\mathrm{atom}} \, \cdot \,  \left[E_\mathrm{GB}-N_\mathrm{GB}E_{bulk}-\sum_i^\mathrm{species}\Delta n_i\mu_i\right] \\ 
        & = \sigma\cdot\frac{ 2S}{N_{atom}} \, ,
        \end{aligned}
    \end{equation}
where $N_{atom}$ is the number of atoms in the GB model.  $\sigma$ and $E_f$ are connected as shown in Eq.~\ref{eq:grainboundaryformation}.

From the quantities defined above, one defines the work of adhesion, $W_{ad}$ of Eq.~\ref{eq:woh} is the energy required to separate two grains to an infinite distance,\cite{zhangFirstprinciplesDeterminationEffect2011} and quantifies the mechanical strength of a GB.\cite{rohrerGrainBoundaryEnergy2011}
    \begin{equation}
        \label{eq:woh}
        W_{ad}=2\gamma-\sigma=\frac{1}{2S} \, \cdot \, \left[ 2E_\mathrm{slab}-E_\mathrm{GB} \right]\, .
    \end{equation} \break

\subsection{Construction of Surfaces and Grain Boundaries of Solid Electrolytes}
Starting from the experimentally reported bulk structures of Table~\ref{tab:materials_involved} of Supporting Information, electrostatically sound slab models of distinct Miller indices were built.\cite{ongPythonMaterialsGenomics2013a,camacho-foreroExploringInterfacialStability2018} 
In general, all slab models are charge-neutral and symmetric, and do not present electrical dipole moments.\cite{taskerStabilityIonicCrystal1979}  
To downsize the vast set of possible Miller indices, we explored only a subset of surfaces up to Miller index 3 (or 2 for LLZO, NaSICON, and \ce{Li_3YCl_6}). 
In polyanion systems, such as \ce{Li_3PS_4} and NaSICON, all the \ce{PS_4^{3-}}, \ce{PO_4^{3-}}, and \ce{SiO_4^{4-}}  moieties must be strictly preserved, which avoids alteration of the underlying chemistry of SEs.\cite{maranaComputationalCharacterizationVLi3PS42022} 
This criterion also ensures that only weak Li(Na)--S or --O bonds are broken, leading typically to low surface energy models.\cite{camacho-foreroElucidatingInterfacialPhenomena2020,tianInterfacialElectronicProperties2019} 
Furthermore, for each Miller index, several distinct surfaces, \emph{i.e.}{\/} chemically different terminations can exist. 
For example, in LLZO, different \emph{d} results in O, La, Li, and Zr terminated surfaces. 
In addition, in polyanionic systems, some surface atoms (Li, Na, O, or S) have to be removed to balance the charge after preserving the integrity of polyanion moieties, such as \ce{PO_4^{3-}} or \ce{SiO_4^{4-}}.  
While implementing these strategies certainly downsizes the pool of surface models to compute, still $~10^5$ distinct surface models may be found.  
To circumvent this, only representative surface models providing carrying the lowest classical Ewald energy were selected for subsequent GB construction.\cite{ewaldBerechnungOptischerUnd1921}

Constructing representative GB models remains a major challenge.\cite{dengAutonomousHighthroughputMultiscale2022} 
Several types of GBs of solid-electrolyte materials can be envisioned; here, we only investigate twinning GB models. 
It has been shown that low-index twinning GBs tend to possess low energy,\cite{zhengMobilePinnedGrain2020} for example, $\Sigma$3(111) in \ce{Li_3ClO}\cite{dawsonAtomicScaleInfluenceGrain2018} (0.34~$\mathrm{J\; m^{-2}}$) or  $\Sigma$2(110) in \ce{Li_{0.16}La_{0.62}TiO_3}\cite{symingtonElucidatingNatureGrain2021} (0.30~$\mathrm{J\; m^{-2}}$).
Experimentally,  twin boundaries have been observed in various perovskite samples including  \ce{BaTiO_3}, \ce{SrTiO_3}, \ce{BaZrO_3} and \ce{CaMnO_3}.\cite{iguchiInfluenceGrainStructures2006,yeandelImpactTiltGrain2018,srivastavaCrystalStructureThermoelectric2015,vonalfthanStructureGrainBoundaries2010,kienzleAtomisticStructure1111998,ernstPreferredGrainOrientation2004}  
Models of twining GBs can be built by joining two grains together. This is achieved by pairing appropriately oriented surface models (slabs) as discussed in Section~\ref{sec:gbmodelconstructure} of Supporting Information.

\subsection{First-principles Calculations}
Except for the calculations of the density of states, all our density functional theory (DFT) calculations were performed using the generalized gradient approximation\cite{perdewGeneralizedGradientApproximation1996} (GGA) for the exchange-correlation functional, as available in the Vienna \emph{ab initio} Simulation Package version 6.3.0. 
A 520 eV energy cutoff is applied for the plane wave expansion.\cite{kresseInitioMolecularDynamics1993,kresseEfficientIterativeSchemes1996} 
Core electrons were described by the projector-augmented wave method by Bl\"{o}chl.\cite{blochlProjectorAugmentedwaveMethod1994,kresseUltrasoftPseudopotentialsProjector1999} 
Dense $\Gamma$-centered, \emph{k}-point grids ensured convergence of DFT calculations within 5 $\mathrm{meV/atom}$.   
DFT total energies and interatomic forces were converged within $10^{-5} \ \mathrm{eV}$ and 0.01 eV~\AA$^{-1}$, respectively.  
In all the bulk, surface, and GB models, the atomic positions were fully relaxed with the approximations stated above. 
In GB models, the \emph{z} (non-periodic) direction of the model was relaxed, while for slab models we fix all the lattice parameters to their respective bulk values. 
 A vacuum of at least 15~\AA\ was employed in the surface models.

\section*{Supplementary Information} 
The supplementary information includes the following sections: 
\begin{itemize}
 \item Section~\ref{sec:material_information} Characteristics of SE bulks examined in this study;
 \item Section~\ref{sec:wulff} Wulff shapes of solid electrolytes;
 \item Section~\ref{sec:dof_gb} Degrees of freedom in grain boundaries; 
 \item Section~{\ref{sec:gbmodelconstructure}} The construction of grain boundary models; 
 \item Section~{\ref{sec:synthesiscondition}}  Synthesis conditions; 
 \item Section~{\ref{sec:data_each_grain_surface}} Energetics of surfaces and grain boundaries,    
 \item Section~{\ref{sec:sigmagamma}} Relationship between grain boundary excess energy and surface energy,
 \item Section~{\ref{sec:llzodos}} Density of states of bulk LLZO, Zr-terminated $\left(001\right)$ surface of LLZO, and a GB, 
 \item Section~{\ref{sec:excess_energy_bandgap}} Relationship between high-energy grain boundary and electronic properties, 
 \item Section~{\ref{sec:phase_diagram}} Phase diagrams of solid electrolytes, 
 \item Section~{\ref{sec:estimated_fracture}} Estimated fracture toughness of Solid-state Electrolytes. 
\end{itemize}
 Models of grain boundaries and surfaces reported in this work, in addition to other relevant physical quantities, are available at \url{https://github.com/caneparesearch/project_grainboundary}.

\begin{acknowledgements}
\noindent We acknowledge funding from the National Research Foundation under NRF Fellowship NRFF12-2020-0012. 
ZD acknowledges the support from his Lee Kuan Yew Postdoctoral Fellowship 22-5930-A0001. 
We acknowledge the support from the NUS IT Research Computing group (\url{https://nusit.nus.edu.sg }). 
We thank the support provided by Dr.\ Miguel Dias Costa and Dr.\ Wang Junhong.  This work used computational resources of the supercomputer Fugaku provided by the RIKEN Centre for Computational Science under the ``Fugaku Projects via National Supercomputing Centre Singapore'', and through the HPCI System Research Project (Project ID: hp230188). 
We would like to acknowledge that cloud resources involved in this research work are partially supported by NUS IT's Cloud Credits for Research Programme.  A proportion of the work was performed at the National Supercomputing Centre, Singapore (\url{https://www.nscc.sg}).  We thank Prof.\ S.\ G.\ Gopalakrishnan at the Indian Institute of Science (Bangalore) for his valuable insights on some aspects of this manuscript.
\end{acknowledgements}


\clearpage
\widetext
\begin{center}
\textbf{\large Supplemental Materials: Effects of Grain Boundaries and Surfaces \\ on Electronic and Mechanical Properties of Solid Electrolytes}
\end{center}

\setcounter{equation}{0}
\setcounter{figure}{0}
\setcounter{table}{0}
\setcounter{page}{1}
\setcounter{section}{0}
\makeatletter
\renewcommand{\theequation}{S\arabic{equation}}
\renewcommand{\thefigure}{S\arabic{figure}}
\renewcommand{\thesection}{S\arabic{section}}
\renewcommand{\thetable}{S\arabic{table}}

\section{Characteristics of SE bulks examined in this study}\label{sec:material_information}

Table~\ref{tab:materials_involved} reports the characteristics of the 590 surfaces and grain boundaries of 11 relevant solid electrolytes investigated.

\begin{table*}[!ht]
  \centering
  \small
  \caption{
Characteristics of SE bulks examined to develop GB models. Structures were obtained from the inorganic chemical structure database (ICSD).\cite{bergerhoff1987crystallographic}   Miller Indices indicate the number of unique Miller indices explored in each type of solid electrolyte, and GBs indicate the total number of surfaces or grain boundaries simulated (more than one surface/GBs can be generated on each Miller index considering different cutting planes $d$ position).
}
  \label{tab:materials_involved}
  \begin{tabular*}{\textwidth}{@{\extracolsep{\fill}}llllll@{}}
      \toprule
      {\bf Compound} &  {\bf ICSD} & {\bf Space Group} & {\bf Disordered} & {\bf  Miller Indices} & {\bf GBs} \\ 
      \midrule
      \ce{Li_2O} &   54368 & $Fm\bar{3}m$ & ordered &5 &5\\ 
      \ce{LiCl} &  53818 &  $Fm\bar{3}m$ & ordered & 6&6 \\ 
      \ce{Li_2S}  &   54396 &  $Fm\bar{3}m$ & ordered &5 &5 \\ 
      \ce{Na_2S} &   60436 & $Fm\bar{3}m$ & ordered &5 & 6\\ 
      $\gamma$-\ce{Li_3PO_4} & 79247 & $Pnma$ & ordered &10 & 54\\ 
      $\beta$-\ce{Li_3PS_4}  & 180319 & $Pnma$ & disordered &13 & 59 \\  
      $\beta$-\ce{Na_3PS_4}  & 230142 & $I\bar{4}3m$ & ordered & 5&5 \\ 
      $\alpha$-\ce{Na_3PS_4} &  113165 & $P\bar{4}2_1c$ & ordered &7 & 19\\ 
      \ce{Li_3YCl_6}  & 29962 & $P\bar{3}m1$ & disordered & 3& 19\\  
      \ce{Li_6PS_5Cl}  & 259200 & $F\bar{4}3m$ & disordered &10 & 31\\ 
      \ce{Li_7La_3Zr_2O_{12}} & 246816 & $I4_1/acd$ & ordered &7 & 37\\ 
     \ce{Na_3Zr_2Si_2PO_{12}} & 15546 & $R\bar{3}c$ & disordered &11 & 46\\
      \bottomrule  
  \end{tabular*}
\end{table*}
\clearpage

\section{Wulff Shapes of Solid Electrolytes}\label{sec:wulff}

\begin{figure}[!ht]
  \centering
  \includegraphics[width=0.8\textwidth]{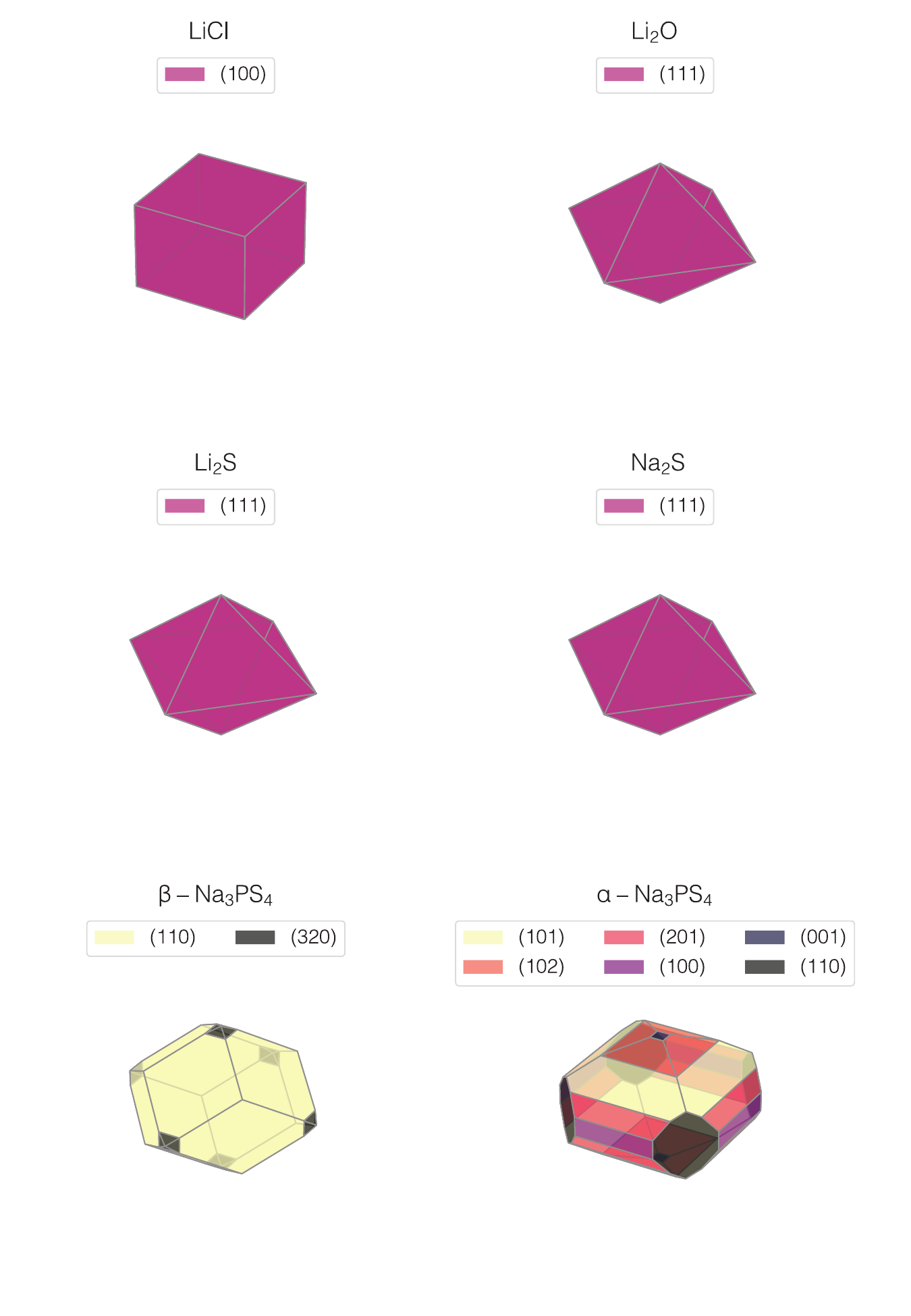}
  \caption{Computed Wulff shapes of \ce{LiCl}, \ce{Li_2O}, \ce{Li_2S}, \ce{Na_2S}, and \ce{Na_3PS_4}.}
  \label{fig:wulff1}
\end{figure}

\begin{figure}[!ht]
  \centering
  \includegraphics[width=0.82\textwidth]{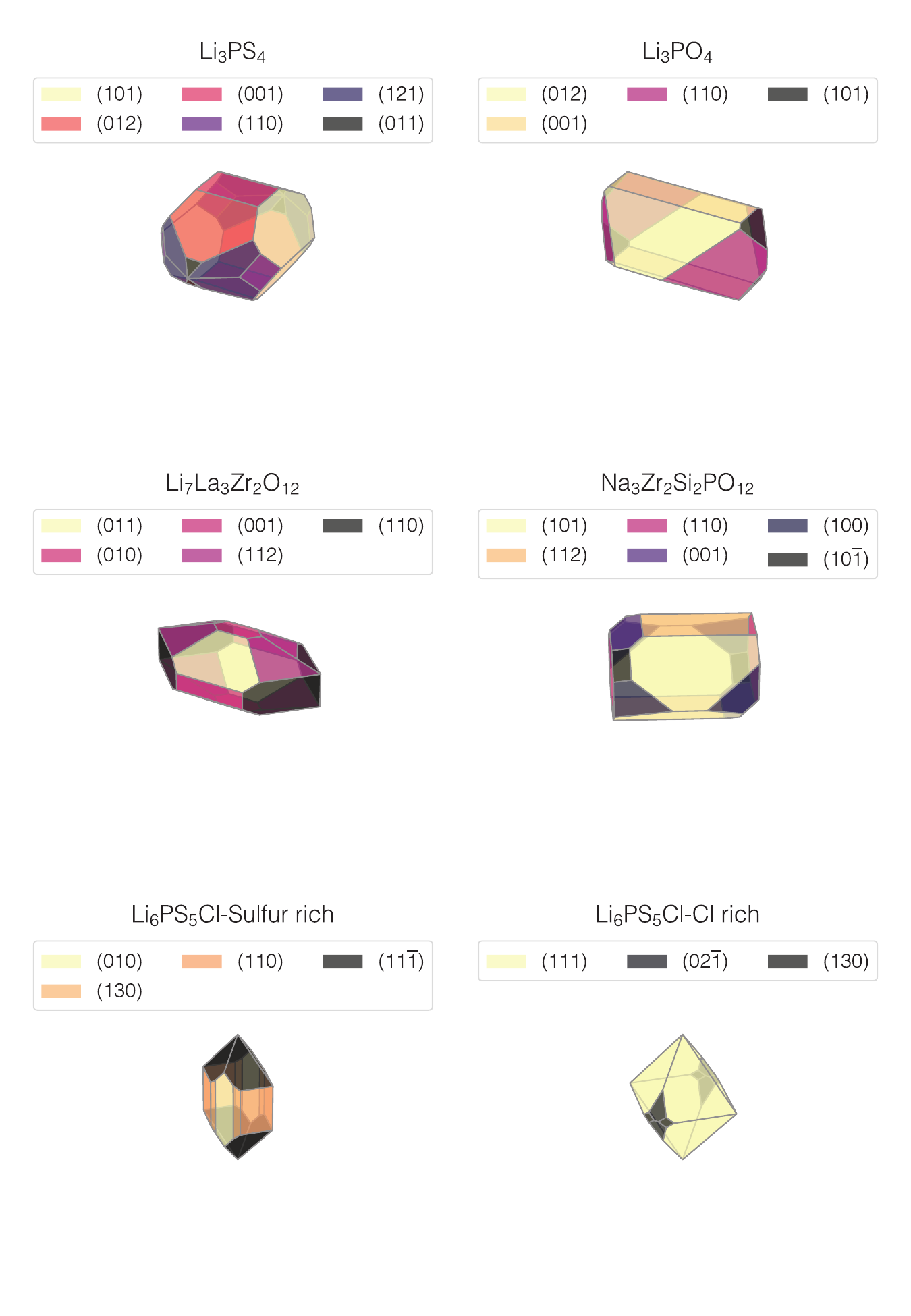}
  \caption{Computed Wulff shapes of \ce{Li_3PS_4},\ce{Li_3PO_4},  \ce{Li_7La_3Zr_2O_{12}}, \ce{Na_3Zr_2Si_2PO_{12}}, and \ce{Li_6PS_5Cl}.}
  \label{fig:wulff2}
\end{figure}

\begin{figure}[!ht]
  \centering
  \includegraphics[width=0.4\columnwidth]{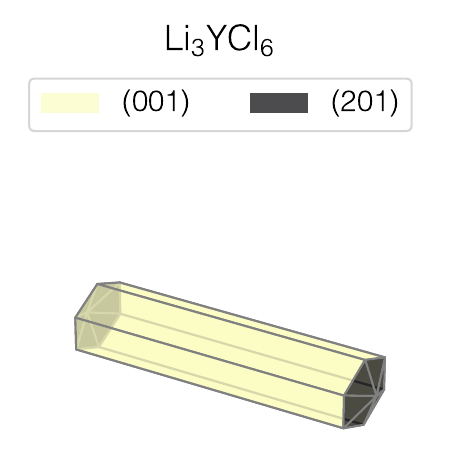}
  \caption{Computed Wulff shape of \ce{Li_3YCl_6}.}
  \label{fig:wulff3}
\end{figure}

The predicted crystal morphologies using the Wulff construction are shown in Figure~\ref{fig:wulff1}, \ref{fig:wulff2}, and \ref{fig:wulff3}.


\clearpage 
\section{Degrees of freedom in grain boundaries}\label{sec:dof_gb}
In a GB, there are in total nine degrees of freedom:\cite{priesterGrainBoundariesTheory2013} five macroscopic parameters characterize the GB and four microscopic parameters, which are null at equilibrium (static conditions). The macroscopic parameters are: 

\begin{enumerate}
    \item Two macroscopic degrees of freedom arise from the rotation axis $\left[uvw\right]$ between two adjacent grains. Here, we explore grain boundaries along different rotation axes up to a specific Miller index.
    \item One macroscopic degree of freedom is the rotation angle $\theta$. Here, we only explore the case that $\theta=180^{\circ}$, which ensures that there is no area mismatch between the grains. 
    \item Two macroscopic degrees of freedom arise from the GB plane's normal $\vec{n}$. Here, we only consider cases where the GB's normal aligns along the rotation axis. These types of GBs are identified as pure \emph{twist} GBs, or \emph{twinnings}.
    \item Three microscopic degrees of freedom involve the in-plane and out-of-plane translation between two grains, i.e., the rigid translation vector $\tau_x,\tau_y, \tau_z$. To avoid confusion, we set $\tau_x,\tau_y$ as two orthogonal vectors in the GB plane, whereas $\tau_z$ points to the GB plane's normal. 
    \item One microscopic degree of freedom indicating the position of cutting plane \emph{d}. 
\end{enumerate}

A clear definition of the degree of freedom in GBs can be found in Ref.~\citenum{priesterGrainBoundariesTheory2013}.

\clearpage

\section{Construction of Grain Boundary Model}\label{sec:gbmodelconstructure}

The GBs investigated here are of twinning type.\cite{bozzoloViewpointFormationEvolution2020} 
Twinning-type GBs are formed by placing together one slab model of a selected Miller index (representing the bottom grain) and its mirror image along the cutting plane (as the top grain), with the exposed surfaces of the two grains in parallel. 
The degrees of freedom explored in these GBs are discussed in Section~S1 of the Supporting Information. Therefore, the top grain is rotated by an angle of 180$^{\circ}$ to the bottom grain. 
This rotation introduces a misorientation between grains. Miller indices of surfaces and the twin GBs are indicated as  $\left(hkl\right)$ and  $\left\{hkl\right\}$, respectively.

\begin{figure*}[!ht]
        \centering
        \includegraphics[width=1.0\textwidth]{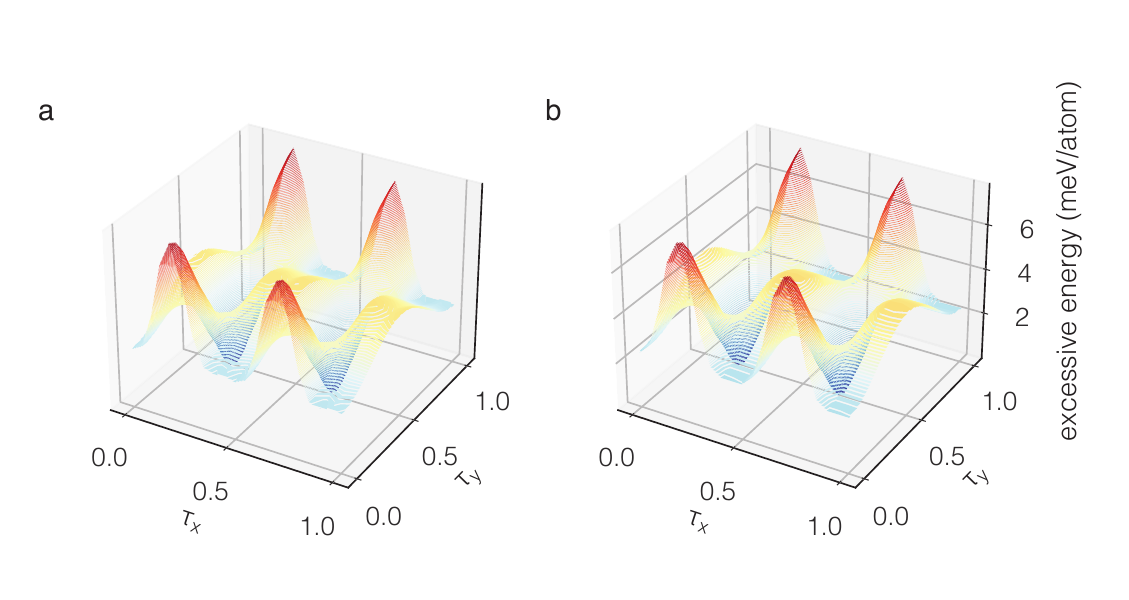}
        \caption{Energy hypersurfaces of the \{012\} \ce{Li_3PO_4}  GB to the in-plane translation between two grains. $\tau_x,\tau_y$ are the axes of in-plane translations shown in terms of fractional coordinates. \textbf{a} shows the energy hypersurfaces $E_\mathrm{Ewald}(\tau_x,\tau_y)$ approximated through the Ewald  method, and \textbf{b}: shows the same quantity but evaluated with DFT, $E_\mathrm{DFT}(\tau_x,\tau_y)$.  The computed $\Sigma$ hypersurfaces with the two methodologies are nearly identical, which motivates the use of the approximated Ewald energy as an accurate method for identifying representative GB models.}
        \label{fig:gamma_surface}
\end{figure*}

To reduce the number of degrees of freedom arising from the rigid translations between two grains (slabs), we search the ``global'' minimum of $\Sigma$-surface (the sigma surface),\cite{kurtzEffectsGrainBoundary2004, terentyevStructureStrength1102010} which is the energy hyperplane formed to in-plane rigid translations of each grain (slab) of the GB model. At every translation step, both $\tau_z$ and the atomic coordinates of the GB model are relaxed to minimize their internal energy. Considering the excessive computational cost encountered for GB models (containing several hundreds of atoms), as well as the sheer amount of possible in-plane translations, we introduced suitable approximations. Using a  fixed $\tau_z$ we found the optimal in-plane translation by searching through a dense grid ($40\times40$ mesh of $[\tau_x,\tau_y]$), and calculated  $\Sigma$-surface using the electrostatic Ewald energy of each GB model.\cite{ewaldBerechnungOptischerUnd1921} The GB model minimizing the $\Sigma$ hypersurface is then optimized with DFT.  Figure~\ref{fig:gamma_surface} compares the $\Sigma$-surface between the approximate Ewald energy and the ground truth (the DFT energy) of a representative GB in $\gamma$-\ce{Li_3PO_4}. The energy minimum from the approximate Ewald energy can quantitatively reproduce the energy minimum from the DFT calculation. 

\begin{figure}[!ht]
  \centering
  \includegraphics[width=0.7\textwidth]{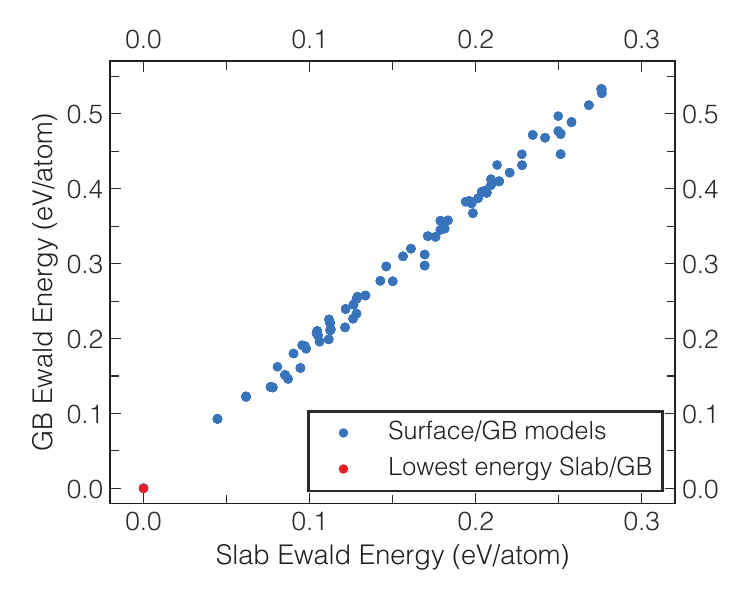}
  \caption{Ewald energies of slab and GB  models of $\gamma$-\ce{Li_3PO_4} are compared in the figure. Relationship between the classical Ewald energy of $(012)$ slab models and $\{012\}$ GB models of $\gamma$-\ce{Li_3PO_4}. 336 slab models are generated parallel to the cutting plane $(012)$. In this context, several cut positions parallel to the Miller plane $(012)$ can be chosen, but these surfaces (or their reconstructions) must be represented by electrostatically meaningful surfaces. The 336 slab models are then used to build 336 GB models.  The ``zero points'' are assigned to the lowest energy slab model and GB models, respectively.}
  \label{fig:ewald_gamma_sigma}
\end{figure}

Here, slab models are constructed according to the Methodology section exposed in the main text. The slab model providing the lowest classical Ewald energy of each Miller index and termination (among various removal combinations of surface atoms) tends to form the GB model with the lowest classical Ewald energy, as shown in Figure~\ref{fig:ewald_gamma_sigma}.

From our calculation, most of GBs possess relatively low excess energy ($<$~50 meV/atom) compared to their respective bulk,\cite{symingtonElucidatingNatureGrain2021} implying that GBs are likely to form with high abundance in these materials, and in agreement with existing experiments.\cite{rohrerGrainBoundaryEnergy2011,ratanaphanFiveParameterGrain2014,liRelativeGrainBoundary2009}

\clearpage
\section{Synthesis Conditions of Solid Electrolytes}\label{sec:synthesiscondition}

Table~\ref{tab:synthesis} summarizes the synthesis conditions of solid electrolytes investigated in this study.  Note that the relative distribution of GBs is insensitive to the synthesis temperature. For example, the thermodynamically most accessible  $\{001\}$ grain boundary of LLZO, is estimated to be 14\% of the total population of GBs at 1503K,  and 18\% of the total population at 298K, respectively.

\begin{table*}[!ht]
  \caption{General synthesis conditions of the solid electrolytes investigated.}
  \label{tab:synthesis}
  \begin{tabular*}{\textwidth}{@{\extracolsep{\fill}}llll@{}}
  \toprule
  {\bf Compound}& {\bf Synthesis Method}&  {\bf Synthesis Temperature (K)}&  {\bf T (K)}\\
   \midrule
   $\beta$-\ce{Li_3PS_4}& heating from glass phase &  573\cite{hommaCrystalStructurePhase2011,kuduReviewStructuralProperties2018} & 573 \\
   $\gamma$-\ce{Li_3PO_4}& heating  &  650\cite{ishigakiRoomTemperatureSynthesis2021}  &  650 \\
   $\beta$-\ce{Na_3PS_4} & liquid-phase/ball milling & 543\cite{uematsuPreparationNa3PS4Electrolyte2018}  &  543\\
   $\alpha$-\ce{Na_3PS_4}& liquid-phase/ball milling & 300\cite{uematsuPreparationNa3PS4Electrolyte2018}  &  300\\     
   \ce{Li_7La_3Zr_2O_{12}}& solid state synthesis/sintering & 1503\cite{muruganFastLithiumIon2007}  &  1503\\     
   \ce{Na_3Zr_2Si_2PO_{12}}& sol-gel/sintering & 873--1173\cite{raoReviewSynthesisDoping2021}  & 873\\     
   \ce{Li_6PS_5Cl}& ball milling & 823\cite{goraiDevilDefectsElectronic2021}   & 823\\    
   \ce{Li_3YCl_6}&solid-state synthesis& 823\cite{sebtiStackingFaultsAssist2022}&823\\
  \bottomrule

  \end{tabular*}
\end{table*}


\clearpage
\clearpage
\section{Relationship between grain boundary excess energy and surface energy}\label{sec:sigmagamma}

\begin{figure}[!ht]
  \centering
  \includegraphics[width=0.7\textwidth]{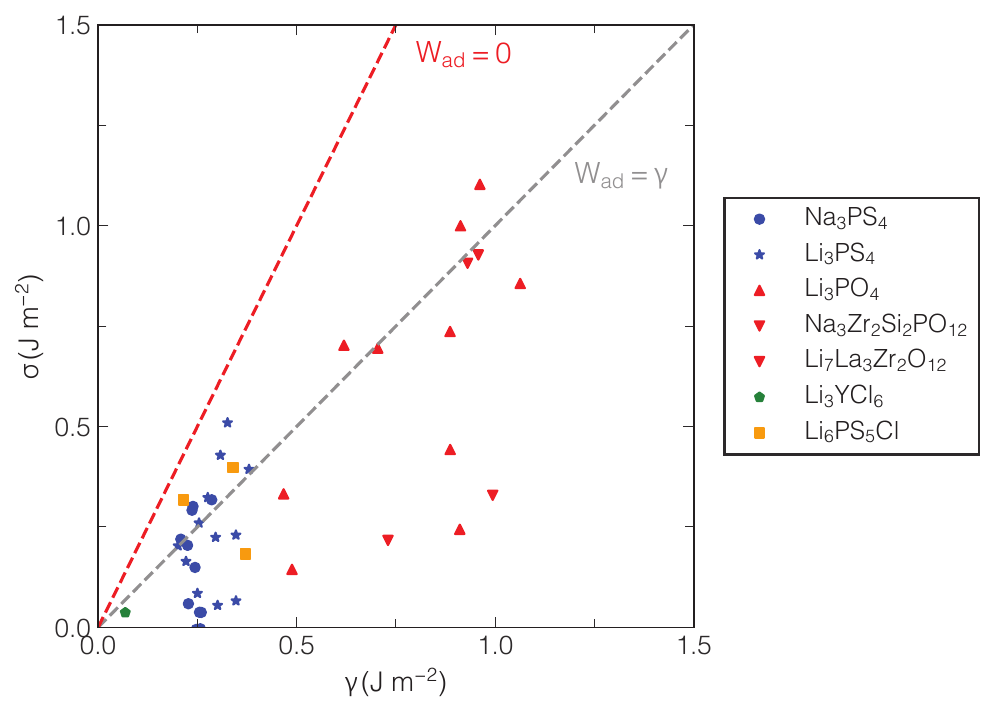}
  \caption{Relationship between surface energy, $\gamma$ ($x$-axis) and GB excess energy, $\sigma$ ($y$-axis) of each Miller index of several SEs.}
  \label{fig:gamma_sigma}
\end{figure}

Figure~\ref{fig:gamma_sigma} shows the surface energy, $\gamma$, correlated to the GB excess energy, $\sigma$ of each Miller index of SEs. High-energy surfaces do not necessarily correspond to high-energy GBs.

\clearpage
\section{Energetics of Surfaces and Grain Boundaries}\label{sec:data_each_grain_surface}

Table~\ref{tab:all_excess_energy_woh} summarizes the energy terms of all surface and grain boundary models calculated by DFT. 
The Miller Indices column indicates the plane of the surface, the twin grain boundary model is built based on this surface cut. 
The Formula column is the chemical formula of the GB model. 
The $\mathbf{\sigma}$ column is the grain boundary excess energy in $\mathrm{J\; m^{-2}}$, whereas $\mathbf{\gamma}$ column is the surface energy in $\mathrm{J\; m^{-2}}$. 
${W_{ad}}$ column is the work of adhesion in $\mathrm{J\; m^{-2}}$. 
Stochiometric indicates whether the surface and the corresponding grain boundary models are stochiometric. 
In the case of the argyrodite solid electrolyte the chemical environment S-rich (Cl--poor) or S-poor (Cl--rich) is also indicated. 

\begin{table*}[!ht]
    \caption{Thermodynamic and mechanical data of representative surfaces and grain boundaries. The surface energy  $\mathbf{\gamma}$, the grain boundary excess energy $\mathbf{\sigma}$, and the work of adhesion $W_{ad}$ are reported in $\mathrm{J\; m^{-2}}$. }
    \label{tab:all_excess_energy_woh}
    \begin{tabular*}{\textwidth}{@{\extracolsep{\fill}}lllcccll@{}}
      \toprule
      {\bf Material} & {\bf Miller Index}& {\bf Formula}  & $\mathbf{\sigma}$  &  $\mathbf{\gamma}$ & $\mathbf{W_{ad}}$ & {\bf Stochiometric}   \\
      
      \midrule
      \multirow{8}{*}{\ce{Li_7La_3Zr_2O_{12}}} & 001 & \ce{Li_{224}La_{88}Zr_{56}O_{356}} &0.63 & 0.97 & 1.30 &  False \\ 
      & 100 & \ce{Li_{160}La_{64}Zr_{40}O_{256}} &0.71 & 0.87 & 1.03 &  False \\ 
      & 101 & \ce{Li_{148}La_{60}Zr_{40}O_{244}} &0.79 & 0.70 & 0.61 &  False \\ 
      & 110 & \ce{Li_{160}La_{72}Zr_{40}O_{268}} &0.81 & 0.99 & 1.17 &  False \\ 
      & 201 & \ce{Li_{168}La_{72}Zr_{40}O_{272}} &0.85 & 0.88 & 0.92 &  False \\ 
      & 112 & \ce{Li_{272}La_{120}Zr_{72}O_{460}} &0.90 & 0.84 & 0.78 &  False \\ 
      & 301 & \ce{Li_{224}La_{96}Zr_{64}O_{384}} &0.91 & 0.93 & 0.95 &  True \\ 
      \midrule
      \multirow{11}{*}{ \ce{Na_3Zr_2Si_2PO_{12}}} & 110 & \ce{Na_{36}Zr_{24}Si_{24}P_{12}O_{144}} &0.12 & 0.87 & 1.63 &  False \\ 
      & 101 & \ce{Na_{48}Zr_{32}Si_{32}P_{16}O_{192}} &0.22 & 0.73 & 1.24 &  True \\ 
      & 10$\mathrm{\bar{1}}$ & \ce{Na_{60}Zr_{40}Si_{40}P_{20}O_{240}} &0.33 & 0.99 & 1.66 &  True \\ 
      & 001 & \ce{Na_{48}Zr_{28}Si_{28}P_{16}O_{176}} &0.88 & 0.92 & 0.96 &  False \\ 
      & 100 & \ce{Na_{24}Zr_{16}Si_{16}P_{8}O_{96}} &0.93 & 0.96 & 0.99 &  True \\ 
      & 112 & \ce{Na_{60}Zr_{40}Si_{32}P_{20}O_{224}} &0.97 & 0.88 & 0.79 &  False \\ 
      & 1$\mathrm{\bar{1}}$0 & \ce{Na_{60}Zr_{48}Si_{40}P_{20}O_{256}} &0.98 & 1.15 & 1.32 &  False \\ 
      & 1$\mathrm{\bar{1}}$2 & \ce{Na_{88}Zr_{64}Si_{52}P_{24}O_{336}} &1.24 & 1.29 & 1.34 &  False \\ 
      & 20$\mathrm{\bar{1}}$ & \ce{Na_{84}Zr_{56}Si_{48}P_{20}O_{300}} &1.29 & 1.13 & 0.96 &  False \\ 
      & 221 & \ce{Na_{68}Zr_{48}Si_{40}P_{20}O_{260}} &1.37 & 1.13 & 0.90 &  False \\ 
      & 11$\mathrm{\bar{2}}$ & \ce{Na_{96}Zr_{64}Si_{56}P_{24}O_{348}} &1.48 & 1.25 & 1.02 &  False \\ 
      \midrule
      \multirow{10}{*}{$\gamma$-\ce{Li_3PO_4}} & 001 & \ce{Li_{60}P_{20}O_{80}} &0.14 & 0.49 & 0.83 &  True \\ 
      & 010 & \ce{Li_{96}P_{32}O_{128}} &0.24 & 0.91 & 1.58 &  True \\ 
      & 012 & \ce{Li_{168}P_{56}O_{224}} &0.33 & 0.47 & 0.60 &  True \\ 
      & 100 & \ce{Li_{132}P_{44}O_{176}} &0.44 & 0.89 & 1.33 &  True \\ 
      & 011 & \ce{Li_{96}P_{32}O_{128}} &0.69 & 0.70 & 0.71 &  True \\ 
      & 110 & \ce{Li_{120}P_{40}O_{160}} &0.70 & 0.62 & 0.54 &  True \\ 
      & 101 & \ce{Li_{144}P_{48}O_{192}} &0.74 & 0.89 & 1.04 &  True \\ 
      & 102 & \ce{Li_{168}P_{56}O_{224}} &0.86 & 1.06 & 1.27 &  True \\ 
      & 111 & \ce{Li_{132}P_{44}O_{176}} &1.00 & 0.91 & 0.83 &  True \\ 
      & 112 & \ce{Li_{192}P_{64}O_{256}} &1.10 & 0.96 & 0.82 &  True \\ 

      \midrule
      \multirow{5}{*}{\ce{Li_2O}} & 110 & \ce{Li_{128}O_{64}} &--0.05 & 0.91 & 1.86 &  True \\ 
      & 100 & \ce{Li_{88}O_{44}} &0.01 & 1.21 & 2.40 &  True \\ 
      & 111 & \ce{Li_{160}O_{80}} &0.58 & 0.50 & 0.41 &  True \\ 
      & 221 & \ce{Li_{280}O_{140}} &0.98 & 0.69 & 0.39 &  True \\ 
      & 210 & \ce{Li_{216}O_{108}} &1.21 & 1.11 & 1.02 &  True \\ 
      \midrule
      \multirow{5}{*}{\ce{Li_2S}} & 110 & \ce{Li_{112}S_{56}} &--0.01 & 0.51 & 1.03 &  True \\ 
      & 100 & \ce{Li_{72}S_{36}} &0.00 & 0.81 & 1.62 &  True \\ 
      & 111 & \ce{Li_{128}S_{64}} &0.31 & 0.33 & 0.35 &  True \\ 
      & 221 & \ce{Li_{216}S_{108}} &0.66 & 0.42 & 0.19 &  True \\ 
      & 210 & \ce{Li_{152}S_{76}} &0.75 & 0.68 & 0.61 &  True \\

      \bottomrule
  
    \end{tabular*}
\end{table*}

\begin{table*}[!ht]

    \begin{tabular*}{\textwidth}{@{\extracolsep{\fill}}lllcccll@{}}
      \toprule
      {\bf Material} & {\bf Miller Index}& {\bf Formula}  & $\mathbf{\sigma}$  &  $\mathbf{\gamma}$ & $\mathbf{W_{ad}}$ & {\bf Stochiometric}   \\
      \midrule
      \multirow{13}{*}{$\beta$-\ce{Li_3PS_4}} & 001 & \ce{Li_{48}P_{16}S_{64}} &--0.02 & 0.23 & 0.48 &  True \\ 
      & 100 & \ce{Li_{84}P_{28}S_{112}} &0.05 & 0.30 & 0.55 &  True \\ 
      & 010 & \ce{Li_{72}P_{24}S_{96}} &0.07 & 0.35 & 0.63 &  True \\ 
      & 110 & \ce{Li_{120}P_{40}S_{160}} &0.08 & 0.25 & 0.42 &  True \\ 
      & 012 & \ce{Li_{120}P_{40}S_{160}} &0.16 & 0.22 & 0.28 &  True \\ 
      & 101 & \ce{Li_{96}P_{32}S_{128}} &0.20 & 0.20 & 0.20 &  True \\ 
      & 111 & \ce{Li_{120}P_{40}S_{160}} &0.22 & 0.30 & 0.37 &  True \\ 
      & 011 & \ce{Li_{84}P_{28}S_{112}} &0.23 & 0.35 & 0.46 &  True \\ 
      & 121 & \ce{Li_{168}P_{56}S_{224}} &0.26 & 0.25 & 0.25 &  True \\ 
      & 112 & \ce{Li_{144}P_{48}S_{192}} &0.32 & 0.28 & 0.23 &  True \\ 
      & 021 & \ce{Li_{168}P_{56}S_{224}} &0.39 & 0.38 & 0.37 &  True \\ 
      & 120 & \ce{Li_{168}P_{56}S_{224}} &0.43 & 0.31 & 0.19 &  True \\ 
      & 102 & \ce{Li_{132}P_{44}S_{176}} &0.51 & 0.33 & 0.14 &  True \\ 

      \midrule
      \multirow{5}{*}{$\beta$-\ce{Na_3PS_4}} & 110 & \ce{Na_{60}P_{20}S_{80}} &--0.04 & 0.18 & 0.41 &  True \\ 
      & 100 & \ce{Na_{42}P_{14}S_{56}} &0.00 & 0.26 & 0.52 &  True \\ 
      & 310 & \ce{Na_{108}P_{36}S_{144}} &0.15 & 0.24 & 0.34 &  True \\ 
      & 210 & \ce{Na_{78}P_{26}S_{104}} &0.20 & 0.23 & 0.25 &  True \\ 
      & 320 & \ce{Na_{162}P_{54}S_{216}} &0.22 & 0.21 & 0.20 &  True \\ 
      \midrule
      \multirow{7}{*}{$\alpha$-\ce{Na_3PS_4}} & 100 & \ce{Na_{42}P_{14}S_{56}} &0.00 & 0.25 & 0.50 &  True \\ 
      & 110 & \ce{Na_{48}P_{16}S_{64}} &0.04 & 0.26 & 0.48 &  True \\ 
      & 001 & \ce{Na_{42}P_{14}S_{56}} &0.04 & 0.26 & 0.47 &  True \\ 
      & 101 & \ce{Na_{48}P_{16}S_{64}} &0.06 & 0.23 & 0.40 &  True \\ 
      & 102 & \ce{Na_{72}P_{24}S_{96}} &0.29 & 0.24 & 0.18 &  True \\ 
      & 201 & \ce{Na_{72}P_{24}S_{96}} &0.30 & 0.24 & 0.18 &  True \\ 
      & 210 & \ce{Na_{78}P_{26}S_{104}} &0.32 & 0.29 & 0.25 &  True \\ 
      \midrule
      \multirow{5}{*}{\ce{Na_2S}} & 110 & \ce{Na_{96}S_{48}} &--0.01 & 0.39 & 0.78 &  True \\ 
      & 100 & \ce{Na_{56}S_{28}} &--0.01 & 0.57 & 1.15 &  True \\ 
      & 111 & \ce{Na_{112}S_{56}} &0.21 & 0.25 & 0.28 &  True \\ 
      & 221 & \ce{Na_{184}S_{92}} &0.47 & 0.32 & 0.17 &  True \\ 
      & 210 & \ce{Na_{152}S_{76}} &0.62 & 0.50 & 0.38 &  True \\ 
      \midrule
      \multirow{6}{*}{\ce{LiCl}} & 100 & \ce{Li_{40}Cl_{40}} &--0.07 & 0.08 & 0.24 &  True \\ 
      & 110 & \ce{Li_{56}Cl_{56}} &0.04 & 0.26 & 0.48 &  True \\ 
      & 210 & \ce{Li_{96}Cl_{96}} &0.06 & 0.16 & 0.26 &  True \\ 
      & 211 & \ce{Li_{192}Cl_{192}} &0.10 & 0.34 & 0.57 &  True \\ 
      & 221 & \ce{Li_{128}Cl_{128}} &0.15 & 0.27 & 0.38 &  True \\ 
      & 111 & \ce{Li_{64}Cl_{64}} &0.44 & 0.40 & 0.35 &  True \\ 
 \midrule
      \multirow{11}{*}{\ce{Li_6PS_5Cl}} & 02$\mathrm{\bar{1}}$ & \ce{Li_{52}P_{10}S_{46}Cl_{10}} &0.09 & 0.29 & 0.48 &  S poor \\ 
      & 130 & \ce{Li_{176}P_{32}S_{152}Cl_{32}} &0.37 & 0.34 & 0.31 &  S poor \\ 
      & 230 & \ce{Li_{256}P_{42}S_{210}Cl_{46}} &0.42 & 0.42 & 0.42 &  S poor \\ 
      & 320 & \ce{Li_{256}P_{42}S_{210}Cl_{46}} &0.54 & 0.36 & 0.19 &  S poor \\ 
      & 11$\mathrm{\bar{1}}$ & \ce{Li_{44}P_{6}S_{32}Cl_{10}} &0.21 & 0.44 & 0.67 &  S poor \\ 
      & 11$\mathrm{\bar{1}}$ & \ce{Li_{52}P_{6}S_{36}Cl_{10}} &0.33 & 0.52 & 0.71 &  S rich \\ 
      & 130 & \ce{Li_{176}P_{24}S_{132}Cl_{32}} &0.23 & 0.36 & 0.50 &  S rich \\ 
      & 20$\mathrm{\bar{1}}$ & \ce{Li_{56}P_{6}S_{38}Cl_{10}} &0.30 & 0.66 & 1.02 &  S rich \\ 
      & 110 & \ce{Li_{96}P_{16}S_{80}Cl_{16}} &0.18 & 0.37 & 0.56 &  True \\ 
      & 111 & \ce{Li_{144}P_{24}S_{120}Cl_{24}} &0.32 & 0.22 & 0.11 &  True \\ 
      & 010 & \ce{Li_{72}P_{12}S_{60}Cl_{12}} &0.40 & 0.34 & 0.28 &  True \\ 
        
      \midrule
 
      \multirow{3}{*}{\ce{Li_3YCl_6}} & 001 & \ce{Li_{42}Y_{14}Cl_{84}} &0.04 & 0.07 & 0.10 &  True \\ 
      & 210 & \ce{Li_{174}Y_{54}Cl_{336}} &0.29 & 0.28 & 0.28 &  False \\ 
      & 201 & \ce{Li_{168}Y_{48}Cl_{312}} &0.32 & 0.26 & 0.19 &  False \\

    \bottomrule
  
    \end{tabular*}
\end{table*}


\clearpage
\section{Density of States of a Selected Grain Boundary and Surface}\label{sec:llzodos}

\begin{figure*}[!h]
    \centering
    \includegraphics[width=1.\textwidth]{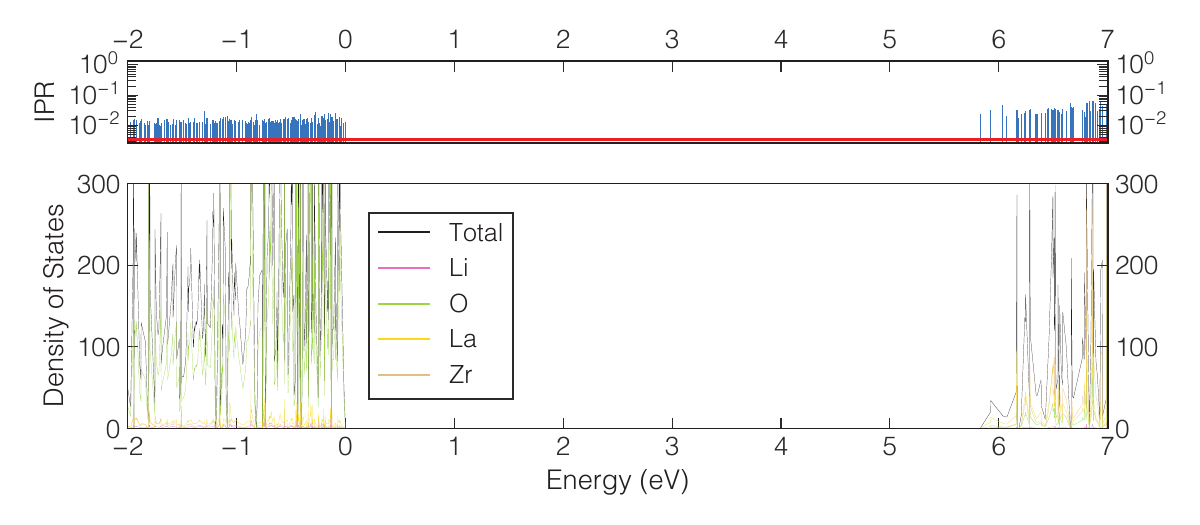}
    \caption{Inversed participation ratio, the density of states, and their projections of the bulk LLZO.}
    \label{fig:iprbulk}
\end{figure*}

\begin{figure*}[!h]
    \centering
    \includegraphics[width=1.\textwidth]{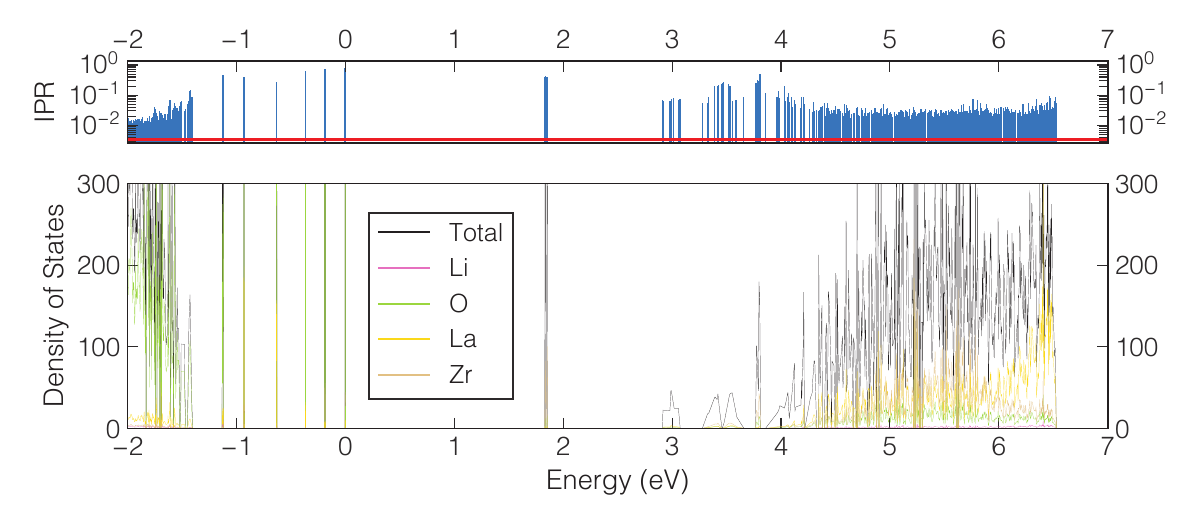}
    \caption{Inversed participation ratio, the density of states, and their projections of a $(001)$ Zr terminated surface of LLZO.}
    \label{fig:ipr}
\end{figure*}

\begin{figure*}
    \centering
    \includegraphics[width=1.\textwidth]{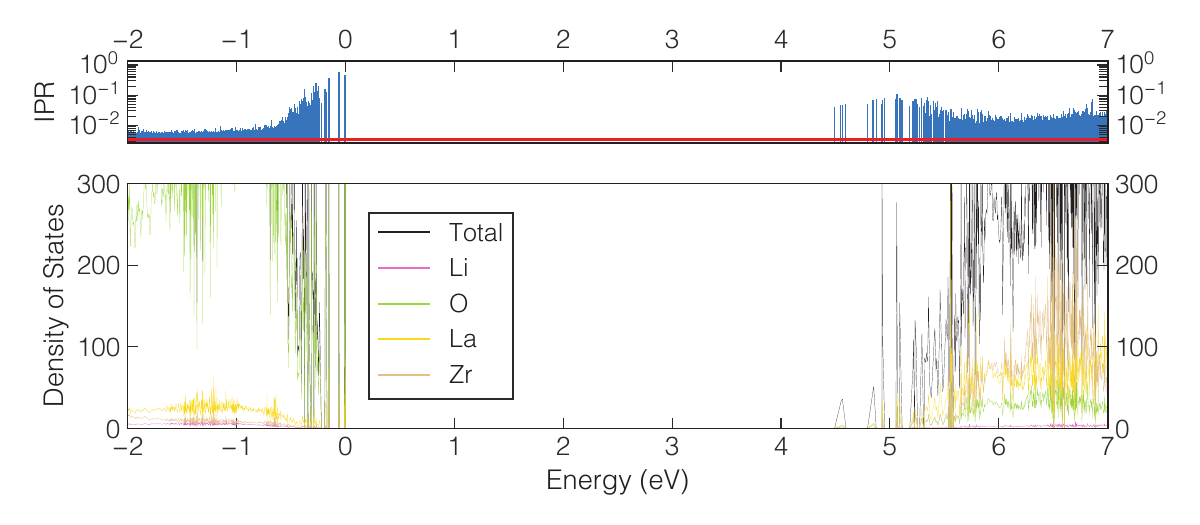}
    \caption{Inversed participation ratio, the density of states, and their projections of a \{001\} GB of LLZO. }
    \label{fig:iprgb}
\end{figure*}

 We investigate the $\left(001\right)$ Zr terminated surface, and the \{001\} GB, which shows the largest change in the bandgap from the bulk reference. The density of states, including the projections on LLZO atoms are shown together with the inversed participation ratio ($IPR$) in Figure~\ref{fig:iprbulk}, Figure~\ref{fig:ipr}, and Figure~\ref{fig:iprgb}. The $IPR$ is defined in Eq.~\ref{eq:ipr}
\begin{equation}
\label{eq:ipr}
IPR = N \, \cdot \, \frac{\sum\limits^{N}_{i=1}\left [p_i \right]^2}{\left(\sum\limits^{N}_{i=1}\left[p_i\right]\right)^2} \, \cdot 
\end{equation}
where  $p_i$ is the electron density at site $i$, and $N$ the total number of atoms in the system.

\clearpage
\section{Relationship of High-Energy Grain Boundary and Electronic Properties}\label{sec:excess_energy_bandgap}
\begin{figure*}[!ht]
\centering
    \includegraphics[width=1.0\textwidth]{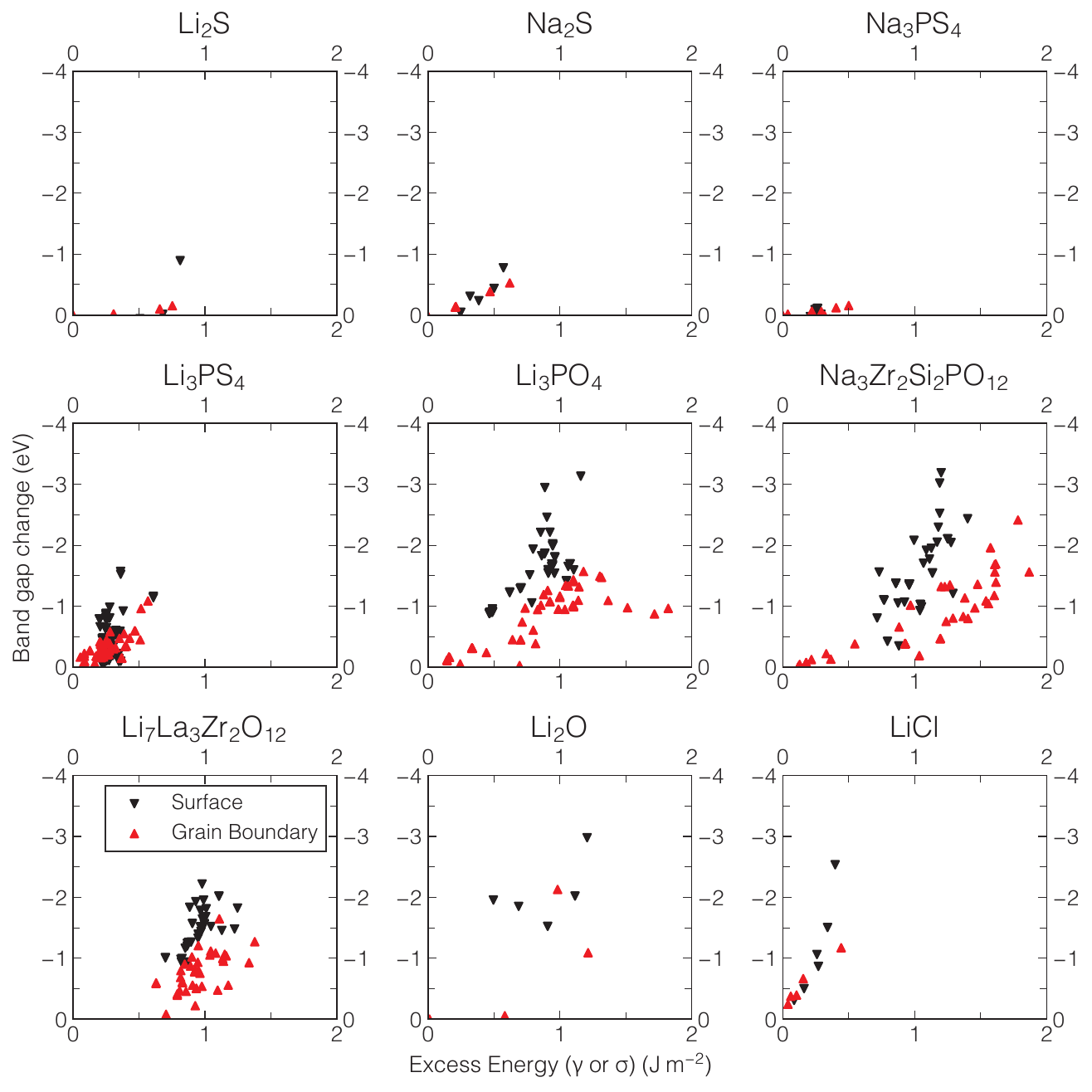}
    \caption{Relationship between excess energy $E_f$ (see Eq.~\ref{eq:grainboundaryformation}) and bandgap change in surfaces and GB of SEs.  }
    \label{fig:energy_bandgap}
\end{figure*}
Figure~\ref{fig:energy_bandgap} shows that high-energy (least stable) GBs and surfaces tend to show larger bandgap changes.





\clearpage
\section{Phase diagrams of solid electrolytes}\label{sec:phase_diagram}

Correct calculation of the values of $\sigma$ and $\gamma$ in off-stochiometric surfaces or grain boundaries would require an explicit definition of the chemical potentials. Therefore, we build the phase diagrams for the materials involved in this study and identify the combination of compounds that are in equilibrium. 

\begin{figure}[!ht]
    \centering
    \includegraphics[width=0.7\textwidth]{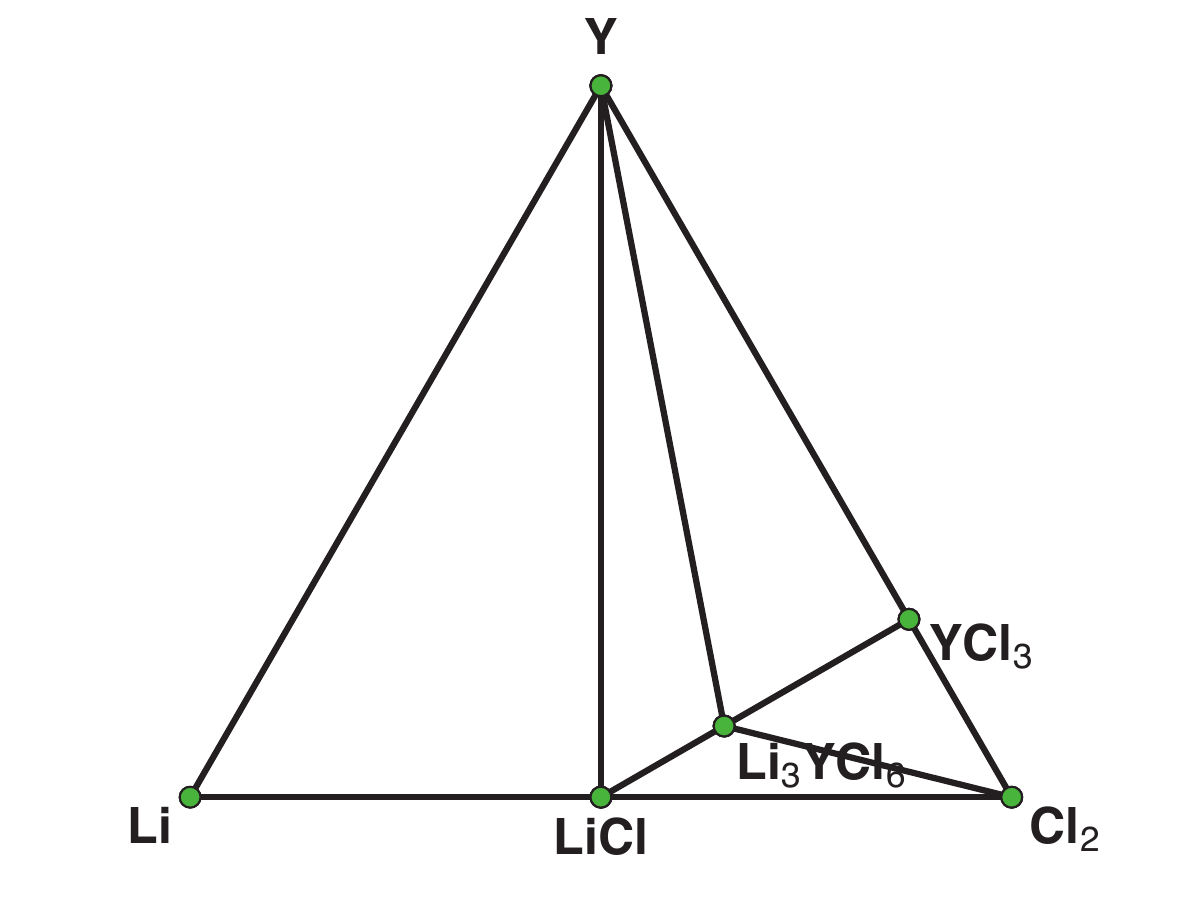}
    \caption{Computed \ce{Li_3YCl_6} phase diagram, where \ce{Li_3YCl_6} are assumed to be stable instead of metastable.}
    \label{fig:liyclphasediagram}
\end{figure}

\ce{Li_3YCl_6} contains Li/Vacancy disorder and we identify several Li/Vacancy orderings to be metastable. Figure~\ref{fig:liyclphasediagram} shows the computed phase diagram of \ce{Li_3YCl_6} where the \ce{Li_3YCl_6} is assumed to be stable, that is it is lowered by its metastability amount on the convex hull. This is a reasonable approximation as \ce{Li_3YCl_6} exists in nature. We found that \ce{LiCl}, \ce{YCl_3}, \ce{Cl_2} gas and \ce{Y} metal are in equilibrium with \ce{Li_3YCl_6}. By setting \ce{LiCl}, \ce{Y}, and \ce{Li_3YCl_6} as the reference points for calculating the chemical potentials of the species, we mimic a reducing environment of the synthesis process. Then the chemical potential can be calculated following the Eq.~\ref{eq:li3ycl6chemicalpotential}, where \textit{E}s are the DFT energies per formula unit, and $\mu$s are the values of chemical potential.

\begin{equation}
\label{eq:li3ycl6chemicalpotential}
\left\{
\begin{aligned}
& E_{\ce{Li_3YCl_6}}&&=3\mu_{\ce{Li}}+\mu_{\ce{Y}}+6\mu_{\ce{Cl}}\, , \\ 
& E_{\ce{LiCl}}&&=\mu_{\ce{Li}}+\mu_{\ce{Cl}} \, ,  \nonumber  \\
& E_{\ce{Y}}&& =\mu_{\ce{Y}} \,.  \nonumber 
\end{aligned} 
\right.
\end{equation}

The procedure elucidated for \ce{Li_3YCl_6} can be extended to other solid electrolytes, with the conditions exposed below.
In the LLZO system, we take Zr metal, \ce{La_2O_3}, \ce{Li_2O} and \ce{Li_7La_3Zr_2O_{12}} as the reference point for calculating chemical potential, following the results of our previously published in Ref.~\citenum{canepaParticleMorphologyLithium2018}. 

In the Argyrodite \ce{Li_6PS_5Cl} system, we have both S-rich and S-deficient grain boundaries. For the S deficient grain boundaries, we take a reported vertice\cite{goraiDevilDefectsElectronic2021} where the Argyrodite is in equilibrium with \ce{Li_4P_2S_6}, \ce{Li_2S}, and \ce{LiCl} and mimicking the S deficient condition. For S-rich grain boundaries, we take the condition that the Argyrodite is in equilibrium with \ce{Li_3PS_4}, \ce{S} and \ce{LiCl}, respectively.

In the NaSICON system, we take a facet in the five-dimensional phase diagram in equilibrium with NaSICON where \ce{ZrO2}, \ce{Na_2ZrSi_2O_7}, \ce{ZrSiO_4}, \ce{O_2} and \ce{Na_3Zr_2Si_2PO_{12}} are taken as reference points.

\clearpage
\section{Estimated fracture toughness of Solid-state Electrolytes}\label{sec:estimated_fracture}

Table~\ref{tab:synthesis} summarizes the mechanical properties and estimated fracture toughness of solid electrolytes. $W_{ad}^{min}$ is the minimum work of adhesion among the grain boundaries of respective materials in this work. The cleavage energy $W_f$ is derived from this work. Young's Modulus and Poisson's ratio are taken from the literature. In the row of fracture toughness, we show the lower limit of fracture toughness at the weakest grain boundary where the critical energy release rate is estimated by the minimum value of work of adhesion while assuming a 50\% reduction in Young's modulus. The upper limit of fracture toughness in the bulk region where the critical energy release rate is estimated by the fracture energy of bulk material. 

\begin{table*}[!ht]
    \caption{Mechanical properties of solid electrolytes, such as $\boldsymbol{W_{ad}^{min}/W_f}$ (in $\mathrm{J\; m^{-2}}$) and fracture toughness (in $\mathrm{\frac{MPa}{ \sqrt{m}}}$). For the estimation of the Young moduli (column {\bf E} in GPa) and Poisson's ratios (column $\pmb{\nu}$). The method of measurement is indicated in the bracket after the numerical value. Exp.\ stands for experimental results. In the case of simulated values, the exchange and correlation functional used in the DFT calculation is indicated. }
    \label{tab:synthesis}
    \begin{tabular*}{\textwidth}{@{\extracolsep{\fill}}lcllc@{}}
    \toprule
     {\bf Material} & $\boldsymbol{W_{ad}^{min}/W_f}$  &  {\bf E} &  $\pmb{\nu}$ & {\bf Fracture Toughness}   \\
     
     \midrule

     $\beta$-\ce{Li_3PS_4} & 0.14/0.40 & 29.5 (PBEsol)\cite{dengElasticPropertiesAlkali2016} & 0.29 (PBEsol)\cite{dengElasticPropertiesAlkali2016}& 0.047/0.114\\

     $\gamma$-\ce{Li_3PO_4}& 0.54/0.93  & 103.4 (PBEsol)\cite{dengElasticPropertiesAlkali2016}  &  0.26 (PBEsol)\cite{dengElasticPropertiesAlkali2016}&0.173/0.321\\

     $\beta$-\ce{Na_3PS_4}& 0.20/0.37 & 32.6 (PBEsol)\cite{dengElasticPropertiesAlkali2016}  &  0.25 (PBEsol)\cite{dengElasticPropertiesAlkali2016}&0.059/0.113\\

     $\alpha$-\ce{Na_3PS_4}& 0.18/0.40 & 33.6 (PBEsol)\cite{dengElasticPropertiesAlkali2016}  &  0.28 (PBEsol)\cite{dengElasticPropertiesAlkali2016}&0.057/0.121\\  

     \ce{Li_7La_3Zr_2O_{12}}& 0.61/1.81 & 163 (PBE){\cite{yuGrainBoundarySoftening2018}}   & 0.26  (Exp.){\cite{wolfenstineMechanicalBehaviorLiionconducting2018,niRoomTemperatureElastic2012}}&0.231/0.563\\    

     \ce{Na_3Zr_2Si_2PO_{12}}& 0.79/1.43 & 97 (Exp.)\cite{wolfenstineMechanicalPropertiesNaSICON2023,goodenoughFastNaIon1976}  & 0.26 (Exp.)\cite{wolfenstineMechanicalPropertiesNaSICON2023,goodenoughFastNaIon1976} &0.202/0.386\\    

     \ce{Li_6PS_5Cl}& 0.11/0.43& 22.1 (PBEsol)\cite{dengElasticPropertiesAlkali2016}   & 0.37 (PBEsol)\cite{dengElasticPropertiesAlkali2016}&0.038/0.105\\    

     \ce{Li_3YCl_6}&0.10/0.14& 41.9 (vdW-DF)\cite{kimMaterialDesignStrategy2021}&0.26 (vdW-DF)\cite{kimMaterialDesignStrategy2021}&0.047/0.079\\

     \ce{Li_2S}&0.19/0.66& 81.6 (PBEsol)\cite{dengElasticPropertiesAlkali2016}&0.19 (PBEsol)\cite{dengElasticPropertiesAlkali2016}&0.127/0.236 \\

     \ce{Li_2O}&0.39/0.99& 169.0 (PBEsol)\cite{dengElasticPropertiesAlkali2016}&0.16 (PBEsol)\cite{dengElasticPropertiesAlkali2016}&0.184/0.414\\

     \ce{LiCl}&0.24/0.17& 49.9 (PBE)\cite{wangStructuralElasticElectronic2020}&0.24 (PBE)\cite{wangStructuralElasticElectronic2020}&0.080/0.095\\

     \ce{Na_2S}&0.17/0.49& 46.2 (PBEsol)\cite{dengElasticPropertiesAlkali2016}&0.25 (PBEsol)\cite{dengElasticPropertiesAlkali2016}&0.090/0.155\\

    \bottomrule
  
    \end{tabular*}
\end{table*}

\clearpage

\bibliographystyle{plain}
\bibliography{biblio}

\begin{thebibliography}{100}

\bibitem{lawnContinuumAspectsCrack1993}
Continuum aspects of crack propagation {{I}}: Linear elastic crack-tip field.
\newblock In Brian Lawn, editor, {\em Fracture of {{Brittle Solids}}},
  Cambridge {{Solid State Science Series}}, pages 16--50. {Cambridge University
  Press}, {Cambridge}, 2 edition, 1993.

\bibitem{CRCHandbookChemistry2008}
{{CRC Handbook}} of {{Chemistry}} and {{Physics}}, 88th ed
  {{Editor-in-Chief}}:\, {{David R}}. {{Lide}} ({{National Institute}} of
  {{Standards}} and {{Technology}}) {{CRC Press}}/{{Taylor}} \& {{Francis
  Group}}:\, {{Boca Raton}}, {{FL}}. 2007. 2640 pp. \$139.95. {{ISBN}}
  0-8493-0488-1.
\newblock {\em Journal of the American Chemical Society}, 130(1):382--382,
  January 2008.

\bibitem{albertusStatusChallengesEnabling2017}
Paul Albertus, Susan Babinec, Scott Litzelman, and Aron Newman.
\newblock Status and challenges in enabling the lithium metal electrode for
  high-energy and low-cost rechargeable batteries.
\newblock {\em Nature Energy}, 3(1):16--21, December 2017.

\bibitem{andersonWorkFunctionLithium1949}
Paul~A. Anderson.
\newblock The {{Work Function}} of {{Lithium}}.
\newblock {\em Physical Review}, 75(8):1205--1207, April 1949.

\bibitem{asanoSolidHalideElectrolytes2018}
Tetsuya Asano, Akihiro Sakai, Satoru Ouchi, Masashi Sakaida, Akinobu Miyazaki,
  and Shinya Hasegawa.
\newblock Solid {{Halide Electrolytes}} with {{High Lithium-Ion Conductivity}}
  for {{Application}} in 4 {{V Class Bulk-Type All-Solid-State Batteries}}.
\newblock {\em Advanced Materials}, 30(44):1803075, November 2018.

\bibitem{athanasiouOperandoMeasurementsDendriteinduced2023d}
Christos~E. Athanasiou, Cole~D. Fincher, Colin Gilgenbach, Huajian Gao,
  W.~Craig Carter, Yet-Ming Chiang, and Brian~W. Sheldon.
\newblock Operando measurements of dendrite-induced stresses in ceramic
  electrolytes using photoelasticity.
\newblock {\em Matter}, November 2023.

\bibitem{bachmanInorganicSolidStateElectrolytes2016}
John~Christopher Bachman, Sokseiha Muy, Alexis Grimaud, Hao-Hsun Chang, Nir
  Pour, Simon~F. Lux, Odysseas Paschos, Filippo Maglia, Saskia Lupart, Peter
  Lamp, Livia Giordano, and Yang {Shao-Horn}.
\newblock Inorganic {{Solid-State Electrolytes}} for {{Lithium Batteries}}:
  {{Mechanisms}} and {{Properties Governing Ion Conduction}}.
\newblock {\em Chemical Reviews}, 116(1):140--162, January 2016.

\bibitem{bakirkandemirModelingAtomicStructure2014}
E.~Bakir~Kandemir, B.~G{\"o}n{\"u}l, G.~T. Barkema, K.~M. Yu, W.~Walukiewicz,
  and L.~W. Wang.
\newblock Modeling of the atomic structure and electronic properties of
  amorphous {{GaN1}}-{{xAsx}}.
\newblock {\em Computational Materials Science}, 82:100--106, February 2014.

\bibitem{baraiMechanicalStressInduced2019}
Pallab Barai, Kenneth Higa, Anh~T. Ngo, Larry~A. Curtiss, and Venkat
  Srinivasan.
\newblock Mechanical {{Stress Induced Current Focusing}} and {{Fracture}} in
  {{Grain Boundaries}}.
\newblock {\em Journal of The Electrochemical Society}, 166(10):A1752, May
  2019.

\bibitem{barsoumFundamentalsCeramics2019}
Michel~W. Barsoum.
\newblock {\em Fundamentals of {{Ceramics}}}.
\newblock {CRC Press}, 2 edition, December 2019.

\bibitem{becherMicrostructuralDesignToughened1991}
Paul~F. Becher.
\newblock Microstructural {{Design}} of {{Toughened Ceramics}}.
\newblock {\em Journal of the American Ceramic Society}, 74(2):255--269,
  February 1991.

\bibitem{bergerhoff1987crystallographic}
G~Bergerhoff, {\relax ID}~Brown, F~Allen, et~al.
\newblock Crystallographic databases.
\newblock {\em International Union of Crystallography, Chester}, 360:77--95,
  1987.

\bibitem{berngesScalingRelationsIonic2022}
Tim Bernges, Thorben B{\"o}ger, Oliver Maus, Paul~S. Till, Matthias~T. Agne,
  and Wolfgang~G. Zeier.
\newblock Scaling {{Relations}} for {{Ionic}} and {{Thermal Transport}} in the
  {{Na}}+ {{Ionic Conductor Na3PS4}}.
\newblock {\em ACS Materials Letters}, 4(12):2491--2498, December 2022.

\bibitem{blochlProjectorAugmentedwaveMethod1994}
P.~E. Bl{\"o}chl.
\newblock Projector augmented-wave method.
\newblock {\em Physical Review B}, 50(24):17953--17979, December 1994.

\bibitem{bozzoloViewpointFormationEvolution2020}
Nathalie Bozzolo and Marc Bernacki.
\newblock Viewpoint on the {{Formation}} and {{Evolution}} of {{Annealing Twins
  During Thermomechanical Processing}} of {{FCC Metals}} and {{Alloys}}.
\newblock {\em Metallurgical and Materials Transactions A}, 51(6):2665--2684,
  June 2020.

\bibitem{broekElementaryEngineeringFracture1982}
David Broek.
\newblock {\em Elementary Engineering Fracture Mechanics}.
\newblock {Springer Netherlands}, {Dordrecht}, 1982.

\bibitem{butlerDesigningInterfacesEnergy2019}
Keith~T. Butler, Gopalakrishnan Sai~Gautam, and Pieremanuele Canepa.
\newblock Designing interfaces in energy materials applications with
  first-principles calculations.
\newblock {\em npj Computational Materials}, 5(1):19, February 2019.

\bibitem{camacho-foreroExploringInterfacialStability2018}
Luis~E. {Camacho-Forero} and Perla~B. Balbuena.
\newblock Exploring interfacial stability of solid-state electrolytes at the
  lithium-metal anode surface.
\newblock {\em Journal of Power Sources}, 396:782--790, August 2018.

\bibitem{camacho-foreroElucidatingInterfacialPhenomena2020}
Luis~E. {Camacho-Forero} and Perla~B. Balbuena.
\newblock Elucidating {{Interfacial Phenomena}} between {{Solid-State
  Electrolytes}} and the {{Sulfur-Cathode}} of {{Lithium}}{\textendash}{{Sulfur
  Batteries}}.
\newblock {\em Chemistry of Materials}, 32(1):360--373, January 2020.

\bibitem{canepaParticleMorphologyLithium2018}
Pieremanuele Canepa, James~A. Dawson, Gopalakrishnan Sai~Gautam, Joel~M.
  Statham, Stephen~C. Parker, and M.~Saiful Islam.
\newblock Particle {{Morphology}} and {{Lithium Segregation}} to {{Surfaces}}
  of the {{Li}} {\textsubscript{7}} {{La}} {\textsubscript{3}} {{Zr}}
  {\textsubscript{2}} {{O}} {\textsubscript{12}} {{Solid Electrolyte}}.
\newblock {\em Chemistry of Materials}, 30(9):3019--3027, May 2018.

\bibitem{chengUnveilingStableNature2020}
Diyi Cheng, Thomas~A. Wynn, Xuefeng Wang, Shen Wang, Minghao Zhang, Ryosuke
  Shimizu, Shuang Bai, Han Nguyen, Chengcheng Fang, Min-cheol Kim, Weikang Li,
  Bingyu Lu, Suk~Jun Kim, and Ying~Shirley Meng.
\newblock Unveiling the {{Stable Nature}} of the {{Solid Electrolyte
  Interphase}} between {{Lithium Metal}} and {{LiPON}} via {{Cryogenic Electron
  Microscopy}}.
\newblock {\em Joule}, 4(11):2484--2500, November 2020.

\bibitem{chengIntergranularLiMetal2017}
Eric~Jianfeng Cheng, Asma Sharafi, and Jeff Sakamoto.
\newblock Intergranular {{Li}} metal propagation through polycrystalline
  {{Li6}}.{{25Al0}}.{{25La3Zr2O12}} ceramic electrolyte.
\newblock {\em Electrochimica Acta}, 223:85--91, January 2017.

\bibitem{culverEvidenceSolidElectrolyteInductive2020}
Sean~P. Culver, Alexander~G. Squires, Nicol{\`o} Minafra, Callum W.~F.
  Armstrong, Thorben Krauskopf, Felix B{\"o}cher, Cheng Li, Benjamin~J. Morgan,
  and Wolfgang~G. Zeier.
\newblock Evidence for a {{Solid-Electrolyte Inductive Effect}} in the
  {{Superionic Conductor Li10Ge1}}{\textendash}{{xSnxP2S12}}.
\newblock {\em Journal of the American Chemical Society}, 142(50):21210--21219,
  December 2020.

\bibitem{damoreSymmetryBreakingBulk2022}
Maddalena D'Amore, Loredana Edith~Daga, Riccardo Rocca, Mauro Francesco~Sgroi,
  Naiara Leticia~Marana, Silvia Maria~Casassa, Lorenzo Maschio, and Anna
  Maria~Ferrari.
\newblock From symmetry breaking in the bulk to phase transitions at the
  surface: A quantum-mechanical exploration of {{Li}} 6 {{PS}} 5 {{Cl}}
  argyrodite superionic conductor.
\newblock {\em Physical Chemistry Chemical Physics}, 24(37):22978--22986, 2022.

\bibitem{dawsonGoingGrainAtomistic2023}
James~A. Dawson.
\newblock Going against the {{Grain}}: {{Atomistic Modeling}} of {{Grain
  Boundaries}} in {{Solid Electrolytes}} for {{Solid-State Batteries}}.
\newblock {\em ACS Materials Au}, October 2023.

\bibitem{dawsonAtomicScaleInfluenceGrain2018}
James~A. Dawson, Pieremanuele Canepa, Theodosios Famprikis, Christian
  Masquelier, and M.~Saiful Islam.
\newblock Atomic-{{Scale Influence}} of {{Grain Boundaries}} on {{Li-Ion
  Conduction}} in {{Solid Electrolytes}} for {{All-Solid-State Batteries}}.
\newblock {\em Journal of the American Chemical Society}, 140(1):362--368,
  January 2018.

\bibitem{deiserothLi7PS6Li6PS5XCl2011}
Hans-J{\"o}rg Deiseroth, Joachim Maier, Katja Weichert, Vera Nickel, Shiao-Tong
  Kong, and Christof Reiner.
\newblock {{Li7PS6}} and {{Li6PS5X}} ({{X}}: {{Cl}}, {{Br}}, {{I}}): {{Possible
  Three-dimensional Diffusion Pathways}} for {{Lithium Ions}} and {{Temperature
  Dependence}} of the {{Ionic Conductivity}} by {{Impedance Measurements}}.
\newblock {\em Zeitschrift f{\"u}r anorganische und allgemeine Chemie},
  637(10):1287--1294, August 2011.

\bibitem{dengAutonomousHighthroughputMultiscale2022}
Zeyu Deng, Vipin Kumar, Felix~T. B{\"o}lle, Fernando Caro, Alejandro~A. Franco,
  Ivano~E. Castelli, Pieremanuele Canepa, and Zhi~Wei Seh.
\newblock Towards autonomous high-throughput multiscale modelling of battery
  interfaces.
\newblock {\em Energy \& Environmental Science}, 15(2):579--594, 2022.

\bibitem{dengPhaseBehaviorRhombohedral2020}
Zeyu Deng, Gopalakrishnan Sai~Gautam, Sanjeev~Krishna Kolli, Jean-No{\"e}l
  Chotard, Anthony~K. Cheetham, Christian Masquelier, and Pieremanuele Canepa.
\newblock Phase {{Behavior}} in {{Rhombohedral NaSiCON Electrolytes}} and
  {{Electrodes}}.
\newblock {\em Chemistry of Materials}, 32(18):7908--7920, September 2020.

\bibitem{dengElasticPropertiesAlkali2016}
Zhi Deng, Zhenbin Wang, Iek-Heng Chu, Jian Luo, and Shyue~Ping Ong.
\newblock Elastic {{Properties}} of {{Alkali Superionic Conductor
  Electrolytes}} from {{First Principles Calculations}}.
\newblock {\em Journal of The Electrochemical Society}, 163(2):A67--A74, 2016.

\bibitem{deysherTransportMechanicalAspects2022}
Grayson Deysher, Phillip Ridley, So-Yeon Ham, Jean-Marie Doux, Yu-Ting Chen,
  Erik~A. Wu, Darren H.~S. Tan, Ashley Cronk, Jihyun Jang, and Ying~Shirley
  Meng.
\newblock Transport and mechanical aspects of all-solid-state lithium
  batteries.
\newblock {\em Materials Today Physics}, 24:100679, May 2022.

\bibitem{dongFastHydrogenDiffusion2017}
Ying Dong, Fei Wang, and Wensheng Lai.
\newblock Fast hydrogen diffusion along {{$\Sigma$13}} grain boundary of
  {$\alpha$}-{{Al2O3}} and its suppression by the dopant {{Cr}}: {{A}}
  first-principles study.
\newblock {\em International Journal of Hydrogen Energy}, 42(15):10124--10130,
  April 2017.

\bibitem{ernstPreferredGrainOrientation2004}
Frank Ernst, Maureen~L. Mulvihill, Oliver Kienzle, and Manfred R{\"u}hle.
\newblock Preferred {{Grain Orientation Relationships}} in {{Sintered
  Perovskite Ceramics}}.
\newblock {\em Journal of the American Ceramic Society}, 84(8):1885--1890,
  December 2004.

\bibitem{evansPerspectiveDevelopmentHighToughness1990}
Anthony~G. Evans.
\newblock Perspective on the {{Development}} of {{High-Toughness Ceramics}}.
\newblock {\em Journal of the American Ceramic Society}, 73(2):187--206, 1990.

\bibitem{ewaldBerechnungOptischerUnd1921}
P.~P. Ewald.
\newblock Die {{Berechnung}} optischer und elektrostatischer
  {{Gitterpotentiale}}.
\newblock {\em Annalen der Physik}, 369(3):253--287, 1921.

\bibitem{famprikisInsightsRichPolymorphism2021}
Theodosios Famprikis, Houssny Bouyanfif, Pieremanuele Canepa, Mohamed Zbiri,
  James~A. Dawson, Emmanuelle Suard, Fran{\c c}ois Fauth, Helen~Y. Playford,
  Damien Dambournet, Olaf~J. Borkiewicz, Matthieu Courty, Oliver Clemens,
  Jean-No{\"e}l Chotard, M.~Saiful Islam, and Christian Masquelier.
\newblock Insights into the {{Rich Polymorphism}} of the {{Na}}
  {\textsuperscript{+}} {{Ion Conductor Na}} {\textsubscript{3}} {{PS}}
  {\textsubscript{4}} from the {{Perspective}} of {{Variable-Temperature
  Diffraction}} and {{Spectroscopy}}.
\newblock {\em Chemistry of Materials}, 33(14):5652--5667, July 2021.

\bibitem{famprikisFundamentalsInorganicSolidstate2019}
Theodosios Famprikis, Pieremanuele Canepa, James~A. Dawson, M.~Saiful Islam,
  and Christian Masquelier.
\newblock Fundamentals of inorganic solid-state electrolytes for batteries.
\newblock {\em Nature Materials}, 18(12):1278--1291, December 2019.

\bibitem{famprikisPressureMechanochemicalEffects2020}
Theodosios Famprikis, {\"O}.~Ula{\c s} Kudu, James~A. Dawson, Pieremanuele
  Canepa, Fran{\c c}ois Fauth, Emmanuelle Suard, Mohamed Zbiri, Damien
  Dambournet, Olaf~J. Borkiewicz, Houssny Bouyanfif, Steffen~P. Emge, Sorina
  Cretu, Jean-No{\"e}l Chotard, Clare~P. Grey, Wolfgang~G. Zeier, M.~Saiful
  Islam, and Christian Masquelier.
\newblock Under {{Pressure}}: {{Mechanochemical Effects}} on {{Structure}} and
  {{Ion Conduction}} in the {{Sodium-Ion Solid Electrolyte Na}}
  {\textsubscript{3}} {{PS}} {\textsubscript{4}}.
\newblock {\em Journal of the American Chemical Society}, 142(43):18422--18436,
  October 2020.

\bibitem{fuchsCurrentDependentLithium2023}
Till Fuchs, Juri Becker, Catherine~G. Haslam, Christian Lerch, Jeff Sakamoto,
  Felix~H. Richter, and J{\"u}rgen Janek.
\newblock Current-{{Dependent Lithium Metal Growth Modes}} in
  ``{{Anode}}-{{Free}}'' {{Solid}}-{{State Batteries}} at the {{Cu}}|{{LLZO
  Interface}}.
\newblock {\em Advanced Energy Materials}, 13(1):2203174, January 2023.

\bibitem{gaoSurfaceDependentStabilityInterface2020}
Bo~Gao, Randy Jalem, and Yoshitaka Tateyama.
\newblock Surface-{{Dependent Stability}} of the {{Interface}} between {{Garnet
  Li7La3Zr2O12}} and the {{Li Metal}} in the {{All-Solid-State Battery}} from
  {{First-Principles Calculations}}.
\newblock {\em ACS Applied Materials \& Interfaces}, 12(14):16350--16358, April
  2020.

\bibitem{garcia-mendezEffectProcessingConditions2017}
Regina {Garcia-Mendez}, Fuminori Mizuno, Ruigang Zhang, Timothy~S. Arthur, and
  Jeff Sakamoto.
\newblock Effect of {{Processing Conditions}} of {{75Li2S-25P2S5 Solid
  Electrolyte}} on its {{DC Electrochemical Behavior}}.
\newblock {\em Electrochimica Acta}, 237:144--151, May 2017.

\bibitem{goodenoughFastNaIon1976}
J.~B. Goodenough, H.~Y-P. Hong, and J.~A. Kafalas.
\newblock Fast {{Na}}+-ion transport in skeleton structures.
\newblock {\em Materials Research Bulletin}, 11(2):203--220, February 1976.

\bibitem{goodenoughChallengesRechargeableLi2010}
John~B. Goodenough and Youngsik Kim.
\newblock Challenges for {{Rechargeable Li Batteries}}.
\newblock {\em Chemistry of Materials}, 22(3):587--603, February 2010.

\bibitem{goodwinProtectiveNaSICONInterlayer2023}
Laura~E. Goodwin, Paul Till, Monika Bhardwaj, Nazia Nazer, Philipp Adelhelm,
  Frank Tietz, Wolfgang~G. Zeier, Felix~H. Richter, and J{\"u}rgen Janek.
\newblock Protective {{NaSICON Interlayer}} between a
  {{Sodium}}{\textendash}{{Tin Alloy Anode}} and {{Sulfide-Based Solid
  Electrolytes}} for {{All-Solid-State Sodium Batteries}}.
\newblock {\em ACS Applied Materials \& Interfaces}, 15(43):50457--50468,
  November 2023.

\bibitem{goraiDevilDefectsElectronic2021}
Prashun Gorai, Theodosios Famprikis, Baltej Singh, Vladan Stevanovi{\'c}, and
  Pieremanuele Canepa.
\newblock Devil is in the {{Defects}}: {{Electronic Conductivity}} in {{Solid
  Electrolytes}}.
\newblock {\em Chemistry of Materials}, 33(18):7484--7498, September 2021.

\bibitem{guoAchievingHighCritical2021}
Haojie Guo, Jianmeng Su, Wenping Zha, Tongping Xiu, Zhen Song, Michael~E.
  Badding, Jun Jin, and Zhaoyin Wen.
\newblock Achieving high critical current density in {{Ta-doped
  Li7La3Zr2O12}}/{{MgO}} composite electrolytes.
\newblock {\em Journal of Alloys and Compounds}, 856:157222, March 2021.

\bibitem{hanHighElectronicConductivity2019}
Fudong Han, Andrew~S. Westover, Jie Yue, Xiulin Fan, Fei Wang, Miaofang Chi,
  Donovan~N. Leonard, Nancy~J. Dudney, Howard Wang, and Chunsheng Wang.
\newblock High electronic conductivity as the origin of lithium dendrite
  formation within solid electrolytes.
\newblock {\em Nature Energy}, 4(3):187--196, March 2019.

\bibitem{hanSuppressingLiDendrite2018}
Fudong Han, Jie Yue, Xiangyang Zhu, and Chunsheng Wang.
\newblock Suppressing {{Li Dendrite Formation}} in {{Li}} {\textsubscript{2}}
  {{S}}-{{P}} {\textsubscript{2}} {{S}} {\textsubscript{5}} {{Solid
  Electrolyte}} by {{LiI Incorporation}}.
\newblock {\em Advanced Energy Materials}, 8(18):1703644, June 2018.

\bibitem{hayashiEffectHydrothermalTemperature2018}
Hiromichi Hayashi and Takeo Ebina.
\newblock Effect of hydrothermal temperature on the tetragonality of
  {{BaTiO}}{\textsubscript{3}} nanoparticles and in-situ {{Raman}} spectroscopy
  under tetragonal{\textendash}cubic transformation.
\newblock {\em Journal of the Ceramic Society of Japan}, 126(3):214--220, 2018.

\bibitem{holmSurfaceFormationEnergy1998}
Elizabeth~A. Holm.
\newblock Surface {{Formation Energy}} for {{Intergranular Fracture}} in
  {{Two-Dimensional Polycrystals}}.
\newblock {\em Journal of the American Ceramic Society}, 81(3):455--459, 1998.

\bibitem{holmCriticalManifoldsPolycrystalline2004}
Elizabeth~A. Holm, J.~H. Meinke, E.~S. McGarrity, and P.~M. Duxbury.
\newblock Critical {{Manifolds}} in {{Polycrystalline Grain Structures}}.
\newblock {\em Materials Science Forum}, 467--470:1039--1044, 2004.

\bibitem{hommaCrystalStructurePhase2011}
Kenji Homma, Masao Yonemura, Takeshi Kobayashi, Miki Nagao, Masaaki Hirayama,
  and Ryoji Kanno.
\newblock Crystal structure and phase transitions of the lithium ionic
  conductor {{Li3PS4}}.
\newblock {\em Solid State Ionics}, 182(1):53--58, February 2011.

\bibitem{hongCrystalStructuresCrystal1976}
H.~Y-P. Hong.
\newblock Crystal structures and crystal chemistry in the system
  {{Na1}}+{{xZr2SixP3}}-{{xO12}}.
\newblock {\em Materials Research Bulletin}, 11(2):173--182, February 1976.

\bibitem{huCarrierGrainBoundary2022}
Chaoliang Hu, Kaiyang Xia, Chenguang Fu, Xinbing Zhao, and Tiejun Zhu.
\newblock Carrier grain boundary scattering in thermoelectric materials.
\newblock {\em Energy \& Environmental Science}, 15(4):1406--1422, April 2022.

\bibitem{huElasticModulusHardness2021}
Shan Hu, Pengyu Xu, Luize~Scalco De~Vasconcelos, Lia Stanciu, Hongwei Ni, and
  Kejie Zhao.
\newblock Elastic {{Modulus}}, {{Hardness}}, and {{Fracture Toughness}} of
  {{Li}} {\textsubscript{6.4}} {{La}} {\textsubscript{3}} {{Zr}}
  {\textsubscript{1.4}} {{Ta}} {\textsubscript{0.6}} {{O}} {\textsubscript{12}}
  {{Solid Electrolyte}}.
\newblock {\em Chinese Physics Letters}, 38(9):098401, October 2021.

\bibitem{iguchiInfluenceGrainStructures2006}
Fumitada Iguchi, Takehisa Yamada, Noriko Sata, Takao Tsurui, and Hiroo Yugami.
\newblock The influence of grain structures on the electrical conductivity of a
  {{BaZr0}}.{{95Y0}}.{{05O3}} proton conductor.
\newblock {\em Solid State Ionics}, 177(26):2381--2384, October 2006.

\bibitem{ishigakiRoomTemperatureSynthesis2021}
Norikazu Ishigaki and Junji Akimoto.
\newblock Room temperature synthesis and phase transformation of lithium
  phosphate {{Li}} {\textsubscript{3}} {{PO}} {\textsubscript{4}} as solid
  electrolyte.
\newblock {\em Journal of Asian Ceramic Societies}, 9(2):452--458, April 2021.

\bibitem{jalemLithiumDynamicsGrain2023}
Randy Jalem, Manas~Likhit Holekevi~Chandrappa, Ji~Qi, Yoshitaka Tateyama, and
  Shyue Ping~Ong.
\newblock Lithium dynamics at grain boundaries of {$\beta$}-{{Li}} 3 {{PS}} 4
  solid electrolyte.
\newblock {\em Energy Advances}, 2(12):2029--2041, 2023.

\bibitem{jiKineticallyStableAnode2021}
Weixiao Ji, Dong Zheng, Xiaoxiao Zhang, Tianyao Ding, and Deyang Qu.
\newblock A kinetically stable anode interface for {{Li3YCl6-based}}
  all-solid-state lithium batteries.
\newblock {\em Journal of Materials Chemistry A}, 9(26):15012--15018, July
  2021.

\bibitem{joklMicroscopicTheoryBrittle1980}
M.~L Jokl, V~Vitek, and C.~J McMahon.
\newblock A microscopic theory of brittle fracture in deformable solids: {{A}}
  relation between ideal work to fracture and plastic work.
\newblock {\em Acta Metallurgica}, 28(11):1479--1488, November 1980.

\bibitem{kalnausSolidstateBatteriesCritical2023}
Sergiy Kalnaus, Nancy~J. Dudney, Andrew~S. Westover, Erik Herbert, and Steve
  Hackney.
\newblock Solid-state batteries: {{The}} critical role of mechanics.
\newblock {\em Science}, 381(6664):eabg5998, September 2023.

\bibitem{kasemchainanCriticalStrippingCurrent2019}
Jitti Kasemchainan, Stefanie Zekoll, Dominic Spencer~Jolly, Ziyang Ning,
  Gareth~O. Hartley, James Marrow, and Peter~G. Bruce.
\newblock Critical stripping current leads to dendrite formation on plating in
  lithium anode solid electrolyte cells.
\newblock {\em Nature Materials}, 18(10):1105--1111, October 2019.

\bibitem{katoXPSSEMAnalysis2018}
Atsutaka Kato, Hiroe Kowada, Minako Deguchi, Chie Hotehama, Akitoshi Hayashi,
  and Masahiro Tatsumisago.
\newblock {{XPS}} and {{SEM}} analysis between {{Li}}/{{Li3PS4}} interface with
  {{Au}} thin film for all-solid-state lithium batteries.
\newblock {\em Solid State Ionics}, 322:1--4, September 2018.

\bibitem{kazyakLiPenetrationCeramic2020}
Eric Kazyak, Regina {Garcia-Mendez}, William~S. LePage, Asma Sharafi, Andrew~L.
  Davis, Adrian~J. Sanchez, Kuan-Hung Chen, Catherine Haslam, Jeff Sakamoto,
  and Neil~P. Dasgupta.
\newblock Li {{Penetration}} in {{Ceramic Solid Electrolytes}}: {{Operando
  Microscopy Analysis}} of {{Morphology}}, {{Propagation}}, and
  {{Reversibility}}.
\newblock {\em Matter}, 2(4):1025--1048, April 2020.

\bibitem{kienzleAtomisticStructure1111998}
O.~Kienzle, M.~Exner, and F.~Ernst.
\newblock Atomistic {{Structure}} of {{$\Sigma$}} = 3, (111) {{Grain
  Boundaries}} in {{Strontium Titanate}}.
\newblock {\em physica status solidi (a)}, 166(1):57--71, 1998.

\bibitem{kimSolidStateLi2021}
Kun~Joong Kim, Moran Balaish, Masaki Wadaguchi, Lingping Kong, and Jennifer
  L.~M. Rupp.
\newblock Solid-{{State Li}}{\textendash}{{Metal Batteries}}: {{Challenges}}
  and {{Horizons}} of {{Oxide}} and {{Sulfide Solid Electrolytes}} and {{Their
  Interfaces}}.
\newblock {\em Advanced Energy Materials}, 11(1):2002689, January 2021.

\bibitem{kimSolidStateLiMetal2021}
Kun~Joong Kim, Moran Balaish, Masaki Wadaguchi, Lingping Kong, and Jennifer
  L.~M. Rupp.
\newblock Solid-{{State Li}}{\textendash}{{Metal Batteries}}: {{Challenges}}
  and {{Horizons}} of {{Oxide}} and {{Sulfide Solid Electrolytes}} and {{Their
  Interfaces}}.
\newblock {\em Advanced Energy Materials}, 11(1):2002689, 2021.

\bibitem{kimMaterialDesignStrategy2021}
Kwangnam Kim, Dongsu Park, Hun-Gi Jung, Kyung~Yoon Chung, Joon~Hyung Shim,
  Brandon~C. Wood, and Seungho Yu.
\newblock Material {{Design Strategy}} for {{Halide Solid Electrolytes Li3MX6}}
  ({{X}} = {{Cl}}, {{Br}}, and {{I}}) for {{All-Solid-State High-Voltage Li-Ion
  Batteries}}.
\newblock {\em Chemistry of Materials}, 33(10):3669--3677, May 2021.

\bibitem{kobayashiDynamicFractureCeramics1991}
Albert~S. Kobayashi.
\newblock Dynamic fracture of ceramics and ceramic composites.
\newblock {\em Materials Science and Engineering: A}, 143(1):111--117,
  September 1991.

\bibitem{koerverCapacityFadeSolidState2017}
Raimund Koerver, Isabel Ayg{\"u}n, Thomas Leichtwei{\ss}, Christian Dietrich,
  Wenbo Zhang, Jan~O. Binder, Pascal Hartmann, Wolfgang~G. Zeier, and
  J{\"u}rgen Janek.
\newblock Capacity {{Fade}} in {{Solid-State Batteries}}: {{Interphase
  Formation}} and {{Chemomechanical Processes}} in {{Nickel-Rich Layered Oxide
  Cathodes}} and {{Lithium Thiophosphate Solid Electrolytes}}.
\newblock {\em Chemistry of Materials}, 29(13):5574--5582, July 2017.

\bibitem{kraftInfluenceLatticePolarizability2017}
Marvin~A. Kraft, Sean~P. Culver, Mario Calderon, Felix B{\"o}cher, Thorben
  Krauskopf, Anatoliy Senyshyn, Christian Dietrich, Alexandra Zevalkink,
  J{\"u}rgen Janek, and Wolfgang~G. Zeier.
\newblock Influence of {{Lattice Polarizability}} on the {{Ionic Conductivity}}
  in the {{Lithium Superionic Argyrodites Li}} {\textsubscript{6}} {{PS}}
  {\textsubscript{5}} {{X}} ({{X}} = {{Cl}}, {{Br}}, {{I}}).
\newblock {\em Journal of the American Chemical Society}, 139(31):10909--10918,
  August 2017.

\bibitem{kraftStatisticalInvestigationEffects2008}
R.~H. Kraft and J.~F. Molinari.
\newblock A statistical investigation of the effects of grain boundary
  properties on transgranular fracture.
\newblock {\em Acta Materialia}, 56(17):4739--4749, October 2008.

\bibitem{krauskopfBottleneckDiffusionInductive2018}
Thorben Krauskopf, Sean~P. Culver, and Wolfgang~G. Zeier.
\newblock Bottleneck of {{Diffusion}} and {{Inductive Effects}} in
  {{Li10Ge1}}{\textendash}{{xSnxP2S12}}.
\newblock {\em Chemistry of Materials}, 30(5):1791--1798, March 2018.

\bibitem{krauskopfLocalTetragonalStructure2018}
Thorben Krauskopf, Sean~P. Culver, and Wolfgang~G. Zeier.
\newblock Local {{Tetragonal Structure}} of the {{Cubic Superionic Conductor
  Na3PS4}}.
\newblock {\em Inorganic Chemistry}, 57(8):4739--4744, April 2018.

\bibitem{krauskopfComparingDescriptorsInvestigating2018}
Thorben Krauskopf, Sokseiha Muy, Sean~P. Culver, Saneyuki Ohno, Olivier
  Delaire, Yang {Shao-Horn}, and Wolfgang~G. Zeier.
\newblock Comparing the {{Descriptors}} for {{Investigating}} the {{Influence}}
  of {{Lattice Dynamics}} on {{Ionic Transport Using}} the {{Superionic
  Conductor Na3PS4}}{\textendash}{{xSex}}.
\newblock {\em Journal of the American Chemical Society}, 140(43):14464--14473,
  October 2018.

\bibitem{kresseEfficientIterativeSchemes1996}
G.~Kresse and J.~Furthm{\"u}ller.
\newblock Efficient iterative schemes for ab initio total-energy calculations
  using a plane-wave basis set.
\newblock {\em Physical Review B}, 54(16):11169--11186, October 1996.

\bibitem{kresseInitioMolecularDynamics1993}
G.~Kresse and J.~Hafner.
\newblock Ab initio molecular dynamics for liquid metals.
\newblock {\em Physical Review B}, 47(1):558--561, January 1993.

\bibitem{kresseUltrasoftPseudopotentialsProjector1999}
G.~Kresse and D.~Joubert.
\newblock From ultrasoft pseudopotentials to the projector augmented-wave
  method.
\newblock {\em Physical Review B}, 59(3):1758--1775, January 1999.

\bibitem{kuduReviewStructuralProperties2018}
{\"O}mer~Ula{\c s} Kudu, Theodosios Famprikis, Benoit Fleutot, Marc-David
  Braida, Thierry Le~Mercier, M.~Sa{\"i}ful Islam, and Christian Masquelier.
\newblock A review of structural properties and synthesis methods of solid
  electrolyte materials in the {{Li2S}} - {{P2S5}} binary system.
\newblock {\em Journal of Power Sources}, 407:31--43, December 2018.

\bibitem{kurtzEffectsGrainBoundary2004}
R.J Kurtz and H.L Heinisch.
\newblock The effects of grain boundary structure on binding of {{He}} in
  {{Fe}}.
\newblock {\em Journal of Nuclear Materials}, 329--333:1199--1203, August 2004.

\bibitem{lepleyStructuresLi2013}
N.~D. Lepley, N.~A.~W. Holzwarth, and Yaojun~A. Du.
\newblock Structures, {{Li}}\$\{\}\^\{+\}\$ mobilities, and interfacial
  properties of solid electrolytes {{Li}}\$\{\}\_\{3\}\${{PS}}\$\{\}\_\{4\}\$
  and {{Li}}\$\{\}\_\{3\}\${{PO}}\$\{\}\_\{4\}\$ from first principles.
\newblock {\em Physical Review B}, 88(10):104103, September 2013.

\bibitem{leungSpatialHeterogeneitiesOnset2017}
Kevin Leung and Katherine~L. Jungjohann.
\newblock Spatial {{Heterogeneities}} and {{Onset}} of {{Passivation
  Breakdown}} at {{Lithium Anode Interfaces}}.
\newblock {\em The Journal of Physical Chemistry C}, 121(37):20188--20196,
  September 2017.

\bibitem{liRelativeGrainBoundary2009}
Jia Li, Shen~J. Dillon, and Gregory~S. Rohrer.
\newblock Relative grain boundary area and energy distributions in nickel.
\newblock {\em Acta Materialia}, 57(14):4304--4311, August 2009.

\bibitem{liProgressPerspectivesHalide2020}
Xiaona Li, Jianwen Liang, Xiaofei Yang, Keegan~R. Adair, Changhong Wang,
  Feipeng Zhao, and Xueliang Sun.
\newblock Progress and perspectives on halide lithium conductors for
  all-solid-state lithium batteries.
\newblock {\em Energy \& Environmental Science}, 13(5):1429--1461, 2020.

\bibitem{linRoleCohesiveZone2017}
Liqiang Lin, Xianqiao Wang, and Xiaowei Zeng.
\newblock The role of cohesive zone properties on intergranular to
  transgranular fracture transition in polycrystalline solids.
\newblock {\em International Journal of Damage Mechanics}, 26(3):379--394,
  April 2017.

\bibitem{liuUnderstandingElectrochemicalPotentials2016}
Chaofeng Liu, Zachary~G. Neale, and Guozhong Cao.
\newblock Understanding electrochemical potentials of cathode materials in
  rechargeable batteries.
\newblock {\em Materials Today}, 19(2):109--123, March 2016.

\bibitem{liuPathwaysPracticalHighenergy2019}
Jun Liu, Zhenan Bao, Yi~Cui, Eric~J. Dufek, John~B. Goodenough, Peter Khalifah,
  Qiuyan Li, Bor~Yann Liaw, Ping Liu, Arumugam Manthiram, Y.~Shirley Meng,
  Venkat~R. Subramanian, Michael~F. Toney, Vilayanur~V. Viswanathan, M.~Stanley
  Whittingham, Jie Xiao, Wu~Xu, Jihui Yang, Xiao-Qing Yang, and Ji-Guang Zhang.
\newblock Pathways for practical high-energy long-cycling lithium metal
  batteries.
\newblock {\em Nature Energy}, 4(3):180--186, March 2019.

\bibitem{liuLocalElectronicStructure2021}
Xiaoming Liu, Regina {Garcia-Mendez}, Andrew~R. Lupini, Yongqiang Cheng,
  Zachary~D. Hood, Fudong Han, Asma Sharafi, Juan~Carlos Idrobo, Nancy~J.
  Dudney, Chunsheng Wang, Cheng Ma, Jeff Sakamoto, and Miaofang Chi.
\newblock Local electronic structure variation resulting in {{Li}} `filament'
  formation within solid electrolytes.
\newblock {\em Nature Materials}, 20(11):1485--1490, November 2021.

\bibitem{maranaComputationalCharacterizationVLi3PS42022}
Naiara~Leticia Marana, Mauro~Francesco Sgroi, Lorenzo Maschio, Anna~Maria
  Ferrari, Maddalena D'Amore, and Silvia Casassa.
\newblock Computational {{Characterization}} of {$\beta$}-{{Li3PS4 Solid
  Electrolyte}}: {{From Bulk}} and {{Surfaces}} to {{Nanocrystals}}.
\newblock {\em Nanomaterials}, 12(16):2795, August 2022.

\bibitem{marinescuDeformationFractureCeramic1999}
Ioan Marinescu and Mariana Pruteanu.
\newblock 2 - {{Deformation}} and {{Fracture}} of {{Ceramic Materials}}.
\newblock In Ioan~D. Marinescu, Hans~K. Tonshoff, and Ichiro Inasaki, editors,
  {\em Handbook of {{Ceramic Grinding}} \& {{Polishing}}}, pages 65--93.
  {William Andrew Publishing}, {Norwich, NY}, January 1999.

\bibitem{mecholskyFractureSurfaceAnalysis1976}
J.~J. Mecholsky, S.~W. Freimam, and R.~W. Rice.
\newblock Fracture surface analysis of ceramics.
\newblock {\em Journal of Materials Science}, 11(7):1310--1319, July 1976.

\bibitem{milanRoleGrainBoundaries2023}
Emily Milan and Mauro Pasta.
\newblock The role of grain boundaries in solid-state {{Li-metal}} batteries.
\newblock {\em Materials Futures}, 2(1):013501, March 2023.

\bibitem{monismithGrainboundaryFractureMechanisms2022}
Scott Monismith, Jianmin Qu, and Remi Dingreville.
\newblock Grain-boundary fracture mechanisms in {{Li7La3Zr2O12}} ({{LLZO}})
  solid electrolytes: {{When}} phase transformation acts as a
  temperature-dependent toughening mechanism.
\newblock {\em Journal of the Mechanics and Physics of Solids}, 160:104791,
  March 2022.

\bibitem{monroeImpactElasticDeformation2005}
Charles Monroe and John Newman.
\newblock The {{Impact}} of {{Elastic Deformation}} on {{Deposition Kinetics}}
  at {{Lithium}}/{{Polymer Interfaces}}.
\newblock {\em Journal of The Electrochemical Society}, 152(2):A396, 2005.

\bibitem{muruganFastLithiumIon2007}
Ramaswamy Murugan, Venkataraman Thangadurai, and Werner Weppner.
\newblock Fast {{Lithium Ion Conduction}} in {{Garnet-Type Li7La3Zr2O12}}.
\newblock {\em Angewandte Chemie International Edition}, 46(41):7778--7781,
  October 2007.

\bibitem{niRoomTemperatureElastic2012}
Jennifer~E. Ni, Eldon~D. Case, Jeffrey~S. Sakamoto, Ezhiyl Rangasamy, and
  Jeffrey~B. Wolfenstine.
\newblock Room temperature elastic moduli and {{Vickers}} hardness of
  hot-pressed {{LLZO}} cubic garnet.
\newblock {\em Journal of Materials Science}, 47(23):7978--7985, December 2012.

\bibitem{nikodimosHalideSolidStateElectrolytes2023}
Yosef Nikodimos, Wei-Nien Su, and Bing~Joe Hwang.
\newblock Halide {{Solid-State Electrolytes}}: {{Stability}} and
  {{Application}} for {{High Voltage All-Solid-State Li Batteries}}.
\newblock {\em Advanced Energy Materials}, 13(3):2202854, 2023.

\bibitem{ningVisualizingPlatinginducedCracking2021}
Ziyang Ning, Dominic~Spencer Jolly, Guanchen Li, Robin De~Meyere, Shengda~D.
  Pu, Yang Chen, Jitti Kasemchainan, Johannes Ihli, Chen Gong, Boyang Liu,
  Dominic L.~R. Melvin, Anne Bonnin, Oxana Magdysyuk, Paul Adamson, Gareth~O.
  Hartley, Charles~W. Monroe, T.~James Marrow, and Peter~G. Bruce.
\newblock Visualizing plating-induced cracking in lithium-anode
  solid-electrolyte cells.
\newblock {\em Nature Materials}, 20(8):1121--1129, August 2021.

\bibitem{ningDendriteInitiationPropagation2023}
Ziyang Ning, Guanchen Li, Dominic L.~R. Melvin, Yang Chen, Junfu Bu, Dominic
  {Spencer-Jolly}, Junliang Liu, Bingkun Hu, Xiangwen Gao, Johann Perera, Chen
  Gong, Shengda~D. Pu, Shengming Zhang, Boyang Liu, Gareth~O. Hartley,
  Andrew~J. Bodey, Richard~I. Todd, Patrick~S. Grant, David E.~J. Armstrong,
  T.~James Marrow, Charles~W. Monroe, and Peter~G. Bruce.
\newblock Dendrite initiation and propagation in lithium metal solid-state
  batteries.
\newblock {\em Nature}, 618(7964):287--293, June 2023.

\bibitem{ongPythonMaterialsGenomics2013a}
Shyue~Ping Ong, William~Davidson Richards, Anubhav Jain, Geoffroy Hautier,
  Michael Kocher, Shreyas Cholia, Dan Gunter, Vincent~L. Chevrier, Kristin~A.
  Persson, and Gerbrand Ceder.
\newblock Python {{Materials Genomics}} (pymatgen): {{A}} robust, open-source
  python library for materials analysis.
\newblock {\em Computational Materials Science}, 68:314--319, February 2013.

\bibitem{ortmannKineticsPoreFormation2023}
Till Ortmann, Simon Burkhardt, Janis~Kevin Eckhardt, Till Fuchs, Ziming Ding,
  Joachim Sann, Marcus Rohnke, Qianli Ma, Frank Tietz, Dina
  Fattakhova-Rohlfing, Christian K{\"u}bel, Olivier Guillon, Christian
  Heiliger, and J{\"u}rgen Janek.
\newblock Kinetics and {{Pore Formation}} of the {{Sodium Metal Anode}} on
  {{NASICON}}-{{Type Na}} {\textsubscript{3.4}} {{Zr}} {\textsubscript{2}}
  {{Si}} {\textsubscript{2.4}} {{P}} {\textsubscript{0.6}} {{O}}
  {\textsubscript{12}} for {{Sodium Solid}}-{{State Batteries}}.
\newblock {\em Advanced Energy Materials}, 13(5):2202712, February 2023.

\bibitem{palSpectrumAtomicExcess2021}
Snehanshu Pal, K.~Vijay Reddy, Tingting Yu, Jianwei Xiao, and Chuang Deng.
\newblock The spectrum of atomic excess free volume in grain boundaries.
\newblock {\em Journal of Materials Science}, 56(19):11511--11528, July 2021.

\bibitem{perdewGeneralizedGradientApproximation1996}
John~P. Perdew, Kieron Burke, and Matthias Ernzerhof.
\newblock Generalized {{Gradient Approximation Made Simple}}.
\newblock {\em Physical Review Letters}, 77(18):3865--3868, October 1996.

\bibitem{pfenningerLowRideProcessing2019}
Reto Pfenninger, Michal Struzik, I{\~n}igo Garbayo, Evelyn Stilp, and Jennifer
  L.~M. Rupp.
\newblock A low ride on processing temperature for fast lithium conduction in
  garnet solid-state battery films.
\newblock {\em Nature Energy}, 4(6):475--483, May 2019.

\bibitem{porzMechanismLithiumMetal2017}
Lukas Porz, Tushar Swamy, Brian~W. Sheldon, Daniel Rettenwander, Till
  Fr{\"o}mling, Henry~L. Thaman, Stefan Berendts, Reinhard Uecker, W.~Craig
  Carter, and Yet-Ming Chiang.
\newblock Mechanism of {{Lithium Metal Penetration}} through {{Inorganic Solid
  Electrolytes}}.
\newblock {\em Advanced Energy Materials}, 7(20):1701003, October 2017.

\bibitem{priesterGrainBoundariesTheory2013}
Louisette Priester.
\newblock {\em Grain Boundaries: From Theory to Engineering}.
\newblock Number v. 172 in Springer Series in Materials Science. {Springer},
  {Dordrecht}, 2013.

\bibitem{qianHighRateStable2015}
Jiangfeng Qian, Wesley~A. Henderson, Wu~Xu, Priyanka Bhattacharya, Mark
  Engelhard, Oleg Borodin, and Ji-Guang Zhang.
\newblock High rate and stable cycling of lithium metal anode.
\newblock {\em Nature Communications}, 6(1):6362, February 2015.

\bibitem{quirkDesignPrinciplesGrain2023}
James~A. Quirk and James~A. Dawson.
\newblock Design {{Principles}} for {{Grain Boundaries}} in {{Solid}}-{{State
  Lithium}}-{{Ion Conductors}}.
\newblock {\em Advanced Energy Materials}, page 2301114, July 2023.

\bibitem{raoReviewSynthesisDoping2021}
Y.~Bhaskara Rao, K.~Kamala Bharathi, and L.~N. Patro.
\newblock Review on the synthesis and doping strategies in enhancing the {{Na}}
  ion conductivity of {{Na3Zr2Si2PO12}} ({{NASICON}}) based solid electrolytes.
\newblock {\em Solid State Ionics}, 366--367:115671, August 2021.

\bibitem{ratanaphanGrainBoundaryCharacter2017}
Sutatch Ratanaphan, Dierk Raabe, Rajchawit Sarochawikasit, David~L. Olmsted,
  Gregory~S. Rohrer, and K.~N. Tu.
\newblock Grain boundary character distribution in electroplated nanotwinned
  copper.
\newblock {\em Journal of Materials Science}, 52(7):4070--4085, April 2017.

\bibitem{ratanaphanFiveParameterGrain2014}
Sutatch Ratanaphan, Yohan Yoon, and Gregory~S. Rohrer.
\newblock The five parameter grain boundary character distribution of
  polycrystalline silicon.
\newblock {\em Journal of Materials Science}, 49(14):4938--4945, July 2014.

\bibitem{rayavarapuVariationStructureLi2012}
Prasada~Rao Rayavarapu, Neeraj Sharma, Vanessa~K. Peterson, and Stefan Adams.
\newblock Variation in structure and {{Li}}+-ion migration in argyrodite-type
  {{Li6PS5X}} ({{X}} = {{Cl}}, {{Br}}, {{I}}) solid electrolytes.
\newblock {\em Journal of Solid State Electrochemistry}, 16(5):1807--1813, May
  2012.

\bibitem{riceCeramicFractureModeintergranular1996}
R.~W. Rice.
\newblock Ceramic fracture mode-intergranular vs transgranular fracture.
\newblock (CONF-950739-), December 1996.

\bibitem{ricePoresFractureOrigins1984}
Roy~W. Rice.
\newblock Pores as fracture origins in ceramics.
\newblock {\em Journal of Materials Science}, 19(3):895--914, March 1984.

\bibitem{richardsInterfaceStabilitySolidState2016}
William~D. Richards, Lincoln~J. Miara, Yan Wang, Jae~Chul Kim, and Gerbrand
  Ceder.
\newblock Interface {{Stability}} in {{Solid-State Batteries}}.
\newblock {\em Chemistry of Materials}, 28(1):266--273, January 2016.

\bibitem{rieggerLithiumMetalAnodeInstability2021}
Luise~M. Riegger, Roman Schlem, Joachim Sann, Wolfgang~G. Zeier, and J{\"u}rgen
  Janek.
\newblock Lithium-{{Metal Anode Instability}} of the {{Superionic Halide Solid
  Electrolytes}} and the {{Implications}} for {{Solid-State Batteries}}.
\newblock {\em Angewandte Chemie (International Ed. in English)},
  60(12):6718--6723, March 2021.

\bibitem{rohrerGrainBoundaryEnergy2011}
Gregory~S. Rohrer.
\newblock Grain boundary energy anisotropy: A review.
\newblock {\em Journal of Materials Science}, 46(18):5881--5895, September
  2011.

\bibitem{sebtiStackingFaultsAssist2022}
Elias Sebti, Hayden~A. Evans, Hengning Chen, Peter~M. Richardson, Kelly~M.
  White, Raynald Giovine, Krishna~Prasad Koirala, Yaobin Xu, Eliovardo
  {Gonzalez-Correa}, Chongmin Wang, Craig~M. Brown, Anthony~K. Cheetham,
  Pieremanuele Canepa, and Rapha{\"e}le~J. Cl{\'e}ment.
\newblock Stacking {{Faults Assist Lithium-Ion Conduction}} in a {{Halide-Based
  Superionic Conductor}}.
\newblock {\em Journal of the American Chemical Society}, 144(13):5795--5811,
  April 2022.

\bibitem{seidelPolymorphsNaIon2020}
Stefan Seidel, Wolfgang~G. Zeier, and Rainer P{\"o}ttgen.
\newblock The polymorphs of the {{Na}}+ ion conductor {{Na3PS4}} viewed from
  the perspective of a group-subgroup scheme.
\newblock {\em Zeitschrift f{\"u}r Kristallographie - Crystalline Materials},
  235(1-2):1--6, February 2020.

\bibitem{sharafiControllingCorrelatingEffect2017}
Asma Sharafi, Catherine~G. Haslam, Robert~D. Kerns, Jeff Wolfenstine, and Jeff
  Sakamoto.
\newblock Controlling and correlating the effect of grain size with the
  mechanical and electrochemical properties of {{Li}} {\textsubscript{7}}
  {{La}} {\textsubscript{3}} {{Zr}} {\textsubscript{2}} {{O}}
  {\textsubscript{12}} solid-state electrolyte.
\newblock {\em J. Mater. Chem. A}, 5(40):21491--21504, 2017.

\bibitem{shenderovaAtomisticModelingFracture2000}
O.~A. Shenderova, D.~W. Brenner, A.~Omeltchenko, X.~Su, and L.~H. Yang.
\newblock Atomistic modeling of the fracture of polycrystalline diamond.
\newblock {\em Physical Review B}, 61(6):3877--3888, February 2000.

\bibitem{shvindlermanUnexploredTopicsPotentials2006}
L.S. Shvindlerman and G.~Gottstein.
\newblock Unexplored topics and potentials of grain boundary engineering.
\newblock {\em Scripta Materialia}, 54(6):1041--1045, March 2006.

\bibitem{songProbingOriginElectronic2019}
Yongli Song, Luyi Yang, Lei Tao, Qinghe Zhao, Zijian Wang, Yanhui Cui, Hao Liu,
  Yuan Lin, and Feng Pan.
\newblock Probing into the origin of an electronic conductivity surge in a
  garnet solid-state electrolyte.
\newblock {\em Journal of Materials Chemistry A}, 7(40):22898--22902, 2019.

\bibitem{srivastavaCrystalStructureThermoelectric2015}
D.~Srivastava, F.~Azough, R.~Freer, E.~Combe, R.~Funahashi, D.~M. Kepaptsoglou,
  Q.~M. Ramasse, M.~Molinari, S.~R. Yeandel, J.~D. Baran, and S.~C. Parker.
\newblock Crystal structure and thermoelectric properties of
  {{Sr}}{\textendash}{{Mo}} substituted {{CaMnO3}}: A combined experimental and
  computational study.
\newblock {\em Journal of Materials Chemistry C}, 3(47):12245--12259, November
  2015.

\bibitem{suttonAnalyticModelGrainboundary1991}
A.~P. Sutton.
\newblock An analytic model for grain-boundary expansions and cleavage
  energies.
\newblock {\em Philosophical Magazine A}, 63(4):793--818, April 1991.

\bibitem{symingtonElucidatingNatureGrain2021}
Adam~R. Symington, Marco Molinari, James~A. Dawson, Joel~M. Statham, John
  Purton, Pieremanuele Canepa, and Stephen~C. Parker.
\newblock Elucidating the nature of grain boundary resistance in lithium
  lanthanum titanate.
\newblock {\em Journal of Materials Chemistry A}, 9(10):6487--6498, 2021.

\bibitem{takadaRecentProgressInterfacial2015}
Kazunori Takada, Narumi Ohta, and Yoshitaka Tateyama.
\newblock Recent {{Progress}} in {{Interfacial Nanoarchitectonics}} in
  {{Solid-State Batteries}}.
\newblock {\em Journal of Inorganic and Organometallic Polymers and Materials},
  25(2):205--213, March 2015.

\bibitem{tanTailoringUniformOrdered2021}
Jian Tan, Mingxin Ye, and Jianfeng Shen.
\newblock Tailoring uniform and ordered grain boundaries in the solid
  electrolyte interphase for dendrite-free lithium metal batteries.
\newblock {\em Materials Today Energy}, 22:100858, December 2021.

\bibitem{taskerStabilityIonicCrystal1979}
P~W Tasker.
\newblock The stability of ionic crystal surfaces.
\newblock {\em Journal of Physics C: Solid State Physics}, 12(22):4977--4984,
  November 1979.

\bibitem{terentyevStructureStrength1102010}
D.~Terentyev, X.~He, A.~Serra, and J.~Kuriplach.
\newblock Structure and strength of {$\langle$}110{$\rangle$} tilt grain
  boundaries in bcc {{Fe}}: {{An}} atomistic study.
\newblock {\em Computational Materials Science}, 49(2):419--429, August 2010.

\bibitem{thompsonElectrochemicalWindowLiIon2017}
Travis Thompson, Seungho Yu, Logan Williams, Robert~D. Schmidt, Regina
  {Garcia-Mendez}, Jeff Wolfenstine, Jan~L. Allen, Emmanouil Kioupakis,
  Donald~J. Siegel, and Jeff Sakamoto.
\newblock Electrochemical {{Window}} of the {{Li-Ion Solid Electrolyte
  Li7La3Zr2O12}}.
\newblock {\em ACS Energy Letters}, 2(2):462--468, February 2017.

\bibitem{tianInterfacialElectronicProperties2019}
Hong-Kang Tian, Zhe Liu, Yanzhou Ji, Long-Qing Chen, and Yue Qi.
\newblock Interfacial {{Electronic Properties Dictate Li Dendrite Growth}} in
  {{Solid Electrolytes}}.
\newblock {\em Chemistry of Materials}, 31(18):7351--7359, September 2019.

\bibitem{tingdongElasticModulusGrainboundary2004}
Xu~Tingdong and Zheng Lei.
\newblock The elastic modulus in the grain-boundary region of polycrystalline
  materials.
\newblock {\em Philosophical Magazine Letters}, 84(4):225--233, April 2004.

\bibitem{tromansFractureToughnessSurface2004}
D.~Tromans and J.~A. Meech.
\newblock Fracture toughness and surface energies of covalent minerals:
  Theoretical estimates.
\newblock {\em Minerals Engineering}, 17(1):1--15, January 2004.

\bibitem{uematsuPreparationNa3PS4Electrolyte2018}
Miwa Uematsu, So~Yubuchi, Kousuke Noi, Atsushi Sakuda, Akitoshi Hayashi, and
  Masahiro Tatsumisago.
\newblock Preparation of {{Na3PS4}} electrolyte by liquid-phase process using
  ether.
\newblock {\em Solid State Ionics}, 320:33--37, July 2018.

\bibitem{uesugiFirstprinciplesCalculationGrain2011}
Tokuteru Uesugi and Kenji Higashi.
\newblock First-principles calculation of grain boundary energy and grain
  boundary excess free volume in aluminum: Role of grain boundary elastic
  energy.
\newblock {\em Journal of Materials Science}, 46(12):4199--4205, June 2011.

\bibitem{vahidiReviewGrainBoundary2021}
Hasti Vahidi, Komal Syed, Huiming Guo, Xin Wang, Jenna~Laurice Wardini, Jenny
  Martinez, and William~John Bowman.
\newblock A {{Review}} of {{Grain Boundary}} and {{Heterointerface
  Characterization}} in {{Polycrystalline Oxides}} by ({{Scanning}})
  {{Transmission Electron Microscopy}}.
\newblock {\em Crystals}, 11(8):878, August 2021.

\bibitem{vannoordenRechargeableRevolutionBetter2014}
Richard Van~Noorden.
\newblock The rechargeable revolution: {{A}} better battery.
\newblock {\em Nature}, 507(7490):26--28, March 2014.

\bibitem{vishnugopiMesoscaleInterrogationReveals2022}
Bairav~S. Vishnugopi, Marm~B. Dixit, Feng Hao, Badri Shyam, John~B. Cook,
  Kelsey~B. Hatzell, and Partha~P. Mukherjee.
\newblock Mesoscale {{Interrogation Reveals Mechanistic Origins}} of {{Lithium
  Filaments}} along {{Grain Boundaries}} in {{Inorganic Solid Electrolytes}}.
\newblock {\em Advanced Energy Materials}, 12(3):2102825, January 2022.

\bibitem{vonalfthanStructureGrainBoundaries2010}
Sebastian {von Alfthan}, Nicole~A. Benedek, Lin Chen, Alvin Chua, David
  Cockayne, Karleen~J. Dudeck, Christian Els{\"a}sser, Michael~W. Finnis,
  Christoph~T. Koch, Behnaz Rahmati, Manfred R{\"u}hle, Shao-Ju Shih, and
  Adrian~P. Sutton.
\newblock The {{Structure}} of {{Grain Boundaries}} in {{Strontium Titanate}}:
  {{Theory}}, {{Simulation}}, and {{Electron Microscopy}}.
\newblock {\em Annual Review of Materials Research}, 40(1):557--599, June 2010.

\bibitem{wangStructuralElasticElectronic2020}
Jingjing Wang, Minghua Deng, Yunhong Chen, Xinghong Liu, Wenyan Ke, Dandan Li,
  Wei Dai, and Kaihua He.
\newblock Structural, elastic, electronic and optical properties of lithium
  halides ({{LiF}}, {{LiCl}}, {{LiBr}}, and {{LiI}}): {{First-principle}}
  calculations.
\newblock {\em Materials Chemistry and Physics}, 244:122733, April 2020.

\bibitem{wangResistiveNatureDecomposing2022}
Juefan Wang, Abhishek~A. Panchal, Gopalakrishnan Sai~Gautam, and Pieremanuele
  Canepa.
\newblock The resistive nature of decomposing interfaces of solid electrolytes
  with alkali metal electrodes.
\newblock {\em Journal of Materials Chemistry A}, 10(37):19732--19742, 2022.

\bibitem{wangLithiumChloridesBromides2019}
Shuo Wang, Qiang Bai, Adelaide~M. Nolan, Yunsheng Liu, Sheng Gong, Qiang Sun,
  and Yifei Mo.
\newblock Lithium {{Chlorides}} and {{Bromides}} as {{Promising Solid}}-{{State
  Chemistries}} for {{Fast Ion Conductors}} with {{Good Electrochemical
  Stability}}.
\newblock {\em Angewandte Chemie International Edition}, 58(24):8039--8043,
  June 2019.

\bibitem{wenzelInterfacialReactivityInterphase2018}
Sebastian Wenzel, Stefan~J. Sedlmaier, Christian Dietrich, Wolfgang~G. Zeier,
  and Juergen Janek.
\newblock Interfacial reactivity and interphase growth of argyrodite solid
  electrolytes at lithium metal electrodes.
\newblock {\em Solid State Ionics}, 318:102--112, May 2018.

\bibitem{wolfenstineMechanicalBehaviorLiionconducting2018}
Jeff Wolfenstine, Jan~L. Allen, Jeff Sakamoto, Donald~J. Siegel, and Heeman
  Choe.
\newblock Mechanical behavior of {{Li-ion-conducting}} crystalline oxide-based
  solid electrolytes: A brief review.
\newblock {\em Ionics}, 24(5):1271--1276, May 2018.

\bibitem{wolfenstineMechanicalPropertiesNaSICON2023}
Jeff Wolfenstine, Wooseok Go, Youngsik Kim, and Jeff Sakamoto.
\newblock Mechanical properties of {{NaSICON}}: A brief review.
\newblock {\em Ionics}, 29(1):1--8, January 2023.

\bibitem{wolfenstinePreliminaryInvestigationFracture2013}
Jeff Wolfenstine, Hyungyung Jo, Yong-Hun Cho, Isabel~N. David, Per Askeland,
  Eldon~D. Case, Hyunjoong Kim, Heeman Choe, and Jeff Sakamoto.
\newblock A preliminary investigation of fracture toughness of {{Li7La3Zr2O12}}
  and its comparison to other solid {{Li-ionconductors}}.
\newblock {\em Materials Letters}, 96:117--120, April 2013.

\bibitem{yeandelImpactTiltGrain2018}
Stephen~R. Yeandel, Marco Molinari, and Stephen~C. Parker.
\newblock The impact of tilt grain boundaries on the thermal transport in
  perovskite {{SrTiO3}} layered nanostructures. {{A}} computational study.
\newblock {\em Nanoscale}, 10(31):15010--15022, August 2018.

\bibitem{yeheskelElasticModuliGrain2005}
Ori Yeheskel, Rachman Chaim, Zhijian Shen, and Mats Nygren.
\newblock Elastic moduli of grain boundaries in nanocrystalline {{MgO}}
  ceramics.
\newblock {\em Journal of Materials Research}, 20(3):719--725, March 2005.

\bibitem{yuGrainBoundaryContributions2017}
Seungho Yu and Donald~J. Siegel.
\newblock Grain {{Boundary Contributions}} to {{Li-Ion Transport}} in the
  {{Solid Electrolyte Li7La3Zr2O12}} ({{LLZO}}).
\newblock {\em Chemistry of Materials}, 29(22):9639--9647, November 2017.

\bibitem{yuGrainBoundarySoftening2018}
Seungho Yu and Donald~J. Siegel.
\newblock Grain {{Boundary Softening}}: {{A Potential Mechanism}} for {{Lithium
  Metal Penetration}} through {{Stiff Solid Electrolytes}}.
\newblock {\em ACS Applied Materials \& Interfaces}, 10(44):38151--38158,
  November 2018.

\bibitem{yuanCoupledCrackPropagation2021}
Chunhao Yuan, Xiang Gao, Yikai Jia, Wen Zhang, Qingliu Wu, and Jun Xu.
\newblock Coupled crack propagation and dendrite growth in solid electrolyte of
  all-solid-state battery.
\newblock {\em Nano Energy}, 86:106057, August 2021.

\bibitem{zhangLithiumWhiskerGrowth2020}
Liqiang Zhang, Tingting Yang, Congcong Du, Qiunan Liu, Yushu Tang, Jun Zhao,
  Baolin Wang, Tianwu Chen, Yong Sun, Peng Jia, Hui Li, Lin Geng, Jingzhao
  Chen, Hongjun Ye, Zaifa Wang, Yanshuai Li, Haiming Sun, Xiaomei Li, Qiushi
  Dai, Yongfu Tang, Qiuming Peng, Tongde Shen, Sulin Zhang, Ting Zhu, and
  Jianyu Huang.
\newblock Lithium whisker growth and stress generation in an in situ atomic
  force microscope{\textendash}environmental transmission electron microscope
  set-up.
\newblock {\em Nature Nanotechnology}, 15(2):94--98, February 2020.

\bibitem{zhangFirstprinciplesDeterminationEffect2011}
Shengjun Zhang, Oleg~Y. Kontsevoi, Arthur~J. Freeman, and Gregory~B. Olson.
\newblock First-principles determination of the effect of boron on aluminum
  grain boundary cohesion.
\newblock {\em Physical Review B}, 84(13):134104, October 2011.

\bibitem{zhangElasticStiffnessGrain1992}
Tong-Yi Zhang and J.~E. Hack.
\newblock {On the elastic stiffness of grain boundaries}.
\newblock {\em Physica Status Solidi (a)}, 131(2):437--443, June 1992.

\bibitem{zhangDurableSafeSolidstate2018}
Wenqiang Zhang, Jinhui Nie, Fan Li, Zhong~Lin Wang, and Chunwen Sun.
\newblock A durable and safe solid-state lithium battery with a hybrid
  electrolyte membrane.
\newblock {\em Nano Energy}, 45:413--419, March 2018.

\bibitem{zhaoHighLithiumIonic2020}
Guowei Zhao, Kota Suzuki, Tomoaki Seki, Xueying Sun, Masaaki Hirayama, and
  Ryoji Kanno.
\newblock High lithium ionic conductivity of {$\gamma$}-{{Li3PO4-type}} solid
  electrolytes in {{Li4GeO4}}-{{Li4SiO4}}{\textendash}{{Li3VO4}} quasi-ternary
  system.
\newblock {\em Journal of Solid State Chemistry}, 292:121651, December 2020.

\bibitem{zhengMobilePinnedGrain2020}
Fangyuan Zheng, Lingli Huang, Lok-Wing Wong, Jin Han, Yuan Cai, Ning Wang,
  Qingming Deng, Thuc~Hue Ly, and Jiong Zhao.
\newblock The {{Mobile}} and {{Pinned Grain Boundaries}} in {{2D Monoclinic
  Rhenium Disulfide}}.
\newblock {\em Advanced Science}, 7(22):2001742, November 2020.

\bibitem{zhuUnderstandingEvolutionLithium2023}
Chao Zhu, Till Fuchs, Stefan A.~L. Weber, Felix~H. Richter, Gunnar Glasser,
  Franjo Weber, Hans-J{\"u}rgen Butt, J{\"u}rgen Janek, and R{\"u}diger Berger.
\newblock Understanding the evolution of lithium dendrites at
  {{Li6}}.{{25Al0}}.{{25La3Zr2O12}} grain boundaries via operando microscopy
  techniques.
\newblock {\em Nature Communications}, 14(1):1300, March 2023.

\bibitem{zhuOriginOutstandingStability2015}
Yizhou Zhu, Xingfeng He, and Yifei Mo.
\newblock Origin of {{Outstanding Stability}} in the {{Lithium Solid
  Electrolyte Materials}}: {{Insights}} from {{Thermodynamic Analyses Based}}
  on {{First-Principles Calculations}}.
\newblock {\em ACS applied materials \& interfaces}, 7(42):23685--23693,
  October 2015.

\end{thebibliography}

\end{document}